\documentclass[prb,twocolumn,amsmath,aps,byrevtex,endfloats,groupaddress,subeqn]{revtex4}
\usepackage{graphicx}
\usepackage{epsfig}
\usepackage[cp1251]{inputenc}
\usepackage[english]{babel}
\usepackage{amsmath}
\usepackage{amssymb}
\usepackage{amstext}
\usepackage{color}
\usepackage[toc,page]{appendix}
\usepackage{comment}

\begin{document}

\title{Quantum transport and the Wigner distribution function for Bloch electrons in spatially homogeneous electric and magnetic fields}

\author{G. J. Iafrate}
\affiliation{Department of Electrical and Computer Engineering,
North Carolina State University, Raleigh, North Carolina, USA}
\author{V. N. Sokolov}
\affiliation{Department of Theoretical Physics,
Institute of Semiconductor Physics, NASU, Kiev, Ukraine.}
\author{J. B. Krieger}
\affiliation{Department of Physics,
Brooklyn College, CUNY, Brooklyn, New York, USA}

\begin{abstract}
The theory of Bloch electron dynamics for carriers in homogeneous electric and magnetic fields of arbitrary time dependence is developed in the framework of the Liouville equation. The Wigner distribution function (WDF) is determined from the single particle density matrix in the ballistic regime, i.e., collision effects are excluded. The single particle transport equation is established with the electric field described in the vector potential gauge, and the magnetic field is treated in the symmetric gauge.

The general approach is to employ the accelerated Bloch state representation (ABR) as a basis so that the dependence upon the electric field, including multiband Zener tunneling, is treated exactly. In the formulation of the WDF, we transform to a new set of variables so that the final WDF is gauge invariant and is expressed explicitly in terms of the position, kinetic momentum, and time.

The methodology for developing the WDF is illustrated by deriving the exact WDF equation for free electrons in homogeneous electric and magnetic fields. The methodology is then extended to the case of electrons described by an effective Hamiltonian corresponding to an arbitrary energy band function. In treating the problem of Bloch electrons in a periodic potential, the methodology for deriving the WDF reveals a multiband character due to the inherent nature of the Bloch states. In examining the single-band WDF, it is found that the collisionless WDF equation matches the equivalent Boltzmann transport equation to first order in the magnetic field. These results are necessarily extended to second order in the magnetic field by employing a unitary transformation that diagonalizes the Hamiltonian using the ABR to second order. The work includes a discussion of the multiband WDF transport analysis and the identification of the combined Zener-magnetic field induced tunneling.

\end{abstract}


\maketitle

\section{Introduction}\label{Sec I}
A central problem in the theory of solids is the question of how to construct the correct quantum-mechanical transport equation for charge carriers under the influence of both electric ${\bf E}$ and magnetic ${\bf B}$ fields.
The first attempt at doing so was provided by Bloch in his fundamental paper~\cite{Bloch} on the application of the quantum theory to transport phenomena. Bloch showed that electrons moving in solids could be treated as quasiparticles having an altered energy-momentum relation, $\varepsilon({\bf k})$, different from the usual free-electron dispersion relation. The corresponding velocity of the electron wave packet constructed from a superposition of states from a single band centered at wave vector ${\bf k}$ is given by
\begin{equation}\label{I1}
  {\bf v}({\bf k}) = \hbar^{-1} \nabla_{{\bf k}} \varepsilon({\bf k}) \,,
\end{equation}
which reduces to the usual expression for the velocity in the free-electron limit.
Bloch then argued that the correct transport equation is the Boltzmann transport equation (BTE) where ${\bf v}$ is given by Eq.~(\ref{I1}) and the scattering rates are calculated quantum mechanically using Fermi's {\it golden rule} of time-dependent perturbation theory. Thus, the BTE for the single-particle distribution function should be written as
\begin{equation}\label{I2}
  \frac{\partial f}{\partial t} + {\bf v}\cdot \nabla_{{\bf x}}f + \hbar^{-1}{\bf F}\cdot \nabla_{{\bf k}}f = \left(\frac{\partial f}{\partial t} \right)_{c}   \,,
\end{equation}
where ${\bf F}$ is given by the Lorentz force
\begin{equation}\label{I3}
  {\bf F} = e\left({\bf E} + \frac{1}{c}{\bf v}\times{\bf B}\right)   \,,
\end{equation}
and ${\bf v}$ is given in Eq.~(\ref{I1}); also, $e$ is the electron charge, $c$ is the speed of light, $\hbar$ is the reduced Planck constant,
$t$ is the time, ${\bf x}$ is the position vector, and $(\partial f/\partial t)_{c}$ represents the collision integral. Equation (\ref{I2}) has been very successful in providing the basic theoretical framework for analyzing low-field galvanomagnetic effects in semiconductors,~\cite{Beer} although its justification is based only on quasiclassical considerations, i.e., it is not directly derived from the fundamental equations of quantum mechanics, but rather from the employment of classical dynamics applied to electron quasiparticles.

Nearly three decades after Bloch's seminal work, the quantum-mechanical derivation of Eq.~(\ref{I2}) from the Liouville equation for the density matrix was given by Kohn and Luttinger~\cite{KohnLatt} for electrons scattered by impurities in a weak uniform electric field; this work was later extended to the case of phonon scattering.~\cite{Argyres} This density matrix approach was generalized to higher electric fields by Levinson~\cite{Levinson} and by Barker,~\cite{Barker} using the effective-mass approximation with the resulting inclusion of the intracollisional field effect. Subsequently, Krieger and Iafrate~\cite{Krieger} have extended the previous results of Levinson~\cite{Levinson} and Calecki and Pottier~\cite{Calecki} to the multiband case for arbitrary electron energy dispersion relations including the effects of Zener tunneling, and demonstrated that the scattering matrix elements entering the collision term are both field and time dependent.

The formal justification of Eq.~(\ref{I2}) for non-zero magnetic field has not been so straightforward. It is well known that even for free electrons, the electron energy eigenvalues are quantized in steps  of $\hbar \omega_c$ (Landau levels) where $\omega_c = eB/mc$ is the electron cyclotron frequency. And, in the high-field limit, i.e., $\hbar \omega_c \gg k_BT$ ($k_B$ is the Boltzmann constant, $T$ is ambient temperature), the electron distribution function can be expected to change significantly from one quantum level to the next with experimentally detectable
consequences. This is the well-known origin of the de Haas-van Alphen oscillations in the magnetic susceptibility of metals in zero electric fields.
In an effort to take into account the existence of quantizing magnetic fields in the transport phenomena for free electrons, Adams and
Holstein~\cite{Adams} employed as a basis the harmonic oscillator states corresponding to Landau levels. They then showed how the
current could be calculated from the Liouville equation in this representation. This formulation significantly departs from the quasiclassical
description given by Eq.~(\ref{I2}) in that the current depends on the off-diagonal elements of the density matrix.

Rhetorically, the questions arise as to whether Eq.~(\ref{I2}) is valid for relatively high magnetic fields, and, further, is it possible to derive a transport equation for a quantum distribution function which is defined within a classical phase space picture. In zero magnetic field, Krieger, Kiselev, and Iafrate~\cite{Kiselev} made use of the effective Hamiltonian to derive the Wigner distribution function (WDF) equation for a random distribution of impurities. However, if one wants to go beyond an effective Hamiltonian approach for Bloch electron dynamics in the electric and magnetic fields
so that both interband tunneling and the proper field- and time-dependent scattering matrix elements may be included as previously done by Krieger and Iafrate~\cite{Krieger} for the case of electric field alone, then a description in terms of a classical phase space WDF approach would be desirable since the exact solution to the Schr\"{o}dinger equation for a Bloch electron in electric and magnetic fields is not known. Therefore, we are motivated~\cite{Krieger1} to address this subject.

In this paper, the theory of Bloch electron transport in homogeneous electric and magnetic fields of arbitrary time dependence is developed within the framework of the Liouville equation. The phase space WDF is determined from the single-particle density matrix within the ballistic regime, i.e., collision effects are excluded, although the methodology for including such effects is straightforward. The electric field is treated in the vector potential gauge and the magnetic field is described in the symmetric gauge. No specific assumptions are adopted concerning the form of the initial distribution in momentum or configuration space. The general approach is to utilize the accelerated Bloch state representation (ABR) as a basis so that the dependence upon the electric field, including the multiband Zener tunneling, is treated exactly. In the formulation of the WDF, we transform to a set of variables based on position, kinetic momentum, and time to insure the gauge invariance of the WDF in our problem.

In Sec. II, the methodology for developing the WDF is described and illustrated by deriving the exact WDF equation for free electrons in homogeneous electric and magnetic fields resulting in the same form as given by the collisionless BTE. In Sec. III, the methodology is extended to the case of electrons described by an effective Hamiltonian corresponding to an arbitrary energy-band function. The exact equation for the WDF is obtained in this case and is shown to approximate the free-electron results when taken to second order in the magnetic field. As a corollary, it is shown that if the WDF of Secs. II and III is a wave packet, then the time rate of change of the electron quasimomentum is given by the Lorentz force. In Sec. IV, the problem of Bloch electrons in a crystal potential in the presence of electric and magnetic fields is treated. The methodology for deriving the WDF reveals a multiband structure due to the inherent nature of the Bloch states. Use is made of the so-called $``$${\bf K}_0$-representation$"$ outlined in Appendix A to express the multiband WDF in a user friendly form.
In order to obtain results beyond first order in the magnetic field, it is necessary to employ a unitary transformation that diagonalizes the Hamiltonian using the ABR to second order in the magnetic field. The unitary transformation process results in an analysis for the single-band WDF equation and a discussion of multiband transport properties leading to the identification of a combined Zener magnetic field induced tunneling. Results also include the explicit development of the multiband WDF to first order in $({\bf K} - {\bf K}_0)$.


\section{Dynamics of free electrons in homogeneous electric and magnetic fields}\label{Sec II}

It has been established previously~\cite{Krieger} that given an arbitrary initial distribution at $t = t_0$, when the fields are turned on, the
equation for the density matrix operator $\hat{f}(t)$ may be written as
\begin{equation}\label{Sec II 4}
  i\hbar\frac{\partial \hat{f}}{\partial t} - [\hat{H},\hat{f}] = C_s\{\hat{f}(t)\}   \,,
\end{equation}
where $C_s\{\hat{f}(t)\}$ involves the scattering Hamiltonian and $\hat{H}$ is the Hamiltonian in the absence of scattering. Equation (\ref{Sec II 4}) was first derived by Levinson~\cite{Levinson} for the case in which electrons initially in thermal equilibrium are interacting with phonons.
Our extension~\cite{Krieger} of his result permits the use of Eq.~(\ref{Sec II 4}) even for initial {\it nonequilibrium distributions and multiband dynamics}.

For free electrons interacting with spatially homogeneous, but arbitrarily time-dependent, electric and magnetic fields, the Hamiltonian is
\begin{equation}\label{Sec II 5}
  \hat{H} = \frac{1}{2m}[{\bf p} - \frac{e}{c}{\bf A}({\bf x},t)]^2    \,,
\end{equation}
where $m$ is the free-electron mass and ${\bf p}$ is the electron momentum. The vector potential ${\bf A}({\bf x},t)$ includes the electric and magnetic field contributions
\begin{equation}\label{Sec II 6}
  {\bf A}({\bf x},t) = {\bf A}_1(t) + {\bf A}_2({\bf x},t)    \,,
\end{equation}
with
\begin{equation}\label{Sec II 7}
  {\bf A}_1(t) = - c \int_0^t {\bf E}(t^{\prime})dt^{\prime} \,.
\end{equation}
Further, for the magnetic vector potential, we choose the symmetric vector potential gauge
\begin{equation}\label{Sec II 8}
  {\bf A}_2({\bf x},t) = \frac{1}{2}{\bf B}(t)\times {\bf x}
\end{equation}
to describe the spatially homogeneous magnetic field ${\bf B}(t) = {\bf \nabla} \times {\bf A}_2({\bf x},t)$. Thus, using Eqs.~(\ref{Sec II 5})  and (\ref{Sec II 6}) and expanding the kinetic term while noting that ${\bf p}$ and ${\bf A}_2$ commute, we see that
\begin{equation}\label{Sec II 9}
 \hat{H} = \hat{H}_0 -\frac{e}{mc}{\bf A}_2 \cdot ({\bf p} - \frac{e}{c}{\bf A}_1) + \frac{e^2}{2mc^2}{\bf A}_2^2  \equiv \hat{H}_0 + V_1 + V_2  \,,
\end{equation}
where the Hamiltonian
\begin{equation}\label{Sec II 10}
 \hat{H}_0 = \frac{1}{2m}[{\bf p} - \frac{e}{c}{\bf A}_1(t)]^2
\end{equation}
describes the free electron in the electric field alone, and the next two terms in Eq.~(\ref{Sec II 9}), $V_1$ and $V_2$, are first and second order in the magnetic field.

To adopt an appropriate basis set with which to evaluate Eq.~(\ref{Sec II 4}), we see in Eqs.~(\ref{Sec II 9}) and (\ref{Sec II 10}) that a natural basis to proceed would be the accelerated state representation which are the instantaneous eigenstates of $\hat{H}_0$ in Eq.~(\ref{Sec II 10}). As such, the accelerated states are
\begin{subequations}
\begin{equation}\label{Sec II 11a}
  \psi_{\bf K}({\bf x}) = \Omega^{-1/2} e^{i{\bf K}\cdot{\bf x}} \equiv |{\bf K} \rangle  \,,
\end{equation}
with eigenvalues
 $\varepsilon^0({\bf k}(t)) = \hbar^2 k^2(t)/2m$,
and where ${\bf k}(t)$ is the time-dependent wave vector due to acceleration by the electric field,
\begin{equation}\label{Sec II 11c}
  {\bf k}(t) = {\bf K} - \frac{e}{\hbar c} {\bf A}_1(t) \equiv {\bf K} + {\bf k}_c(t)   \,,
\end{equation}
\end{subequations}
with ${\bf k}_c(t) = (e/\hbar)\int_0^t {\bf E}(t^{\prime})dt^{\prime}$. Here, $\Omega$ is the normalization volume, $k = |{\bf k}|$, and ${\bf K}$ is chosen such that $\psi_{\bf K}({\bf x})$ satisfies periodic boundary conditions.
It is noted that the choice of $|{\bf K} \rangle$ as a basis allows us to work in a representation in which the electron motion is indexed by the momentum ${\bf p} =  \hbar{\bf K}$;  if we had chosen to work in the representation based on the instantaneous eigenstates of the full Hamiltonian in Eq.~(\ref{Sec II 5}), we would then have oscillator states which are inconvenient in that they are not eigenfunctions of the momentum operator.

In obtaining the WDF for the density matrix, we note that the WDF, $f({\bf x},{\bf p})$, is fundamentally defined~\cite{Imre} as the off-diagonal matrix elements  of the density matrix operator, $\hat{f}(t)$, in Eq.~(\ref{Sec II 4}). As such,
\begin{equation}\label{Sec II 12}
 f({\bf x},{\bf p}) = (2\pi\hbar)^{-3} \int d{\bf y} \langle{\bf x} - {\bf y}/2|\hat{f}|{\bf x} + {\bf y}/2\rangle
 e^{i{\bf p}\cdot{\bf y}/\hbar}  \,.
\end{equation}
Then, for the complete set of basis states defined in Eq.~(\ref{Sec II 11a}), we see that Eq.~(\ref{Sec II 12}) can be expressed as
\begin{eqnarray}\label{Sec II 13}
 \nonumber
 f({\bf x},{\bf p}) = \sum_{{\bf K}_1 {\bf K}_2} \langle{\bf K}_1|\hat{f}|{\bf K}_2\rangle (2\pi\hbar)^{-3}\times  \\
  \int d{\bf y} \psi_{{\bf K}_2}^{\ast}({\bf x} + {\bf y}/2) \psi_{{\bf K}_1}({\bf x} - {\bf y}/2) e^{i{\bf p}\cdot{\bf y}/\hbar}  \,.
\end{eqnarray}
Using the explicit spectral dependence for $\psi_{\bf K}({\bf x})$ in the integral over ${\bf y}$ of Eq.~(\ref{Sec II 13}), and utilizing
$(2\pi\hbar)^{-3} \int d{\bf y} e^{i{\bf p}\cdot{\bf y}/\hbar} = \delta({\bf p})$, we obtain
\begin{eqnarray}\label{Sec II 14}
 \nonumber
 f({\bf x},{\bf p}) = \Omega^{-1} \sum_{{\bf K}_1 {\bf K}_2} e^{i({\bf K}_1 - {\bf K}_2)\cdot{\bf x}} \langle{\bf K}_1|\hat{f}|{\bf K}_2\rangle \\ \times \delta[{\bf p} - \hbar ({\bf K}_1 + {\bf K}_2)/2]   \,.
\end{eqnarray}
Letting
\begin{equation}\label{Sec II 15}
 {\bf K}_1 = {\bf K} + \frac{{\bf u}}{2},\;\;\;\;   {\bf K}_2 = {\bf K} - \frac{{\bf u}}{2}
\end{equation}
in Eq.~(\ref{Sec II 14}), the WDF becomes
\begin{subequations}
\begin{equation}\label{Sec II 16a}
 f({\bf x},{\bf p}) = \sum_{{\bf K}} f^0({\bf x},{\bf K}) \delta({\bf p} - \hbar{\bf K}) = f^0({\bf x},{\bf K})|_{{\bf p} = \hbar{\bf K}}   \,,
\end{equation}
where
\begin{equation}\label{Sec II 16b}
 f^0({\bf x},{\bf K}) = \Omega^{-1} \sum_{{\bf u}} \langle {\bf K} + \frac{{\bf u}}{2} |\hat{f}| {\bf K} - \frac{{\bf u}}{2} \rangle   e^{i{\bf u} \cdot {\bf x}} \,;
\end{equation}
\end{subequations}
here, ${\bf p} =  \hbar{\bf K}$, and $\langle{\bf K}_1|\hat{f}|{\bf K}_2\rangle$ are the momentum matrix elements of the density matrix operator in Eq.~(\ref{Sec II 4}) evaluated at ${\bf K}_1, {\bf K}_2$ of Eq.~(\ref{Sec II 15}). Finally, in Eq.~(\ref{Sec II 16b}), we make the change of variables~\cite{Serimea}
\begin{equation}\label{Sec II 17}
 {\bf k}({\bf x},t) = {\bf K} - \frac{e}{\hbar c}{\bf A}({\bf x},t)    \,,
\end{equation}
where the Jacobian of the transformation from ${\bf k}$ to ${\bf K}$ is unity, and where $\hbar {\bf k}({\bf x},t)$ is the ({\bf x},t)-dependent kinetic momentum; then we obtain
\begin{equation}\label{Sec II 18}
 F({\bf x},{\bf k},t) \equiv f({\bf x},{\bf K},t)    \,,
\end{equation}
the gauge invariant WDF~\cite{Levinson,Serimea} in the ${\bf k}$ representation for an electron subjected to the vector potential ${\bf A}({\bf x},t)$ of Eq.~(\ref{Sec II 6}).

For the Hamiltonian of Eq.~(\ref{Sec II 9}), we determine the equation of motion for the WDF as outlined in Eqs.~(\ref{Sec II 13})-(\ref{Sec II 18}) considering, for simplicity, the case of ballistic or collisionless transport for which $C_s\{\hat{f}\} \equiv 0$ in Eq.~(\ref{Sec II 4}). Then, we basically start with
\begin{equation}\label{Sec II 19}
 i\hbar \langle {\bf K}_1 |\frac{\partial\hat{f}}{\partial t}| {\bf K}_2 \rangle = \langle {\bf K}_1 |[\hat{H},\hat{f}]| {\bf K}_2 \rangle   \,.
\end{equation}
Since $|{\bf K} \rangle$ is independent of time, we note that $\langle {\bf K}_1 |\partial\hat{f}/\partial t| {\bf K}_2 \rangle =
\partial \langle {\bf K}_1 |\hat{f}| {\bf K}_2 \rangle /\partial t$; as well the Hamiltonian in Eq.~(\ref{Sec II 9}) is reexpressed as
\begin{equation}\label{Sec II 20}
 \hat{H} = \hat{H}_0 - \frac{e}{2mc}({\bf p} - \frac{e}{c}{\bf A}_1) \cdot ({\bf B} \times {\bf x}) +
 \frac{e^2}{8mc^2} ({\bf B}\times {\bf x})^2   \,.
\end{equation}
Using the vector identity ${\bf a}\cdot({\bf b}\times {\bf c}) = ({\bf a}\times {\bf b})\cdot {\bf c}$ in the second term, we can write
\begin{equation}\label{Sec II 22}
 \hat{H} = \hat{H}_0 - \frac{e}{2mc} [({\bf p} - \frac{e}{c} {\bf A}_1)\times {\bf B}] \cdot {\bf x} +
 \frac{e^2}{8mc^2} ({\bf B}\times {\bf x})^2   \,.
\end{equation}
Noting that $({\bf p} - (e/c) {\bf A}) | {\bf K}\rangle = (\hbar{\bf K} - (e/c){\bf A}_1) | {\bf K}\rangle$
and ${\bf x} |{\bf K}\rangle =\frac{1}{i} \nabla_{\bf K} |{\bf K}\rangle$,
we obtain the term on right-hand side of Eq.~(\ref{Sec II 19}) as
 \begin{equation}\label{Sec II 24a}
  \langle {\bf K}_1 |[\hat{H},\hat{f}]| {\bf K}_2 \rangle = \big[\varepsilon_{+}({\bf K}_1) - \varepsilon_{-}({\bf K}_2)\big] f({\bf K}_1,{\bf K}_2,t) \,,
 \end{equation}
where
 \begin{eqnarray}\label{Sec II 24b}
  \nonumber
  \varepsilon({\bf K})_{\pm} = \frac{1}{2m} \Big\{(\hbar{\bf K} - \frac{e}{c}{\bf A}_1)^2 \pm  \\
  \frac{e}{c}[(\hbar{\bf K} - \frac{e}{c}{\bf A}_1)\times{\bf B}] \cdot \frac{1}{i}\nabla_{{\bf K}}
  + \frac{e^2}{4c^2}({\bf B} \times \frac{1}{i}\nabla_{{\bf K}})^2 \Big\} \nonumber
 \end{eqnarray}
and $f({\bf K}_1,{\bf K}_2,t) \equiv \langle {\bf K}_1 |\hat{f}(t)| {\bf K}_2 \rangle$.

Next, we change the variables as prescribed by Eq.~(\ref{Sec II 15}) while noting that
\begin{equation}\label{Sec II 25a}
  \nabla_{{\bf K}_1} f({\bf K}_1,{\bf K}_2,t) = (\frac{1}{2}\nabla_{\bf K}
  + \nabla_{\bf u}) f({\bf K} + \frac{\bf u}{2},{\bf K} - \frac{\bf u}{2},t),  \nonumber
\end{equation}
\begin{equation}\label{Sec II 25b}
  \nabla_{{\bf K}_2} f({\bf K}_1,{\bf K}_2,t) = (\frac{1}{2}\nabla_{\bf K}
  - \nabla_{\bf u}) f({\bf K} + \frac{\bf u}{2},{\bf K} - \frac{\bf u}{2},t)       \,.
\end{equation}
Thus, using (\ref{Sec II 15}) and (\ref{Sec II 25b}), Eq.~(\ref{Sec II 24a}) becomes
\begin{multline}\label{Sec II 26}
  \langle {\bf K} + \frac{\bf u}{2} |[\hat{H},\hat{f}]| {\bf K} - \frac{\bf u}{2} \rangle = \frac{1}{m} \Big\{(\hbar{\bf K} - \frac{e}{c}
  {\bf A}_1)\cdot \hbar {\bf u} + \\
  \frac{e}{2c}[(\hbar{\bf K} - \frac{e}{c}{\bf A}_1)\times{\bf B}] \cdot \frac{1}{i}\nabla_{{\bf K}} + \frac{e}{2c}(\hbar {\bf u}\times{\bf B}) \cdot \frac{1}{i}\nabla_{{\bf u}} +  \\
  \frac{e^2}{4c^2}({\bf B} \times \frac{1}{i}\nabla_{{\bf K}}) \cdot ({\bf B} \times \frac{1}{i}\nabla_{{\bf u}})
  \Big\} f({\bf K} + \frac{\bf u}{2},{\bf K} - \frac{\bf u}{2},t)   \,.
\end{multline}
Noting that
  $({\bf B} \times \nabla_{{\bf K}}) \cdot ({\bf B} \times \nabla_{{\bf u}})
  = [({\bf B} \times \nabla_{{\bf u}}) \times {\bf B}] \cdot \nabla_{{\bf K}}$
and  $({\bf u} \times {\bf B}) \cdot \nabla_{{\bf u}} = {\bf u} \cdot ({\bf B} \times \nabla_{{\bf u}})$,
then Eq.~(\ref{Sec II 26}) becomes
\begin{multline}\label{Sec II 28}
  \langle {\bf K} + \frac{\bf u}{2} |[\hat{H},\hat{f}]| {\bf K} - \frac{\bf u}{2} \rangle =
  \frac{1}{m} \Big\{\hbar {\bf u}\cdot (\hbar{\bf K} - \frac{e}{c}{\bf A}_1 + \\
  \frac{e}{2c}{\bf B} \times \frac{1}{i}\nabla_{{\bf u}})
  + \frac{e}{2c}[(\hbar{\bf K} - \frac{e}{c}{\bf A}_1 + \frac{e}{2c} {\bf B} \times \frac{1}{i}\nabla_{\bf u})\times {\bf B}]
  \cdot \frac{1}{i}\nabla_{{\bf K}} \Big\} \\
  \times f({\bf K} + \frac{\bf u}{2},{\bf K} - \frac{\bf u}{2},t)   \,.
\end{multline}
In following the definition of the WDF in (\ref{Sec II 16a}) and (\ref{Sec II 16b}), we multiply Eq.~(\ref{Sec II 28}) by $e^{i{\bf u}\cdot{\bf x}}/\Omega$ and sum over ${\bf u}$; we then integrate the $\nabla_{{\bf u}}$ term by parts, noting that
 $\sum_{{\bf u}} e^{i{\bf u}\cdot{\bf x}} i \nabla_{{\bf u}} f({\bf K},{\bf u}) =
 \sum_{{\bf u}} {\bf x} e^{i{\bf u}\cdot{\bf x}}  f({\bf K},{\bf u})$
where the surface term goes to zero as ${\bf u}$ tends to infinity;
we also use ${\bf u} e^{i{\bf u}\cdot{\bf x}} = -i {\bf \nabla}_{{\bf x}} e^{i{\bf u}\cdot{\bf x}}$ and ${\bf \nabla} \cdot ({\bf B} \times {\bf x}) = 0$
to obtain for Eq.~(\ref{Sec II 28})
\begin{multline}\label{Sec II 31}
  \Omega^{-1} \sum_{{\bf u}} \langle {\bf K} + \frac{\bf u}{2} |[\hat{H},\hat{f}]| {\bf K} - \frac{\bf u}{2} \rangle e^{i{\bf u}\cdot{\bf x}} = \\
    \frac{\hbar}{m} [\hbar{\bf K} - \frac{e}{c}{\bf A}({\bf x},t)] \cdot \frac{1}{i} \nabla_{{\bf x}} f({\bf x},{\bf K},t)   \\
  +   \Big\{ \frac{e}{2mc}[\hbar{\bf K} - \frac{e}{c}{\bf A}({\bf x},t)] \times {\bf B}\Big\} \cdot \frac{1}{i} \nabla_{{\bf K}} f({\bf x},{\bf K},t)  \,.
\end{multline}
For $i\hbar (\partial \hat{f}/\partial t)$ in Eq.~(\ref{Sec II 4}), we obtain, using (\ref{Sec II 16a}), that
\begin{equation}\label{Sec II 32}
  \Omega^{-1} \sum_{{\bf u}} i\hbar \langle {\bf K} + \frac{\bf u}{2} |\frac{\partial \hat{f}}{\partial t}| {\bf K} - \frac{\bf u}{2} \rangle
  e^{i{\bf u}\cdot{\bf x}} =  i\hbar \frac{\partial f({\bf x},{\bf K},t)}{\partial t}  \,.
\end{equation}
Thus, the WDF in variables $({\bf x},{\bf K},t)$ is determined by the equation
\begin{multline}\label{Sec II 33}
  \Omega^{-1} \sum_{{\bf u}} \langle {\bf K} + \frac{{\bf u}}{2} |i\hbar \frac{\partial \hat{f}}{\partial t} - [\hat{H},\hat{f}]| {\bf K} - \frac{{\bf u}}{2} \rangle e^{i{\bf u}\cdot{\bf x}} \equiv \\
  i\hbar \Big\{\frac{\partial}{\partial t} +
    \frac{1}{m} [\hbar{\bf K} - \frac{e}{c}{\bf A}({\bf x},t)] \cdot \nabla_{{\bf x}}    \\
  + \frac{e}{2\hbar mc}[(\hbar{\bf K} - \frac{e}{c}{\bf A}({\bf x},t)) \times {\bf B}] \cdot \nabla_{{\bf K}} \Big\} f({\bf x},{\bf K},t)  = 0     \,.
\end{multline}
Lastly, we change variables in Eq.~(\ref{Sec II 33}) as noted in (\ref{Sec II 17})-(\ref{Sec II 18}) and use the following transformation properties
\begin{subequations}
\begin{equation}\label{Sec II 34a}
  \frac{\partial f}{\partial t} = \frac{\partial F}{\partial t} + \nabla_{{\bf k}}F \cdot \dot{{\bf k}} =
  \frac{\partial F}{\partial t} + \hbar^{-1} e{\bf E}(t) \cdot \nabla_{{\bf k}}F     \,,
\end{equation}
\begin{equation}\label{Sec II 34b}
  \nabla_{{\bf K}} f({\bf x},{\bf K},t) = \nabla_{{\bf k}} F({\bf x},{\bf k},t)       \,,
\end{equation}
and
\begin{equation}\label{Sec II 34c}
  \nabla_{x_i} f = \nabla_{x_i} F + \frac{\partial F}{\partial {\bf k}} \cdot \frac{\partial {\bf k}}{\partial x_i}  =
  \nabla_{x_i} F + \frac{e}{2\hbar c} ({\bf B}\times \nabla_{{\bf k}} F)_i   \,.
\end{equation}
\end{subequations}
Therefore, the equation for the WDF from Eq.~(\ref{Sec II 33}) becomes
\begin{eqnarray}\label{Sec II 35}
  \nonumber
  \frac{\partial }{\partial t} F({\bf x},{\bf k},t) + {\bf v}({\bf k}) \cdot \nabla_{{\bf x}} F({\bf x},{\bf k},t)   \\
  + \hbar^{-1} \Big[e {\bf E}(t) + \frac{e}{c}{\bf v}({\bf k}) \times {\bf B}\Big] \cdot \nabla_{{\bf k}} F({\bf x},{\bf k},t) = 0     \,,
\end{eqnarray}
where ${\bf v}({\bf k}) = \hbar {\bf k}/m$. Equation (\ref{Sec II 35}) for $F({\bf x},{\bf k},t)$ is the exact equation for the collisionless WDF obtained with the Hamiltonian of
Eq.~(\ref{Sec II 9}); it also is the identical form of the collisionless BTE for the same problem.

We note that in the analysis above, we have tacitly assumed that ${\bf B}$ was constant, independent of time. If we had chosen to assume that
${\bf B}$ were time dependent, then in Eq.~(\ref{Sec II 35}), ${\bf E}(t)$ would have to be replaced by
\begin{equation}\label{Sec II 36}
  {\bf {\cal E}}({\bf x},t) = - \frac{1}{c} \dot{{\bf A}}({\bf x},t) = {\bf E}(t) - \frac{1}{2c} \dot{{\bf B}}(t) \times {\bf x}  \,. \nonumber
\end{equation}

We point out that Levinson~\cite{Levinson} asserts Eq.~(\ref{Sec II 35}) without proof. He also only considers the case in which the system is initially in thermal equilibrium, and under such conditions, the WDF, $F = F({\bf k},t)$, is a function of $({\bf k},t)$ alone so that the term ${\bf v}\cdot \nabla_{{\bf x}} F({\bf x},{\bf k},t)$ in Eq.~(\ref{Sec II 35}) is missing. In our derivation, the initial condition $F({\bf x},{\bf k},t=t_0)$ is completely arbitrary for an admissible Wigner distribution function, so we can discuss the motion of wave packets in the presence of arbitrarily large electric and magnetic fields in the absence of collisions.
This follows if one multiplies Eq.~(\ref{Sec II 35}) by $k_i(t)$, the $i$th component of ${\bf k}(t)$, and integrate by parts to get
\begin{equation}\label{}
  \frac{d }{dt} \langle \hbar k_i \rangle = \int \big(e {\bf E} + \frac{e}{c}{\bf v} \times {\bf B}\big)_i F({\bf x},{\bf k},t) d{\bf x} d{\bf k}    \,,
\end{equation}
where
\begin{equation}\label{}
  \langle \hbar k_i \rangle = \hbar \int k_i F({\bf x},{\bf k},t) d{\bf x} d{\bf k}     \,. \nonumber
\end{equation}
Thus, if the function
\begin{equation}\label{}
   F({\bf k},t) = \int F({\bf x},{\bf k},t) d{\bf x}  \nonumber
\end{equation}
is peaked about some ${\bf k}(t)$, then it follows from $\langle \hbar k_i \rangle$ that
\begin{equation}\label{}
  \hbar \frac{d }{dt} {\bf k}(t) = e {\bf E}(t) + \frac{e}{c}{\bf v}({\bf k}(t)) \times {\bf B}
\end{equation}
for arbitrary strengths of ${\bf E}$ and ${\bf B}$. Thus, not only is the classical phase space description possible, but the results are exactly the same as those given by the quasiclassical approach. However, unlike the Boltzmann distribution function, the WDF, $F({\bf x},{\bf k},t)$, need not be positive everywhere; its exact structure depends on the initial conditions, and $F({\bf x},{\bf k},t)$ may be negative in certain regions of configuration and momentum space due to quantum effects.

\section{Extension to electrons in solids described by an effective Hamiltonian}\label{Sec III}

We now proceed as described in Sec. II, except here the Hamiltonian of Eq.~(\ref{Sec II 5}) is replaced by the effective Hamiltonian
\begin{equation}\label{Sec III 37}
  \hat{H} = \varepsilon ([{\bf p} - (e/c){\bf A}({\bf x},t)]/\hbar )    \,,
\end{equation}
and the zero magnetic field Hamiltonian is replaced by
\begin{equation}\label{Sec III 38}
  \hat{H}_0 = \varepsilon ([{\bf p} - (e/c){\bf A}_1(t)]/\hbar )    \,.
\end{equation}
Here, it is assumed that $\varepsilon({\bf K})$ is a physical single energy band and therefore $\hat{H}$ comes from a properly symmetrized  Hermitian operator. We once again make use of the instantaneous eigenstates $|{\bf K}\rangle$ of $\hat{H}_0$ given by
\begin{equation}\label{Sec III 39}
  \hat{H}_0 |{\bf K}\rangle  = \varepsilon ({\bf K} - (e/\hbar c){\bf A}_1(t)) |{\bf K}\rangle   \,,
\end{equation}
with eigenvalues $\varepsilon ({\bf K} - (e/\hbar c){\bf A}_1(t))$, which are still given by $|{\bf K}\rangle$ of Eq.~(\ref{Sec II 11a}); they also satisfy the properties $\hat{{\bf p}} |{\bf K}\rangle  = \hbar {\bf K} |{\bf K}\rangle$ and
$\frac{1}{i} \nabla_{{\bf K}} |{\bf K}\rangle  = {\bf x} |{\bf K}\rangle$ so that
${\bf A}({\bf x},t) |{\bf K}\rangle  = {\bf A}(\frac{1}{i} \nabla_{{\bf K}},t) |{\bf K}\rangle$.
Thus, following the previous prescription for calculating the WDF from Eq.~(\ref{Sec II 19}) for $\hat{H}$ in Eq.~(\ref{Sec III 37}),
we have
\begin{eqnarray}\label{Sec III 40a}
 \nonumber
 \langle {\bf K}_1|[\hat{H},\hat{f}]| {\bf K}_2\rangle f({\bf K}_1,{\bf K}_2,t)
 &=& \big[\varepsilon_{+}({\bf K}_1) - \varepsilon_{-}({\bf K}_2)\big]   \\
 &\times& f({\bf K}_1,{\bf K}_2,t)   \,,
\end{eqnarray}
where
\begin{equation}\label{Sec III 40b}
 \varepsilon_{\pm}({\bf K}) = \varepsilon [{\bf K} + {\bf k}_c(t) \pm \frac{e}{2\hbar c}
  ({\bf B} \times \frac{1}{i}\nabla_{{\bf K}})]  \nonumber
\end{equation}
and ${\bf k}_c(t)$ is defined in Eq.~(\ref{Sec II 11c}).
Changing variables from ${\bf K}_{1,2}$ to ${\bf K} \pm {\bf u}/2$ and using (\ref{Sec II 25b}), we obtain
\begin{subequations}
 \begin{eqnarray}\label{Sec III 41a}
  \nonumber
  \langle {\bf K} + \frac{{\bf u}}{2}|[\hat{H},\hat{f}]| {\bf K} - \frac{{\bf u}}{2} \rangle &=&
  \Big[\varepsilon_{+}({\bf K} + \frac{{\bf u}}{2}) - \varepsilon_{-}({\bf K} - \frac{{\bf u}}{2})\Big]  \\
  &\times& f({\bf K} + \frac{{\bf u}}{2},{\bf K} - \frac{{\bf u}}{2},t)   \,,
 \end{eqnarray}
where
 \begin{equation}
  \varepsilon_{\pm}({\bf K} \pm \frac{{\bf u}}{2}) = \varepsilon\Big[{\bf K} \pm \frac{{\bf u}}{2} + {\bf k}_c(t) \pm \frac{e}{2\hbar c}
  {\bf B} \times \frac{1}{i}(\frac{1}{2}\nabla_{\bf K} \pm \nabla_{\bf u}) \Big]  \,.  \nonumber
\end{equation}
We multiply Eq.~(\ref{Sec III 41a}) by $e^{i{\bf u}\cdot{\bf x}}/\Omega$ and sum over ${\bf u}$; then using the relation ${\bf u}e^{i{\bf u}\cdot{\bf x}} = \frac{1}{i} \nabla_{\bf x} e^{i{\bf u}\cdot{\bf x}}$ and integrating over ${\bf u}$ by parts, we obtain
\begin{multline}\label{Sec III 41b}
 \Omega^{-1} \sum_{{\bf u}} e^{i{\bf u}\cdot{\bf x}} \langle {\bf K} + \frac{{\bf u}}{2}|[\hat{H},\hat{f}]| {\bf K} - \frac{{\bf u}}{2} \rangle = \\ \Big\{\varepsilon [{\bf K} - \frac{e}{\hbar c}{\bf A} + \frac{1}{2i}(\nabla_{\bf x} + \frac{e}{2\hbar c}
  {\bf B} \times \nabla_{\bf K})]  \\
  - \varepsilon [{\bf K} - \frac{e}{\hbar c}{\bf A} - \frac{1}{2i}(\nabla_{\bf x} + \frac{e}{2\hbar c}
  {\bf B} \times \nabla_{\bf K})]\Big\} f({\bf x},{\bf K},t) \,.
\end{multline}
\end{subequations}
Again, changing variables to ${\bf k}$ and $F({\bf x},{\bf k},t)$ as in Eqs.~(\ref{Sec II 17})-(\ref{Sec II 18}) and using the relations (\ref{Sec II 34a})-(\ref{Sec II 34c}), we get
\begin{multline}\label{Sec III 42}
  \Omega^{-1} \sum_{{\bf u}} e^{i{\bf u}\cdot{\bf x}} \langle {\bf K} + \frac{{\bf u}}{2}|[\hat{H},\hat{f}]| {\bf K} - \frac{{\bf u}}{2} \rangle = \\ \big\{\varepsilon ({\bf k} + {\bf q}) - \varepsilon ({\bf k} - {\bf q})\big\} F({\bf x},{\bf k},t) \,,
\end{multline}
where ${\bf q} = \frac{1}{2i}(\nabla_{\bf x} + \frac{e}{\hbar c} {\bf B} \times \nabla_{\bf k})$ and $\nabla_{\bf x}$ commutes with $\nabla_{\bf k}$.
Now, it follows from Eqs.~(\ref{Sec II 19}), (\ref{Sec III 42}) as well as from the use of Eq.~(\ref{Sec II 34a}) that the equation for the WDF takes the form
\begin{multline}\label{Sec III 43}
  i\hbar\Big\{\frac{\partial}{\partial t} + \hbar^{-1} e{\bf E}(t) \cdot \nabla_{{\bf k}} \Big\}F({\bf x},{\bf k},t) = \\
  \big\{\varepsilon ({\bf k} + {\bf q})
  - \varepsilon ({\bf k} - {\bf q}) \big\} F({\bf x},{\bf k},t)    \,.
\end{multline}
This is the {\it exact equation} for the collisionless quantum transport WDF for the effective Hamiltonian in homogeneous electric and magnetic fields. In noting that for a single band, the periodic function $\varepsilon ({\bf k})$ can be represented by the Fourier expansion
\begin{equation}\label{Sec III 44}
  \varepsilon ({\bf k}) = \sum_{{\bf l}} \varepsilon ({\bf l}) \, e^{i{\bf k}\cdot {\bf l}}    \,,  \nonumber
\end{equation}
so that $\varepsilon ({\bf k} \pm {\bf q})$ becomes
\begin{equation}\label{}
  \varepsilon ({\bf k} \pm {\bf q}) = \sum_{{\bf l}} \varepsilon ({\bf l}) \, e^{i({\bf k} \pm {\bf q}) \cdot {\bf l}}
  \,;  \nonumber
\end{equation}
we can then write the energy difference on the right-hand side of Eq.~(\ref{Sec III 43}) as
\begin{multline}\label{Sec III 45}
 \varepsilon ({\bf k} + {\bf q}) - \varepsilon ({\bf k} - {\bf q}) = 2i \sum_{{\bf l}} \varepsilon ({\bf l}) \,
  e^{i{\bf k} \cdot {\bf l}} \sin({\bf q}\cdot {\bf l}) \\
  \simeq 2\hbar {\bf v}({\bf k}) \cdot {\bf q} + O({\bf q}^3)     \,,
\end{multline}
where ${\bf v}({\bf k}) = \hbar^{-1} \nabla_{\bf k} \varepsilon ({\bf k})$. Making use of (\ref{Sec III 45}) in (\ref{Sec III 43}) results in
\begin{multline}\label{}
  \Big\{\frac{\partial}{\partial t} + {\hbar}^{-1}e{\bf E}(t) \cdot \nabla_{{\bf k}} \Big\}F({\bf x},{\bf k},t) = \\
  - {\bf v}({\bf k})\cdot (\nabla_{\bf x} + \frac{e}{\hbar c}{\bf B} \times \nabla_{\bf k}) F({\bf x},{\bf k},t) +
  O({\bf B}^3).  \nonumber
\end{multline}
Then to $O({\bf B}^2)$, the equation to the WDF is found to be
\begin{multline}\label{Sec III 46}
  \frac{\partial}{\partial t}F({\bf x},{\bf k},t) + {\bf v}({\bf k})\cdot \nabla_{{\bf x}} F({\bf x},{\bf k},t)  \\
  + \hbar^{-1}[e{\bf E}(t) + \frac{e}{c}({\bf v} \times {\bf B})] \cdot \nabla_{\bf k} F({\bf x},{\bf k},t) = 0 \,,
\end{multline}
where we have used the relations ${\bf v}\cdot ({\bf B}\times \nabla_{\bf k})F = ({\bf v} \times {\bf B})\cdot \nabla_{\bf k}F$ and $({\bf B} \times \nabla_{\bf k})\cdot{\bf v} = 0$.

The equation~(\ref{Sec III 46}) is the analog of Eq.~(\ref{Sec II 35}) obtained for free electrons. The major difference between the two results is that for free electrons with ${\bf v} = \hbar {\bf k}/m$, the derived quantum transport equation for the WDF is {\it exact}, whereas for the general energy-band function $\varepsilon ({\bf k})$, the result is {\it approximate}, good to order $O({\bf B}^2)$, and where ${\bf v} = \hbar^{-1} \nabla_{\bf k} \varepsilon ({\bf k})$. Also, in keeping with the discussion in Sec. II on the WDF as a wave packet, again with no scattering, the rate of change of the electron quasimomentum is given by the Lorentz force even to $O({\bf B}^3)$, since the term of $O({\bf B}^3)$ (if it is present) in Eq.~(\ref{Sec III 46}) does not contribute to the $d \langle \hbar {\bf k} \rangle/dt$.

\section{Bloch electrons in homogeneous electric and magnetic fields; single-band results and multi-band considerations}\label{Sec IV}

\subsection{Development of multiband Wigner distribution function}\label{SubecA IV}

For Bloch electrons interacting with spatially homogeneous, but arbitrarily time-dependent, electric and magnetic fields, the Hamiltonian is
\begin{equation}\label{Sec IV 47}
  \hat{H} = \frac{1}{2m}[{\bf p} - \frac{e}{c}{\bf A}({\bf x},t)]^2 + V_c({\bf x})     \,,
\end{equation}
where $V_c({\bf x})$ is the periodic potential of the crystal. The vector potential ${\bf A}({\bf x},t)$ includes the electric and magnetic field contributions given in Eqs.~(\ref{Sec II 6})-(\ref{Sec II 8}). Thus, as in Eq.~(\ref{Sec II 9}), expanding the kinetic term while noting that ${\bf p}$ and ${\bf A}_2$ commute, we arrive at
\begin{subequations}
\begin{equation}\label{Sec IV 48a}
 \hat{H} = \hat{H}_0 -\frac{e}{mc}{\bf A}_2 \cdot ({\bf p} - \frac{e}{c}{\bf A}_1) + \frac{e^2}{2mc^2} {\bf A}_2^2    \,,
\end{equation}
where the Hamiltonian term
\begin{equation}\label{Sec IV 48b}
 \hat{H}_0 = \frac{1}{2m}[{\bf p} - \frac{e}{c}{\bf A}_1(t)]^2 + V_c({\bf x})
\end{equation}
\end{subequations}
describes the Bloch electron in the electric field alone, and the next two terms are first and second order in the magnetic field
[Here, we note that the energy shifts due to the two magnetic field terms are generally small compared to the $\hat{H}_0$ term for applicable laboratory field strengths; therefore, throughout this discourse, we consider only changes induced by the magnetic field to second order, although higher orders can be necessarily obtained with effort (See J. Callaway~\cite{Callaway})].

To adopt an appropriate basis set with which to evaluate Eq.~(\ref{Sec II 4}), we see in Eqs.~(\ref{Sec IV 48a}) and (\ref{Sec IV 48b}) that a natural basis with which to proceed here is the ABR which are the instantaneous eigenstates of $\hat{H}_0$,
\begin{equation}\label{Sec IV 49}
  \psi_{n{\bf K}}({\bf x},t) = \Omega^{-1/2} e^{i{\bf K}\cdot{\bf x}} u_{n{\bf k}(t)}({\bf x}) \equiv |n,{\bf K};t \rangle    \,,
\end{equation}
with instantaneous Bloch eigenvalues $\varepsilon_n^0({\bf k}(t)) = \varepsilon_{n{\bf k}}^0$ and ${\bf k}(t)$ defined in Eq.~(\ref{Sec II 11c}). Following the WDF analysis from Eqs.~(\ref{Sec II 12})-(\ref{Sec II 14}), we insert the complete set of ABR states of Eq.~(\ref{Sec IV 49}) into Eq.~(\ref{Sec II 12}) to obtain
\begin{multline}\label{Sec IV 50a}
  f({\bf x},{\bf p},t) = \sum_{n_1 {\bf K}_1,n_2 {\bf K}_2} \langle n_1, {\bf K}_1;t|\hat{f}|n_2, {\bf K}_2; t\rangle  \\
  \times T_{n_2 {\bf K}_2 n_1  {\bf K}_1}({\bf x},{\bf p},t)  \,,  \\
  T_{n_2 {\bf K}_2 n_1  {\bf K}_1}({\bf x},{\bf p},t) =   \\
  (2\pi\hbar)^{-3} \int d{\bf y} \psi_{n_2 {\bf K}_2}^{\ast}({\bf x} + \frac{{\bf y}}{2}) \psi_{n_1 {\bf K}_1}({\bf x} - \frac{{\bf y}}{2})
  e^{i{\bf p}\cdot{\bf y}/\hbar}  \,;
\end{multline}
the $T_{n_2 {\bf K}_2 n_1  {\bf K}_1}({\bf x},{\bf p},t)$ are commonly referred to as the transition functions,~\cite{Groenewold} and they
form a complete orthonormal set in phase space. The general properties of $T_{n_2 {\bf K}_2 n_1  {\bf K}_1}({\bf x},{\bf p},t)$ are reviewed by Moyal.~\cite{Groenewold} The function $f({\bf x},{\bf p},t)$ can be presented in the form of a multiband WDF~\cite{Deneio}
\begin{subequations}
\begin{equation}\label{Sec IV 51a}
  f({\bf x},{\bf p},t) = \sum_{n_1n_2} f_{n_1 n_2}({\bf x},{\bf p},t) \,,
\end{equation}
where the multiband components are
\begin{multline}\label{Sec IV 51b}
  f_{n_1 n_2}({\bf x},{\bf p},t) = \sum_{{\bf K}_1{\bf K}_2} \langle n_1, {\bf K}_1;t|\hat{f}|n_2, {\bf K}_2;t \rangle   \\
  \times T_{n_2 {\bf K}_2 n_1 {\bf K}_1}({\bf x},{\bf p},t)  \,.
\end{multline}
\end{subequations}
Using the explicit form of Eq.~(\ref{Sec IV 49}) for $\psi_{n {\bf K}}$ and using ${\bf K}_{1,2}$ as defined in
Eq.~(\ref{Sec II 15}), the multiband components can be expressed in a form comparable to Eqs.~(\ref{Sec II 16a}), (\ref{Sec II 16b}) as
\begin{subequations}
\begin{multline}\label{Sec IV 52a}
  f_{n_1 n_2}({\bf x},{\bf p},t) =  \\
  \Omega^{-1} \sum_{{\bf K} {\bf u}} e^{i{\bf u}\cdot{\bf x}}
  \langle n_1, {\bf K} + \frac{{\bf u}}{2};t|\hat{f}|n_2, {\bf K} - \frac{{\bf u}}{2};t \rangle  \\
  \times I_{n_2 n_1}({\bf x},{\bf p};{\bf u},{\bf K},t),
\end{multline}
where
\begin{multline}\label{Sec IV 52b}
  I_{n_2 n_1}({\bf x},{\bf p};{{\bf u},\bf K},t) = (2\pi\hbar)^{-3}   \\
  \times \int d{\bf y} u_{n_2, {\bf k}(t) - {\bf u}/2}^{\ast}({\bf x} + \frac{{\bf y}}{2}) u_{n_1, {\bf k}(t) + {\bf u}/2}({\bf x} - \frac{{\bf y}}{2}) \times   \\
  e^{i({\bf p} - \hbar{\bf K})\cdot{\bf y}/\hbar}   \,.
\end{multline}
\end{subequations}


Unlike the WDF of previous cases discussed in Secs. II and III, namely, Eqs.~(\ref{Sec II 16a}) and (\ref{Sec II 16b}), which were based on plane-wave instantaneous eigenstates, the WDF of Eq.~(\ref{Sec IV 52a}) is more complex in that it reflects the multiband character of the ABR, including the explicit time dependence contained in the cellular components of the Bloch wave functions. Therefore, $f_{n_1 n_2}({\bf x},{\bf p},t)$ of Eq.~(\ref{Sec IV 52a}) generally manifests its time dependence from both the matrix elements of $\hat{f}$ and $I_{n_2 n_1}$ of Eq.~(\ref{Sec IV 52b}).
It follows from Eq.~(\ref{Sec IV 52a}) that the time derivative of $f_{n_1 n_2}({\bf x},{\bf p},t)$ will depend on the product derivative of
$\langle n_1, {\bf K} + {\bf u}/2;t|\hat{f}|n_2, {\bf K} - {\bf u}/2;t\rangle$ and $I_{n_2 n_1}$. The time evolution of the matrix elements of $\hat{f}$ is governed by the Liouville equation as discussed in Secs. II, III, and will be continued further in this section; the derivative of $I_{n_2 n_1}$ will depend upon the time derivatives of the cellular Bloch functions, and using $i{\bf \nabla}_{\bf k} u_{n{\bf k}}({\bf x}) = \sum_{n^{\prime} \neq n} {\bf R}_{n^{\prime}n}({\bf k}) u_{n^{\prime}{\bf k}}({\bf x})$, where ${\bf R}_{n^{\prime}n}({\bf k})$
is given in Eq.~(\ref{Sec IV 58b}), we see that $\partial I_{n_2 n_1}/\partial t = \dot{{\bf k}} \cdot {\bf \nabla}_{\bf k} I_{n_2 n_1}$
promotes tunneling to states beyond $n_1$ and $n_2$. Generally, this makes the time development of $f_{n_1 n_2}({\bf x},{\bf p},t)$ quite complex.

Here, in our approach, we unfold $f_{n_1 n_2}({\bf x},{\bf p},t)$ of Eqs.~(\ref{Sec IV 52a}), (\ref{Sec IV 52b}) so as to reflect $I_{n_2 n_1}$ in a relatively useful fashion. In this regard, use is made of the well-known fact~\cite{LuttKohn}
that $\{u_{n{\bf K}}\}$, for any ${\bf K} = {\bf K}_0$, span a complete set of orthonormal functions for any function periodic in the unit cell. Therefore, we expand $u_{n {\bf K}}({\bf x})$ in terms of the set $\{u_{n {\bf K}_0}({\bf x})\}$ in the
${\bf K}_0$-{\it representation}~\cite{Kane} as
\begin{equation}\label{Sec IV 53}
 u_{n\bf K}({\bf x}) = \sum_{n^{\prime}} c_{nn^{\prime}}({\bf K}-{\bf K}_0)u_{n^{\prime}{\bf K}_0}({\bf x})   \,,
\end{equation}
where the coefficients $c_{nn^{\prime}}({\bf K}-{\bf K}_0)$ are determined by the method described in Appendix A, and the ${\bf K}_0$ values are chosen conveniently to suit the problem at hand (usually, ${\bf K}_0$ is chosen to be zero thus defining the band edges).
Using this representation, we can express $f_{n_1 n_2}({\bf x},{\bf p},t)$ of Eq.~(\ref{Sec IV 52a}) as
\begin{subequations}
\begin{multline}\label{Sec IV 54a}
 f_{n_1 n_2}({\bf x},{\bf p},t) = \sum_{{\bf K}}  \sum_{n^{\prime} n^{\prime\prime}}
 {\cal I}_{n^{\prime\prime}n^{\prime}}({\bf x},{\bf p};{\bf K},{\bf K}_0)  \times  \\
 \Omega^{-1} \sum_{\bf u} c^{\ast}_{n_2 n^{\prime\prime}}({\bf k}(t)-{\bf K}_0 - \frac{{\bf u}}{2})
 c_{n_1n^{\prime}}({\bf k}(t)-{\bf K}_0 + \frac{{\bf u}}{2})   \\
 \times \langle n_1, {\bf K} + \frac{\bf u}{2};t|\hat{f}|n_2, {\bf K} - \frac{\bf u}{2};t\rangle e^{i{\bf u} \cdot {\bf x}}  \,;
\end{multline}
here,
\begin{multline}\label{Sec IV 54bb}
 {\cal I}_{n^{\prime\prime}n^{\prime}}({\bf x},{\bf p};{\bf K},{\bf K}_0) = (2\pi\hbar)^{-3} \\
 \times \int_{\Omega_c} u^{\ast}_{n^{\prime\prime}{\bf K}_0}({\bf x} + \frac{{\bf y}}{2}) u_{n^{\prime}{\bf K}_0}({\bf x} - \frac{{\bf y}}{2}) e^{i({\bf p} - \hbar{\bf K})\cdot{\bf y}/\hbar} d{\bf y}    \,,
\end{multline}
independent of time, and ${\bf k}(t)$ is defined in Eq.~(\ref{Sec II 11c}).
This is exact provided we know the exact solution to the matrix equation for $c_{nn^{\prime}}({\bf K}-{\bf K}_0)$ in Appendix A, Eq.~(\ref{App-B10a}).
Using ${\bf u} e^{i{\bf u}\cdot{\bf x}} = - i \nabla_{{\bf x}}e^{i{\bf u}\cdot{\bf x}}$, we note that
\begin{multline}\label{Sec IV 54b}
 c^{\ast}_{n_2 n^{\prime\prime}}({\bf k}(t)-{\bf K}_0 - \frac{{\bf u}}{2})
 c_{n_1n^{\prime}}({\bf k}(t)-{\bf K}_0 + \frac{{\bf u}}{2}) e^{i{\bf u} \cdot {\bf x}} =   \\
 \hat{c}^{\ast}_{n_2 n^{\prime\prime}}({\bf k}(t)-{\bf K}_0 + \frac{i}{2}\nabla_{{\bf x}})
 \hat{c}_{n_1n^{\prime}}({\bf k}(t)-{\bf K}_0 - \frac{i}{2}\nabla_{{\bf x}}) e^{i{\bf u} \cdot {\bf x}}     \,;
\end{multline}
\end{subequations}
then, $f_{n_1 n_2}({\bf x},{\bf p},t)$ in Eq.~(\ref{Sec IV 54a}) can be expressed as
\begin{subequations}
\begin{multline}\label{Sec IV 55a}
  f_{n_1 n_2}({\bf x},{\bf p,t}) = \sum_{{\bf K}}
  \hat{\Gamma}_{n_1 n_2}({\bf x},{\bf p};{\bf k}(t),{\bf K}_0,i{\bf \nabla}_{\bf x})   \\
  \times f^0_{n_1n_2}({\bf x},{\bf K},t)     \,,
\end{multline}
where
\begin{equation}\label{Sec IV 55b}
  f_{n_1 n_2}^0({\bf x},{\bf K},t) = \Omega^{-1} \sum_{\bf u}
  \langle n_1, {\bf K} + \frac{\bf u}{2};t|\hat{f}|n_2, {\bf K} - \frac{\bf u}{2};t\rangle e^{i{\bf u}\cdot{\bf x}}
\end{equation}
and
\begin{multline}\label{Sec IV 55c}
 \hat{\Gamma}_{n_1 n_2}({\bf x},{\bf p};{\bf k}(t),{\bf K}_0,i{\bf \nabla}_{\bf x}) = \sum_{n^{\prime} n^{\prime\prime}}
 {\cal I}_{n^{\prime\prime}n^{\prime}}({\bf x},{\bf p};{\bf K},{\bf K}_0)   \\
 \times \hat{c}^{\ast}_{n_2 n^{\prime\prime}}({\bf k}(t)-{\bf K}_0 + \frac{i}{2}\nabla_{{\bf x}})
 \hat{c}_{n_1n^{\prime}}({\bf k}(t)-{\bf K}_0 - \frac{i}{2}\nabla_{{\bf x}})  \,.
\end{multline}
We see that $f_{n_1 n_2}({\bf x},{\bf p},t)$ of Eqs.~(\ref{Sec IV 55a}) and (\ref{Sec IV 55b}) exhibits a comparable form to the plane-wave based $f({\bf x},{\bf p},t)$ of Eqs.~(\ref{Sec II 16a}) and (\ref{Sec II 16b}), although here, $\hat{\Gamma}_{n_1 n_2}$ reflects the role of interband cellular Bloch envelope components and $f_{n_1 n_2}^0({\bf x},{\bf K},t)$ serves as the {\it reduced multiband} WDF. It is seen that the exact multiband WDF is composed of a momentum superposition of $f_{n_1 n_2}^0({\bf x},{\bf K},t)$ and the coefficient $\hat{\Gamma}_{n_1 n_2}$ of Eq.~(\ref{Sec IV 55c}), where $f_{n_1 n_2}^0({\bf x},{\bf K},t)$ is the multiband generalization of the plane-wave WDF found in Eqs.~(\ref{Sec II 16a}), (\ref{Sec II 16b}).
$f_{n_1 n_2}({\bf x},{\bf p},t)$ in Eq.~(\ref{Sec IV 55a}) is a key representation of the multiband components of the WDF for Bloch dynamics in the ABR representation, and shows the importance of the so-called reduced multiband WDF, $f^0_{n_1n_2}({\bf x},{\bf K},t)$.
$\hat{\Gamma}_{n_1 n_2}$ in Eq.~(\ref{Sec IV 55c}), which is determined from the ${\bf K}_0$-representation, fully accounts for the presence of $I_{n_2n_1}$ in Eq.~(\ref{Sec IV 52a}); therefore, the WDF dependence on the complete electric and magnetic field will be reflected in the quantum behavior of $f_{n_1 n_2}^0$ of Eq.~(\ref{Sec IV 55b}).

Finally, from the definition of ${\cal I}_{n^{\prime\prime}n^{\prime}}$ in Eq.~(\ref{Sec IV 54bb}),
we note that in expressing $u_{n {\bf K}_0}({\bf x} \pm {\bf y}/2)$ as
a Taylor series in $(\pm{\bf y}/2)$, we can therefore express the integrand of ${\cal I}_{n^{\prime\prime} n^\prime}$ as a term by term explicit function of ${\bf y}$ and then integrate over $d {\bf y}$ to obtain an infinite series of delta functions in $\delta ({\bf p} - \hbar{\bf K})$ as
\begin{multline}\label{}
  {\cal I}_{n^{\prime\prime} n^\prime}({\bf x},{\bf p};{\bf K},{\bf K}_0) = \sum_{n=0}^{\infty} \Big(\frac{i\hbar}{2}\Big)^n
  \sum_{m=0}^n \frac{(-1)^{n-m}}{m!(n-m)!}  \times  \\
  [(\nabla_{\bf p}\cdot\nabla_{\bf x})^{n-m} u^{\ast}_{n^{\prime\prime}{\bf K}_0}({\bf x})]
  [(\nabla_{\bf p}\cdot\nabla_{\bf x})^m u_{n^{\prime}{\bf K}_0}({\bf x})] \delta ({\bf p} - \hbar{\bf K}).  \nonumber
\end{multline}
This becomes
\begin{multline}\label{Sec IV 55d}
  {\cal I}_{n^{\prime\prime} n^\prime}({\bf x},{\bf p};{\bf K},{\bf K}_0) = {\cal I}^{(0)}_{n^{\prime\prime} n^\prime}({\bf x},{\bf K}_0)\delta ({\bf p} - \hbar{\bf K}) +  \\
  {\bf {\cal I}}^{(1)}_{n^{\prime\prime} n^\prime}({\bf x},{\bf K}_0) \cdot \nabla_{{\bf K}} \delta ({\bf p} - \hbar{\bf K}) + \ldots   \,,
\end{multline}
\end{subequations}
where
\begin{multline}\label{}
  {\cal I}^{(0)}_{n^{\prime\prime} n^\prime}({\bf x},{\bf K}_0) =  u_{n^{\prime\prime} {\bf K}_0}^{\ast}({\bf x})
  u_{n^{\prime} {\bf K}_0}({\bf x})    \,,     \\
  {\bf {\cal I}}^{(1)}_{n^{\prime\prime} n^\prime}({\bf x},{\bf K}_0) = \frac{1}{2i} \big[u_{n^{\prime\prime} {\bf K}_0}^{\ast}({\bf x})
  \nabla_{\bf x} u_{n^{\prime} {\bf K}_0}({\bf x}) - \\
  u_{n^{\prime} {\bf K}_0}({\bf x}) \nabla_{\bf x} u^{\ast}_{n^{\prime\prime} {\bf K}_0}({\bf x}) \big],  \nonumber
\end{multline}
and so forth. Thus, using ${\cal I}_{n^{\prime\prime} n^\prime}({\bf x},{\bf p};{\bf K},{\bf K}_0)$ of Eq.~(\ref{Sec IV 55d}) in Eq.~(\ref{Sec IV 55c}) allows for the integral over ${\bf K}$ in Eq.~(\ref{Sec IV 55a}) to be evaluated directly.
In subsection F, $\hat{\Gamma}_{n_1 n_2}({\bf x},{\bf p};{\bf k}(t),{\bf K}_0,i{\bf \nabla}_{\bf x})$ of Eq.~(\ref{Sec IV 55c}) is established explicitly to first order in $({\bf K} - {\bf K}_0)$ [Eq.~(\ref{App-B22})] although methodology is outlined for easily extending the approximation to higher orders; all coefficients are determined at a specific choice of ${\bf K}_0$.

It follows from Eq.~(\ref{Sec IV 55a}) that the integration of the time derivative of $f_{n_1 n_2}({\bf x},{\bf p},t)$ allows for the introduction of initial conditions so that we can write
\begin{subequations}
\begin{multline}\label{Sec IV 56a}
  f_{n_1 n_2}({\bf x},{\bf p},t) = f_{n_1 n_2}({\bf x},{\bf p},t_0)    \\
  + \sum_{{\bf K}} \Big[\hat{\Gamma}_{n_1 n_2}({\bf x},{\bf p};{\bf k}(t),{\bf K}_0,i{\bf \nabla}_{\bf x}) f^0_{n_1n_2}({\bf x},{\bf K},t)   \\
  - \hat{\Gamma}_{n_1 n_2}({\bf x},{\bf p};{\bf K},{\bf K}_0,i{\bf \nabla}_{\bf x}) f^0_{n_1n_2}({\bf x},{\bf K},t_0) \Big]   \,.
\end{multline}
Thus, $f_{n_1 n_2}({\bf x},{\bf p},t)$, as expressed in Eq.~(\ref{Sec IV 56a}), is the complete formal result for the multiband WDF in terms of the {\it reduced multiband} WDF, $f^0_{n_1n_2}({\bf x},{\bf K},t)$, the operator coefficient $\hat{\Gamma}_{n_1 n_2}$, and their initial conditions. Further, to transform $f_{n_1 n_2}({\bf x},{\bf p},t)$ of Eq.~(\ref{Sec IV 56a}) to a gauge invariant form as was done in Secs. II and III, we make use of the transformation of variables from ${\bf K}$ to ${\bf k}({\bf x},t)$ with $f^0({\bf x},{\bf K},t) = F^0({\bf x},{\bf k},t)$ as noted in Eqs.~(\ref{Sec II 17}) and (\ref{Sec II 18}), and then make use of transformations specified in Eqs.~(\ref{Sec II 34a})-(\ref{Sec II 34c}) to find that
\begin{multline}\label{}
  \big({\bf K} + {\bf k}_c(t) - {\bf K}_0 \pm \frac{i}{2}{\bf \nabla}_{\bf x}\big) f^0_{n_1n_2}({\bf x},{\bf K},t) \rightarrow \\
  \big[{\bf k}({\bf x},t) - {\bf K}_0 \pm \frac{i}{2}{\bf \nabla}_{\bf x} + \frac{e}{\hbar c} {\bf A}_2({\bf x} \pm \frac{i}{2}{\bf \nabla}_{\bf k}\big)] F^0_{n_1n_2}({\bf x},{\bf k},t)   \,.  \nonumber
\end{multline}
Then, Eq.~(\ref{Sec IV 56a}) becomes
\begin{multline}\label{Sec IV 56c}
  f_{n_1 n_2}({\bf x},{\bf p},t) = f_{n_1 n_2}({\bf x},{\bf p},t_0)    \\
  + \sum_{{\bf k}} \Big[{\tilde{\Gamma}}_{n_1 n_2}({\bf x},{\bf p};{\bf k}({\bf x},t),{\bf K}_0,i{\bf \nabla}_{\bf x}) F^0_{n_1n_2}({\bf x},{\bf k},t)   \\
  - {\tilde{\Gamma}}_{n_1 n_2}({\bf x},{\bf p};{\bf k}({\bf x},t_0),{\bf K}_0,i{\bf \nabla}_{\bf x}) F^0_{n_1n_2}({\bf x},{\bf k},t_0)\Big],
\end{multline}
where
\begin{multline}\label{Sec IV 56d}
 {\tilde{\Gamma}}_{n_1 n_2}({\bf x},{\bf p};{\bf k}({\bf x},t),{\bf K}_0,i{\bf \nabla}_{\bf x}) =    \\
 \sum_{n^{\prime} n^{\prime\prime}}
 {\cal I}_{n^{\prime\prime}n^{\prime}}({\bf x},{\bf p} - \frac{e}{c}{\bf A}-\hbar {\bf k},{\bf K}_0)   \\
 \times \hat{c}^{\ast}_{n_2 n^{\prime\prime}}\big({\bf k}({\bf x},t)-{\bf K}_0 + \frac{i}{2}\nabla_{{\bf x}} +
 \frac{e}{\hbar c}{\bf A}_2({\bf x} + \frac{i}{2}{\bf \nabla}_{\bf k})\big)    \\
 \times \hat{c}_{n_1n^{\prime}}\big({\bf k}({\bf x},t)-{\bf K}_0 - \frac{i}{2}\nabla_{{\bf k}} +
 \frac{e}{\hbar c}{\bf A}_2({\bf x} - \frac{i}{2}{\bf \nabla}_{\bf k})\big)  \,.
\end{multline}
\end{subequations}

\subsection{The reduced Wigner distribution function to $O({\bf B}^2)$}\label{SubecB IV}

Given the fundamental role of $f^0_{n_1 n_2}({\bf x},{\bf K},t)$ as noted in Eqs.~(\ref{Sec IV 55a}), (\ref{Sec IV 55b}), we now proceed by treating the matrix elements of $\hat{f}$ and $f^0_{n_1 n_2}({\bf x},{\bf K},t)$ as the essential components in examining the WDF in Bloch electron analysis.

In order to obtain the lowest order, nontrivial single-band WDF using the ABR, we assume that the fields $|{\bf E}|$ and $|{\bf B}|$ are sufficiently small so that we neglect the interband matrix elements of $f^0_{n_1n_2}$. Thus, we consider the matrix elements of Eq.~(\ref{Sec II 4}) [with $C_s\{\hat{f}\} = 0$] as
\begin{equation}\label{Sec IV 57}
 i\hbar \langle n,{\bf K}_1;t |\frac{\partial\hat{f}}{\partial t}| {n,\bf K}_2;t \rangle = \langle n,{\bf K}_1;t |[\hat{H},\hat{f}]|
 n,{\bf K}_2;t \rangle   \,,
\end{equation}
where $\langle n,{\bf K};t|$ are the time-dependent ABR of Eq.~(\ref{Sec IV 49}) for the energy band $n$. We can show~\cite{Krieger} that
\begin{subequations}
\begin{equation}\label{Sec IV 58a}
 i\hbar \frac{\partial}{\partial t}\psi_{n{\bf K}} = {\bf F}(t) \cdot \sum_{n^{\prime} \neq n} {\bf R}_{n^{\prime}n}({\bf k}(t)) \psi_{n^{\prime}{\bf K}}   \,,
\end{equation}
where ${\bf R}_{n^{\prime}n}({\bf K}) = {\bf R}^{\ast}_{nn^{\prime}}({\bf K})$ is the usual band mixing integral,
\begin{equation}\label{Sec IV 58b}
 {\bf R}_{n^{\prime}n}({\bf K}) = \frac{i}{\Omega_c} \int_{\Omega_c}
 u^{\ast}_{{n^{\prime}}{\bf K}}({\bf x}) {\bf \nabla}_{\bf K} u_{n{\bf K}}({\bf x}) d{\bf x}  \,,
\end{equation}
\end{subequations}
and where the phases of $\psi_{n{\bf K}}$ are chosen~\cite{Zak} so that ${\bf R}_{nn}({\bf k}) = 0$, a provision which assumes the crystal possesses an inversion symmetry. If inversion symmetry is broken, then ${\bf R}_{nn}({\bf k})$ is non-zero and therefore needs to be retained; this gives rise to significant Berry phase effects~\cite{Xiao} which will be considered in a future study. For relatively weak fields,
${\bf R}_{n^{\prime}n}({\bf k}) \simeq 0$ so that $\partial \psi_{n{\bf K}}/\partial t \simeq 0$. Thus, in the weak field, single-band limit,
$\langle n,{\bf K}_1;t |\partial\hat{f}/\partial t| {n,\bf K}_2;t \rangle = \partial\langle n,{\bf K}_1;t |\hat{f}| {n,\bf K}_2;t \rangle/\partial t $ and Eq.~(\ref{Sec IV 57}) can be written as
\begin{equation}\label{Sec IV 59a}
 i\hbar \frac{\partial}{\partial t} f_n({\bf K}_1,{\bf K}_2;t) = \langle n,{\bf K}_1;t |[\hat{H},\hat{f}]|n,{\bf K}_2;t \rangle  \,,
\end{equation}
where  $f_n({\bf K}_1,{\bf K}_2;t) \equiv \langle n,{\bf K}_1;t |\hat{f}| n,{\bf K}_2;t \rangle$.
Using $\hat{H}$ in Eq.~(\ref{Sec IV 48a}) along with $\langle n,{\bf K};t|$ from Eq.~(\ref{Sec IV 49}), the term $\langle n,{\bf K}_1;t
|[\hat{H},\hat{f}]|n,{\bf K}_2;t \rangle$ becomes
\begin{multline}\label{Sec IV 60}
  \langle n,{\bf K}_1;t |[\hat{H},\hat{f}]|n,{\bf K}_2;t \rangle = \Big\{\varepsilon_n^0({\bf K}_1 - \frac{e}{\hbar c}{\bf A}_1) - \\ \varepsilon_n^0({\bf K}_2 - \frac{e}{\hbar c}{\bf A}_1) + \frac{e}{2c}\Big[({\bf B} \times \frac{1}{i}\nabla_{{\bf K}_1}) \cdot {\bf v}_n ({\bf K}_1 - \frac{e}{\hbar c}{\bf A}_1)  \\
  - ({\bf B} \times \frac{1}{i}\nabla_{{\bf K}_2}) \cdot {\bf v}_n ({\bf K}_2 - \frac{e}{\hbar c}{\bf A}_1)\Big]\Big\}
  f_n({\bf K}_1,{\bf K}_2,t) + O({\bf B}^2),
\end{multline}
where ${\bf v}_n ({\bf k}) = \hbar^{-1} \nabla_{\bf k} \varepsilon_n^0 ({\bf k})$ (necessary matrix elements of $[\hat{H},\hat{f}]$ can be found
in Appendix C). Here, a contributing term of order $O({\bf B}^2)$ would come from the Hamiltonian term $\frac{e^2}{2mc^2}{\bf A}_2^2$, but there are additional terms of $O({\bf B}^2)$ which have been excluded because of the interband dependence of $\hat{f}$ in Eq.~(\ref{Sec IV 59a}). Therefore, a more rigorous approach for obtaining terms of order $O({\bf B}^2)$ and higher would be to proceed by employing a unitary transformation~\cite{Callaway} of Eq.~(\ref{Sec II 4}) which diagonalizes the Hamiltonian (\ref{Sec IV 48a}) to the desired order, here to $O({\bf B}^2)$ in the magnetic field and to all orders in the electric field by utilizing the ABR. We note that Eq.~(\ref{Sec IV 60}) is the same result that we obtained for the effective Hamiltonian case of Eq.~(\ref{Sec III 40a}) when this equation is taken to $O({\bf B})$. To $O({\bf B})$, the quantum transport equation for the single-band WDF is
\begin{multline}\label{Sec IV 61}
  \frac{\partial}{\partial t}F_n({\bf x},{\bf k},t) + {\bf v}_n({\bf k})\cdot \nabla_{{\bf x}} F_n({\bf x},{\bf k},t)   \\
  + \hbar^{-1} \Big[e{\bf E}(t) + \frac{e}{c}({\bf v}_n \times {\bf B})\Big] \cdot \nabla_{\bf k} F_n({\bf x},{\bf k},t) + O({\bf B}^2) = 0 \,.
\end{multline}


If the Hamiltonian of Eq.~(\ref{Sec IV 48a}) were diagonal in the $|n,{\bf K};t \rangle$ representation, it would be trivial to calculate the matrix elements $[\hat{H},\hat{f}]_{n^{\prime}{\bf K}^{\prime}n{\bf K}}$ in Eq.~(\ref{Sec II 4}), and, in this case, only the intraband matrix elements of
$\hat{f}$ would enter into the problem. However, since this is not the case, we seek a unitary transformation, $e^{i\hat{U}}$, with the Hermitian operator $\hat{U} = \hat{U}^{\dag}$, such that the Hamiltonian transforms as
\begin{equation}\label{Sec IV 62}
 \overline{\hat{H}} = e^{-i\hat{U}}\hat{H} e^{i\hat{U}}    \,,
\end{equation}
where $\overline{\hat{H}}$ is diagonal in the ABR to order of $O({\bf B}^2)$. Then applying the same unitary transformation to
Eq.~(\ref{Sec II 4}) results in
\begin{multline}\label{Sec IV 63}
  i\hbar\frac{\partial \overline{\hat{f}}}{\partial t} + i\hbar \Big[e^{-i\hat{U}}(\frac{\partial}{\partial t}e^{i\hat{U}})\overline{\hat{f}}
  + \overline{\hat{f}}(\frac{\partial}{\partial t}e^{-i\hat{U}}) e^{i\hat{U}} \Big] - [\overline{\hat{H}},\overline{\hat{f}}]   \\
  = \overline{C_s}\{\overline{\hat{f}}\}   \,,
\end{multline}
where $\overline{\hat{f}} = e^{-i\hat{U}}\hat{f} e^{i\hat{U}}$
and $\overline{C_s}$ is similarly defined. While operators transform by the unitary transformation defined by Eq.~(\ref{Sec IV 62}), it follows equivalently that the ABR state vectors transform as $\overline{|n,{\bf K};t \rangle} = e^{i\hat{U}} |n,{\bf K};t \rangle$
These state vectors could have been also utilized to establish the transformation of Eq.~(\ref{Sec II 4}). An outline of the methodology for
diagonalization of the Hamiltonian in Eq.~(\ref{Sec IV 48a}) and the determination of the matrix elements of the operator $\hat{U}$ to the
desired order of approximation can be found in Appendix B. In the derivations, we express $\hat{H}$ of Eq.~(\ref{Sec IV 48a}) as
\begin{equation}\label{Sec IV 66}
  \hat{H} = \hat{H}_0 + \beta V_1 + \beta^2 V_2  \,,
\end{equation}
where $\beta$ (dimensionless) refers to the order of the magnetic field associated with ${\bf A}_2$ (in the final results, we set $\beta = 1$); we also look for $\hat{U}$ as a perturbation expansion in magnetic field
\begin{equation}\label{Sec IV 67}
  \hat{U} =  \beta \hat{U}_1 + \beta^2 \hat{U}_2 + O(\beta^3) \,.
\end{equation}

The diagonal matrix elements of $\overline{\hat{H}}$ of Eq.~(\ref{Sec IV 62}) are represented as
\begin{equation}\label{Sec II 68a}
  (\overline{\hat{H}})_{n{\bf K}n{\bf K}} \equiv \varepsilon_{n{\bf K}}(\beta) = \varepsilon_n^{0} + \beta \varepsilon_{n,1} + \beta^2 \varepsilon_{n,2} + O(\beta^3)    \,,  \nonumber
\end{equation}
and we find, to $O(\beta^2)$,
\begin{subequations}
\begin{multline}\label{Sec IV 68b}
  \varepsilon_{n{\bf K}} (\beta) = \varepsilon_n^{0}({\bf k}) + \beta  (V_1)_{n{\bf K}n{\bf K}}
  + \beta^2 \Big[(V_2)_{n{\bf K}n{\bf K}}     \\
  + \sum_{n^{\prime}\neq n} \frac{|(V_1)_{n{\bf K}n^{\prime}{\bf K}}|^2}{\varepsilon^{0}_{n{\bf k}} -
  \varepsilon^{0}_{n^{\prime}{\bf k}}}\Big]  \,.
\end{multline}
We note that in Eq.~(\ref{Sec IV 68b}), the term of $O(\beta^2)$ includes not only $(V_2)_{n{\bf K}n{\bf K}}$ which corresponds to
$(e^2/2mc^2){\bf A}_2^2$, but also includes an additional term which depends on states $n^{\prime}\neq n$; this completes the correction to and including terms of order ${\bf B}^2$. In using the calculated matrix elements for $V_{1,2}$, which have been derived in Appendix C, we see that $\varepsilon_{n{\bf K}} (\beta = 1) \equiv \varepsilon_{n{\bf K}}$ of Eq.~(\ref{Sec IV 68b}) reduces to
\begin{multline}\label{Sec IV 68c}
 \varepsilon_{n{\bf K}} = \varepsilon_n^0({\bf k}) + \frac{e}{2\hbar c} \frac{\partial \varepsilon_n^0({\bf k})}{\partial {\bf k}}\Big|_{{\bf k}={\bf k}(t)} \cdot ({\bf B}\times \frac{1}{i}\nabla_{{\bf K}})    \\
 + \frac{1}{2}\left(\frac{e}{2\hbar c}\right)^2 \sum_{l\,, m =1}^3
 \frac{\partial^2 \varepsilon_n^0({\bf k})}{\partial k_l \partial k_m}\Big|_{{\bf k}={\bf k}(t)}
 ({\bf B} \times \frac{1}{i}\nabla_{{\bf K}})_l    \\
 \times ({\bf B} \times \frac{1}{i}\nabla_{{\bf K}})_m \,,
\end{multline}
\end{subequations}
where ${\bf k}(t) = {\bf K} - (e/\hbar c){\bf A}_1(t) = {\bf K} + {\bf k}_c(t)$. The operator dependence of $\varepsilon_{n{\bf K}}$ arises typically in the crystal momentum representation.~\cite{Callaway}

\subsection{The Liouville equation and unitary transformations}\label{SubSecC IV}

Having established in Appendix B the $\hat{U}_{1,2}$ that diagonalizes $\overline{\hat{H}}$ through $O({\bf B}^2)$, we now focus on the specific
form and character of the Liouville equation in Eq.~(\ref{Sec IV 63}) while using the ABR. We note from Eq.~(\ref{Sec IV 63}) the transformed
Liouville equation of Eq.~(\ref{Sec II 4}) can be expressed in compact form as
\begin{subequations}
\begin{equation}\label{Sec IV 69a}
  i\hbar\frac{\partial \overline{\hat{f}}}{\partial t} - [\overline{\hat{H}},\overline{\hat{f}}]
  = \overline{C}\{\overline{\hat{f}}\} + \overline{C_s}  \,,
\end{equation}
where
\begin{equation}\label{Sec IV 69b}
    \overline{C}\{\overline{\hat{f}}\} = - i\hbar \left[e^{-i\hat{U}}\big(\frac{\partial}{\partial t}e^{i\hat{U}}\big)\overline{\hat{f}}
  + \overline{\hat{f}}\big(\frac{\partial}{\partial t}e^{-i\hat{U}}\big) e^{i\hat{U}} \right]      \,,
\end{equation}
\end{subequations}
and $\overline{C_s}$ includes the explicit scattering from phonons.~\cite{Levinson}  The term $\overline{C}\{\overline{\hat{f}}\}$, which
originates from the unitary transformation of $\partial \hat{f}/\partial t$ in Eq.~(\ref{Sec II 4}), represents an internal
pseudo-collision term which strongly influences the interband mixing effects.
In this particular work, we focus heavily on the canonical kinematics, so we suppress the term $\overline{C_s}$ here and consider only
ballistic transport; a thorough analysis of the general behavior of the term $C_s$ has been given in context with phonons
previously,~\cite{Krieger} and $\overline{C_s}$, including phonons and impurity scattering,~\cite{Kiselev} will be considered in terms of Wigner transport in the electric and magnetic fields in a future companion paper.

For the analysis of $\overline{C}\{\overline{\hat{f}}\}$, it suffices to calculate the quantity in question to $O(\hat{U}^2)$; this insures that expansion terms up to and including $O({\bf B}^2)$ are included. It has been previously established~\cite{Lutzky}
that the operator terms in Eq.~(\ref{Sec IV 69b}) can be reduced to
\begin{equation}\label{}
  e^{-i\hat{U}} \frac{\partial}{\partial t}e^{i\hat{U}} = i \Big\{\frac{\partial \hat{U}}{\partial t}, G(i\hat{U})  \Big\} \,, \nonumber
\end{equation}
\begin{equation}\label{}
  (\frac{\partial}{\partial t}e^{-i\hat{U}}) e^{i\hat{U}} = \big(e^{-i\hat{U}} \frac{\partial}{\partial t} e^{i\hat{U}}\big)^{\dag}
  = -i \Big\{G(-i\hat{U}), \frac{\partial \hat{U}}{\partial t} \Big\}   \,;    \nonumber
\end{equation}
here $\{\hat{a},\hat{b}\}$ denotes anti-commutation. This allows us to expess $\overline{C}\{\overline{\hat{f}}\}$ as
\begin{subequations}
\begin{equation}\label{Sec IV 70a}
 \overline{C}\{\overline{\hat{f}}\} = \hbar \Big[\{\frac{\partial \hat{U}}{\partial t},G(i\hat{U}) \}\overline{\hat{f}} - \overline{\hat{f}}\{G(-i\hat{U}),\frac{\partial \hat{U}}{\partial t}  \}  \Big]   \,,
\end{equation}
where
\begin{equation}\label{Sec IV 70b}
  G(\pm i\hat{U}) \equiv \frac{e^{\pm i\hat{U}} - 1}{\pm i\hat{U}} = 1 + \frac{1}{2!} (\pm i\hat{U}) + \frac{1}{3!} (\pm i\hat{U})^2 + \ldots    \,.
\end{equation}
\end{subequations}
Now, $\overline{C}\{\overline{\hat{f}}\}$ in Eq.~(\ref{Sec IV 70a}) is an exact expression in terms of the operator $\hat{U}$. To obtain $\overline{C}\{\overline{\hat{f}}\}$ to order $\hat{U}^2$, we use Eq.~(\ref{Sec IV 70b}) in Eq.~(\ref{Sec IV 70a}) and note that
\begin{equation}\label{}
  \Big\{\frac{\partial \hat{U}}{\partial t},G(\pm i\hat{U}) \Big\} = 2 \frac{\partial \hat{U}}{\partial t} \pm
  \frac{i}{2}\Big\{\frac{\partial \hat{U}}{\partial t},\hat{U} \Big\} + O(\hat{U}^3)   \,.   \nonumber
\end{equation}
We see that to $O(\hat{U}^2)$, $\overline{C}\{\overline{f}\}$ in Eq.~(\ref{Sec IV 70a}) reduces to
\begin{subequations}
\begin{equation}\label{Sec IV 71a}
  \overline{C}\{\overline{\hat{f}}\} = \hbar [(\hat{h}_1 + i \hat{h}_2)\overline{\hat{f}} - \overline{\hat{f}}
  (\hat{h}_1 - i \hat{h}_2)   ]   \,,
\end{equation}
where
\begin{equation}\label{Sec IV 71b}
  \hat{h}_1 = 2 \frac{\partial \hat{U}}{\partial t}, \;\;\;\;   \hat{h}_2 = \frac{1}{2} \big\{ \hat{U},\frac{\partial \hat{U}}{\partial t} \big\}
   \,,
\end{equation}
\end{subequations}
with $\hat{h}^{\dag}_1 = \hat{h}_1$ and $\hat{h}^{\dag}_2 = \hat{h}_2$. Thus, the Liouville equation of Eq.~(\ref{Sec IV 69a}) with $\overline{C}_s = 0$ becomes, to second order in $\hat{U}$,
\begin{subequations}
\begin{equation}\label{Sec IV 72a}
 i\hbar\frac{\partial \overline{\hat{f}}}{\partial t} - [\overline{\hat{H}},\overline{\hat{f}}] = \hat{H}^{\prime}\overline{\hat{f}} - \overline{\hat{f}} (\hat{H}^{\prime})^{\dag} ,
\end{equation}
where
\begin{equation}\label{Sec IV 72b}
 \hat{H}^{\prime} = \hbar(\hat{h}_1 + i \hat{h}_2), \;\; (\hat{H}^{\prime})^{\dag} = \hbar(\hat{h}_1 - i \hat{h}_2) .
\end{equation}
\end{subequations}
In taking the matrix elements of Eq.~(\ref{Sec IV 72a}) with the ABR, while remembering that the $|n,{\bf K};t \rangle$ are time dependent from
Eq.~(\ref{Sec IV 58a}), and $\overline{\hat{H}}$ is diagonal in $|n,{\bf K};t \rangle$ to second order in ${\bf B}$, we obtain
\begin{subequations}
\begin{multline}\label{Sec IV 73a}
 i\hbar\frac{\partial }{\partial t} \overline{f}_{n_1{\bf K}_1n_2{\bf K}_2} = (\varepsilon_{n_1{\bf K}_1}
 - \varepsilon_{n_2{\bf K}_2}) \overline{f}_{n_1{\bf K}_1n_2{\bf K}_2}     \\
 + \sum_{n^{\prime} {\bf K}^{\prime}} \big(H^{\prime\prime}_{n_1{\bf K}_1n^{\prime}{\bf K}^{\prime}}
 \overline{f}_{n^{\prime}{\bf K}^{\prime}n_2{\bf K}_2} - \overline{f}_{n_1{\bf K}_1n^{\prime}{\bf K}^{\prime}}
 \widetilde{H^{\prime\prime}}_{n^{\prime}{\bf K}^{\prime}n_2{\bf K}_2} \big)   ,
\end{multline}
where
\begin{multline}\label{Sec IV 73c}
 H^{\prime\prime}_{n_1{\bf K}_1n^{\prime}{\bf K}^{\prime}} =
 H^{\prime}_{n_1{\bf K}_1n^{\prime}{\bf K}^{\prime}} - {\bold F}(t)\cdot{\bold R}_{n_1n^{\prime}}
 ({\bf k}_1)\delta_{{\bf K}_1{\bf K}^{\prime}},     \\
 \widetilde{H^{\prime\prime}}_{n_1{\bf K}_1n^{\prime}{\bf K}^{\prime}} =
 (H^{\prime})^{\dag}_{n_1{\bf K}_1n^{\prime}{\bf K}^{\prime}} - {\bold F}(t)\cdot{\bold R}_{n_1n^{\prime}}({\bf k}_1)\delta_{{\bf K}_1{\bf K}^{\prime}}          \,,
\end{multline}
\end{subequations}
and $\varepsilon_{n{\bf K}}$ is given in Eq.~(\ref{Sec IV 68c}). Here, the effect of the additional ${\bold F}(t)$-dependent electric field term
on the right-hand side of expressions in Eq.~(\ref{Sec IV 73c}) is due to the time dependence of the ABR and simply adds to the matrix elements of $\hat{H}^{\prime}$ and $(\hat{H}^{\prime})^{\dag}$; the ${\bf F}(t)$-dependent terms generally promote Zener interband tunneling stimulated by the electric field ${\bold F}(t)$. In examining the off-diagonal second order contributions of $\hat{H}^{\prime}$ and $(\hat{H}^{\prime})^{\dag}$ to
the total transition matrices, we will show that these terms contribute a magnetic component of $O({\bf B}^2)$ to the interband
tunneling.

Now, it is clear that Eq.~(\ref{Sec IV 73a}) describes all possible matrix elements of $\overline{f}_{n_1{\bf K}_1n_2{\bf K}_2}$ correct to order
 $\hat{U}^2$. In an effort to reduce Eq.~(\ref{Sec IV 73a}) to a more tractable form, one which retains essential information, we proceed in the
spirit of the Wigner-Weisskopf approximation (WWA) by retaining from the term $\sum_{n^{\prime},{\bf K}^{\prime}}(...)$ on the right-hand
side of Eq.~(\ref{Sec IV 73a}) the terms corresponding to $n_1{\bf K}_1,n_2{\bf K}_2$ while ignoring all others; this will result in an approximate expression for the diagonal and off-diagonal matrix elements of $\overline{f}_{n_1{\bf K}_1n_2{\bf K}_2}$. So, for the sum $\sum_{n^{\prime}{\bf K}^{\prime}}(...)$ in Eq.~(\ref{Sec IV 73a}), we get
\begin{multline}\label{Sec IV 74}
 \sum_{n^{\prime} {\bf K}^{\prime}}(...) = H^{\prime\prime}_{n_1{\bf K}_1n_1{\bf K}_1} \overline{f}_{n_1{\bf K}_1n_2{\bf K}_2}  \\
 - \overline{f}_{n_1{\bf K}_1n_2{\bf K}_2} \widetilde{H^{\prime\prime}}_{n_2{\bf K}_2n_2{\bf K}_2}
 + H^{\prime\prime}_{n_1{\bf K}_1n_2{\bf K}_2}\overline{f}_{n_2{\bf K}_2n_2{\bf K}_2}      \\
 - \overline{f}_{n_1{\bf K}_1n_1{\bf K}_1} \widetilde{H^{\prime\prime}}_{n_1{\bf K}_1n_2{\bf K}_2}
 + \sum_{n^{\prime}{\bf K}^{\prime} \neq n_1{\bf K}_1,n_2{\bf K}_2}(...)    \,.
\end{multline}
Dropping the sum on the right-hand side of
 Eq.~(\ref{Sec IV 74}), and inserting the remainder into Eq.~(\ref{Sec IV 73a}), we get, for $n_1{\bf K}_1 \neq n_2{\bf K}_2$,
\begin{multline}\label{Sec IV 75}
 i\hbar\frac{\partial }{\partial t} \overline{f}_{n_1{\bf K}_1n_2{\bf K}_2} = (\varepsilon_{n_1{\bf K}_1}
 - \varepsilon_{n_2{\bf K}_2}) \overline{f}_{n_1{\bf K}_1n_2{\bf K}_2}    \\
 + \hbar \big[(h_1 + ih_2)_{n_1{\bf K}_1n_1{\bf K}_1} \overline{f}_{n_1{\bf K}_1n_2{\bf K}_2}    \\
  - \overline{f}_{n_1{\bf K}_1n_2{\bf K}_2} (h_1 - ih_2)_{n_2{\bf K}_2n_2{\bf K}_2}\big]      \\
 + H^{\prime\prime}_{n_1{\bf K}_1n_2{\bf K}_2}
 \overline{f}_{n_2{\bf K}_2n_2{\bf K}_2} - \overline{f}_{n_1{\bf K}_1n_1{\bf K}_1}
  \widetilde{H^{\prime\prime}}_{n_1{\bf K}_1n_2{\bf K}_2}       \,;
\end{multline}
here, $\varepsilon_{n{\bf K}}$ is given by Eq.~(\ref{Sec IV 68c}), and the off-diagonal elements of $\hat{H}^{\prime\prime}$,
$\widetilde{\hat{H}^{\prime\prime}}$ are given by Eq.~(\ref{Sec IV 73c}).
Since Eq.~(\ref{Sec IV 75}) contains a mix of off-diagonal and diagonal elements of $\overline{\hat{f}}$, closure is reached for the diagonal elements of the system by setting $n_1{\bf K}_1 = n_2{\bf K}_2 = n{\bf K}$ in Eq.~(\ref{Sec IV 73a}) to obtain
\begin{multline}\label{Sec IV 76}
 i\hbar\frac{\partial }{\partial t} \overline{f}_{n_{\bf K}n{\bf K}} =
 \hbar \big[(h_1 + ih_2)_{n{\bf K}n{\bf K}} \overline{f}_{n{\bf K}n{\bf K}}    \\
  - \overline{f}_{n{\bf K}n{\bf K}} (h_1 - ih_2)_{n{\bf K}n{\bf K}}\big]       \\
 + \sum_{n^{\prime} {\bf K}^{\prime} \neq n {\bf K}} \big(H^{\prime\prime}_{n{\bf K}n^{\prime}{\bf K}^{\prime}}
 \overline{f}_{n^{\prime}{\bf K}^{\prime}n{\bf K}} - \overline{f}_{n{\bf K}n^{\prime}{\bf K}^{\prime}}
 \widetilde{H^{\prime\prime}}_{n^{\prime}{\bf K}^{\prime}n{\bf K}} \big)       \,;
\end{multline}
here, the diagonal term $n^{\prime}{\bf K}^{\prime} = n{\bf K}$ has been extracted from the term
$\sum_{n^{\prime}{\bf K}^{\prime}}(...)$ to display the diagonal matrix elements of $\overline{\hat{f}}$ explicitly. Thus, Eqs.~(\ref{Sec IV 75}), (\ref{Sec IV 76}) give a closed set of equations whereby, in principle, one can self-consistently solve for the approximate diagonal and off-diagonal elements of $\overline{\hat{f}}$, consistent with their respective initial conditions. Moreover, these equations contain the multiband generalization of the Liouville equation valid to $O(\hat{U}^2)$; the $\hat{U}$ dependence is expressed in the matrix elements of $\hat{H}^{\prime}$ and $(\hat{H}^{\prime})^{\dag}$ established in Eq.~(\ref{Sec IV 72b}). The detail calculations of the matrix elements of $\hat{H}^{\prime}$ and $(\hat{H}^{\prime})^{\dag}$ can be found in Appendix E.

\subsection{Single-band analysis}\label{SubSecD IVb}

We now consider the specific case of single-band analysis for Eqs.~(\ref{Sec IV 75}), (\ref{Sec IV 76}); specifically, we consider the case where
$n_1 = n_2 = n$. Then, letting $\overline{f}_{n{\bf K}_1n{\bf K}_2} \equiv \overline{f}_n({\bf K}_1,{\bf K}_2)$, Eq.~(\ref{Sec IV 75}) becomes
\begin{multline}\label{Sec IV 77}
 i\hbar\frac{\partial }{\partial t} \overline{f}_n({\bf K}_1,{\bf K}_2) = (\varepsilon_{n{\bf K}_1}
 - \varepsilon_{n{\bf K}_2}) \overline{f}_n({\bf K}_1,{\bf K}_2)    \\
 + \hbar [(h_1 + ih_2)_{n{\bf K}_1n{\bf K}_1} \overline{f}_n({\bf K}_1,{\bf K}_2)    \\
  - (h_1 - ih_2)_{n{\bf K}_2n{\bf K}_2} \overline{f}_n({\bf K}_1,{\bf K}_2)]         \\
 + \hbar [(h_1 + ih_2)_{n{\bf K}_1n{\bf K}_2} \overline{f}_n({\bf K}_2,{\bf K}_2)     \\
 - \overline{f}_n({\bf K}_1,{\bf K}_1) (h_1 - ih_2)_{n{\bf K}_1n{\bf K}_2}]    \,.
\end{multline}
We show in Appendix E that the required diagonal matrix elements for $(\hat{h}_1 \pm i\hat{h}_2)$ in Eq.~(\ref{Sec IV 77}) are of $O({\bf B}^2)$ and have dependences only beyond $n$ for $n^{\prime} \neq n$ [See Eqs.~(\ref{AppF6}), (\ref{AppF8a})]; thus, in the
single-band limit, with ${\bf R}_{nn^{\prime}} \simeq 0$, we see that
\begin{equation}\label{Sec IV 78}
 i\hbar\frac{\partial }{\partial t} \overline{f}_n({\bf K}_1,{\bf K}_2) = (\varepsilon_{n{\bf K}_1}
 - \varepsilon_{n{\bf K}_2}) \overline{f}_n({\bf K}_1,{\bf K}_2)        \,,
\end{equation}
where $\varepsilon_{n{\bf K}}$ is given by Eq.~(\ref{Sec IV 68c}).

In developing the WDF equation from Eq.~(\ref{Sec IV 78}), we first explicitly write down the expression for $\varepsilon_{n{\bf K}}$ in Eq.~(\ref{Sec IV 68c}). Then in finding the equation for $f^0_n({\bf x},{\bf K},t)$, we proceed as in Secs. II and III; let ${\bf K}_1 = {\bf K} + {\bf u}/2,\; {\bf K}_2 = {\bf K} - {\bf u}/2$, and form
\begin{multline}\label{Sec IV 79}
  f^0_{nn}({\bf x},{\bf K},t) \equiv f^0_n({\bf x},{\bf K},t) =    \\
  \frac{1}{\Omega} \sum_{{\bf u}}
  f_n({\bf K} + \frac{{\bf u}}{2},{\bf K} - \frac{{\bf u}}{2},t) e^{i{\bf u}\cdot{\bf x}} \,.
\end{multline}
Using Eqs.~(\ref{Sec II 25b}) and (\ref{Sec II 34a})-(\ref{Sec II 34c}) in the same algebraic procedure as previously applied, that is, transforming from $({\bf K}_1,{\bf K}_2)$ to $({\bf K},{\bf u})$ variables, using
\begin{equation}
  \frac{1}{i}\nabla_{{\bf u}} e^{i{\bf u}\cdot{\bf x}} = {\bf x} e^{i{\bf u}\cdot{\bf x}}  \,,\;\;\;
  \frac{1}{i}\nabla_{{\bf x}} e^{i{\bf u}\cdot{\bf x}} = {\bf u} e^{i{\bf u}\cdot{\bf x}}  \,,  \nonumber
\end{equation}
integrating $f^0_n({\bf x},{\bf K},t)$ over ${\bf u}$ by parts, and transforming to ${\bf k}({\bf x},t) = {\bf K} - (e/\hbar c){\bf A}({\bf x},t)$ with $F^0_n({\bf x},{\bf k},t) = f^0_n({\bf x},{\bf K},t)$, we obtain the gauge invariant equation for $f^0_n({\bf x},{\bf K},t)$ as
\begin{multline}\label{Sec IV 80}
  \frac{\partial }{\partial t} F^0_n({\bf x},{\bf k},t) + {\bf v}_n({\bf k}) \cdot \nabla_{{\bf x}} F^0_n({\bf x},{\bf k},t)   \\
  + \Big[e {\bf E}(t) + \frac{e}{c}{\bf v}_n({\bf k}) \times {\bf B}\Big] \cdot \frac{1}{\hbar} \nabla_{{\bf k}} F^0_n({\bf x},{\bf k},t) + O({\bf B}^3) = 0  \,.
\end{multline}
Therefore, it follows that if we neglect the interband tunneling terms that arise from both the usual Zener and
magnetic-induced interband tunneling, the residual Liouville equation of Eq.~(\ref{Sec IV 78}), using $\varepsilon_{n{\bf K}}$ of Eq.~(\ref{Sec IV 68c}) so that ${\bf v}_n ({\bf k}) = \hbar^{-1} \nabla_{\bf k} \varepsilon_{n{\bf K}} ({\bf k})$, transforms into the analogous WDF equation of Eq.~(\ref{Sec IV 61}), but valid through $O({\bf B}^3)$ in the single-band, collisionless approximation.
One points out here that Eq.~(\ref{Sec IV 68c}) for $\varepsilon_{n{\bf K}}$ is exactly what we would have obtained if we assumed an effective Hamiltonian $\hat{H} = \varepsilon [({\bf p} - (e/c){\bf A})/\hbar] = \varepsilon [({\bf p} - (e/c){\bf A}_1 - (e/c){\bf A}_2)/\hbar]$ and taken the matrix elements with respect to plane waves where ${\bf A}_2 = \frac{1}{2}{\bf B}\times {\bf x}$, and then expanded the result about ${\bf B} = 0$ to $O({\bf B}^2)$, and replacing ${\bf x}$ by $\frac{1}{i}\nabla_{{\bf k}}$. It then follows that through terms to order ${\bf B}^2$, the interband matrix elements of $[\overline{\hat{H}},\overline{\hat{f}}]$ with respect to the ABR are diagonal in band, and are given by the same expression as one would obtain using an effective Hamiltonian given by Sec. III.

\subsection{Multiband considerations}\label{SubSecE IV}

The multiband consideration requires the analysis of Eqs.~(\ref{Sec IV 75}), (\ref{Sec IV 76}), a closed set of equations for the diagonal and off-diagonal matrix elements of $\overline{\hat{f}}$ derived from the Liouville equation using the WWA. As observed in Eqs.~(\ref{Sec IV 75}), (\ref{Sec IV 76}), the diagonal elements of $(\hat{h}_1 \pm i\hat{h}_2)$ present in these equations give rise to terms of $O({\bf B}^2)$
and a multiband dependence as determined in Appendix E. On the other hand, the off-diagonal matrix elements of $\hat{H}^{\prime\prime}$ and $\widetilde{\hat{H}^{\prime\prime}}$ present in (\ref{Sec IV 75}), (\ref{Sec IV 76}) as defined by Eq.~(\ref{Sec IV 73c})
give rise to the presence of interband tunneling promoted by the electric field dependent Zener tunneling as represented by ${\bold F}(t)\cdot{\bold R}_{nn^{\prime}}({\bf k})\delta_{{\bf K}{\bf K}^{\prime}}$, and the magnetic component of interband tunneling that is implicitly contained in the off-diagonal matrix elements of $(\hat{h}_1 \pm i\hat{h}_2)$.

As defined in Eq.~(\ref{Sec IV 73c}), the interband terms, with $n \neq n^{\prime}$, are
\begin{multline}\label{Sec IV 81b}
 H^{\prime\prime}_{n_{\bf K}n^{\prime}{\bf K}^{\prime}} =
 \hbar (h_1 + ih_2)_{n{\bf K}n^{\prime}{\bf K}^{\prime}} - {\bold F}(t)\cdot{\bold R}_{nn^{\prime}}
 ({\bf k})\delta_{{\bf K}{\bf K}^{\prime}},  \\
 \widetilde{H^{\prime\prime}}_{n{\bf K}n^{\prime}{\bf K}^{\prime}} =
 \hbar (h_1 - ih_2)_{n{\bf K}n^{\prime}{\bf K}^{\prime}} - {\bold F}(t)\cdot{\bold R}_{nn^{\prime}}({\bf k})\delta_{{\bf K}{\bf K}^{\prime}}          \,.
\end{multline}
Using $(h_1 \pm ih_2)_{n{\bf K}n^{\prime}{\bf K}^{\prime}}$ from Appendix E [Eq.~(\ref{AppF16})], we see that Eq.~(\ref{Sec IV 81b}) becomes
\begin{multline}\label{Sec IV 82b}
 H^{\prime\prime}_{n_{\bf K}n^{\prime}{\bf K}^{\prime}} = \hbar \frac{\partial}{\partial t}
 \Big[2(\beta U_1 + \beta^2 U_2)_{n{\bf K}n^{\prime}{\bf K}^{\prime}} +    \\
 \frac{i}{2} \beta^2 \delta_{{\bf K}{\bf K}^{\prime}} N_{nn^{\prime}}({\bf K})\Big]
 - {\bold F}(t)\cdot{\bold R}_{nn^{\prime}}({\bf k})\delta_{{\bf K}{\bf K}^{\prime}}
 [1 - 2\beta^2 {\cal G}_{nn^{\prime}}({\bf K})],   \\
 \widetilde{H^{\prime\prime}}_{n{\bf K}n^{\prime}{\bf K}^{\prime}} = \hbar \frac{\partial}{\partial t}
 \Big[2(\beta U_1 + \beta^2 U_2)_{n{\bf K}n^{\prime}{\bf K}^{\prime}} -    \\
 \frac{i}{2} \beta^2 \delta_{{\bf K}{\bf K}^{\prime}} N_{nn^{\prime}}({\bf K})\Big]
 - {\bold F}(t)\cdot{\bold R}_{nn^{\prime}}({\bf k})\delta_{{\bf K}{\bf K}^{\prime}}
 [1 - 2\beta^2 {\cal G}_{nn^{\prime}}({\bf K})].
\end{multline}
As observed in Eq.~(\ref{Sec IV 82b}), the first terms are time derivatives which allow for an integrating factor in
Eqs.~(\ref{Sec IV 75})-(\ref{Sec IV 76}), and the second term has an explicit magnetic interband tunneling contribution of
$O({\bf B}^2)$ to the Zener tunneling term (See E. J. Blount~\cite{Callaway}). Both ${\cal G}_{nn^{\prime}}({\bf K})$ and $N_{nn^{\prime}}({\bf K})$ are defined in terms of $\hat{U}$ in Eqs.~(\ref{AppF13b}) and (\ref{AppF15}), respectively. A full treatment of the properties of Eqs.~(\ref{Sec IV 75}), (\ref{Sec IV 76}), including the derivation of the WDF equation, will be discussed in a companion paper; but here we present an outline of this  with salient features.

The reduction of Eqs.~(\ref{Sec IV 75}), (\ref{Sec IV 76}), and the resulting WDF equation can be obtained to $O({\bf B}^2)$ by retaining the coefficients in these equations up to $O({\bf B}^2)$. To this end, making use of the commutation properties of $\hat{U}_1^2$ noted in Appendix E, we find that Eq.~(\ref{Sec IV 75}), in combination with Eq.~(\ref{Sec IV 76}), can be expressed to $O({\bf B}^2)$ as
\begin{subequations}
\begin{multline}\label{Sec IV a81a}
 i\hbar\frac{\partial }{\partial t} \overline{f}_{n_1{\bf K}_1n_2{\bf K}_2}(t) = ({\varepsilon}_{n_1{\bf K}_1}
 - {\varepsilon}_{n_2{\bf K}_2}) \overline{f}_{n_1{\bf K}_1n_2{\bf K}_2}(t)   \\
 + W_{n_1{\bf K}_1n_2{\bf K}_2}\overline{f}_{n_2{\bf K}_2}(t) - \overline{f}_{n_1{\bf K}_1}(t)W_{n_1{\bf K}_1n_2{\bf K}_2}   \\
 - Z_{n_1{\bf K}_1n_2{\bf K}_2}\overline{f}_{n_1{\bf K}_1n_2{\bf K}_2}(t_0)   \\
 + (X_{n_1{\bf K}_1n_2{\bf K}_2} - Y_{n_1{\bf K}_1n_2{\bf K}_2})\overline{f}_{n_1{\bf K}_1}(t_0)  \\
 + (X_{n_1{\bf K}_1n_2{\bf K}_2} + Y_{n_1{\bf K}_1n_2{\bf K}_2})\overline{f}_{n_2{\bf K}_2}(t_0)       \,,
\end{multline}
with $\overline{f}_{n_1{\bf K}_1n_2{\bf K}_2}(t_0)$ assumed to be a constant. Here,
\begin{equation}\label{Sec IV a81b}
 W_{n_1{\bf K}_1n_2{\bf K}_2} = 2\hbar \frac{\partial}{\partial t} U_{n_1{\bf K}_1n_2{\bf K}_2}
 - {\bold F}(t)\cdot{\bold R}_{n_1n_2}({\bf K}_1)\delta_{{\bf K}_1{\bf K}_2}           \,,
\end{equation}
with $\hat{U} = \beta \hat{U}_1 + \beta^2 \hat{U}_2$, and
\begin{multline}\label{Sec IV a81c}
 X_{n_1{\bf K}_1n_2{\bf K}_2} =  \beta^2 \frac{i\hbar}{2} \delta_{{\bf K}_1{\bf K}_2} \frac{\partial}{\partial t}
 N_{n_1n_2}({\bf K}_1), \\
 Y_{n_1{\bf K}_1n_2{\bf K}_2} =  \beta^2 {\bold F}(t)\cdot{\bold R}_{n_1n_2}({\bf K}_1)\delta_{{\bf K}_1{\bf K}_2}
 {\cal G}_{n_1n_2}({\bf K}_1),   \\
 Z_{n_1{\bf K}_1n_2{\bf K}_2} = \beta^2 \frac{i\hbar}{2} \frac{\partial}{\partial t}
 \big[(U_1^2)_{n_1{\bf K}_1n_1{\bf K}_1} -3(U_1^2)_{n_2{\bf K}_2n_2{\bf K}_2}\big].
\end{multline}
\end{subequations}
In Eq.~(\ref{Sec IV a81a}), the following key points are noted: 1) the multiband equation for $\overline{f}_{n_1{\bf K}_1n_2{\bf K}_2}(t)
\equiv \overline{f}^{0}_{n_1{\bf K}_1n_2{\bf K}_2}(t)$ [$\overline{f}^{0}$ refers to the reduced WDF of Eq.~(\ref{Sec IV 55b})] is an inhomogeneous equation, with the inhomogeneity dependent upon $\overline{f}_{n{\bf K}n{\bf K}}(t) \equiv \overline{f}_{n{\bf K}}^{\,0}(t)$,
the instantaneous time-dependent diagonal matrix elements, as well as the initial conditions for $\overline{f}^{\,0}_{n{\bf K}}(t_0)$ and $\overline{f}^{\,0}_{n{\bf K}n^{\prime}{\bf K}^{\prime}}(t_0)$; 2) through the definition of $W_{n_1{\bf K}_1n_2{\bf K}_2}$ in Eq.~(\ref{Sec IV a81b}), we observe the presence of the electric Zener tunneling term,
\begin{subequations}
\begin{equation}\label{Sec IV a82a}
  {\bold F}(t)\cdot{\bold R}_{n_1n_2}({\bf K}_1)\delta_{{\bf K}_1{\bf K}_2}
  \big(\overline{f}_{n_1{\bf K}_1}(t) - \overline{f}_{n_2{\bf K}_2}(t)\big),  \nonumber
\end{equation}
which depends on the instantaneous behavior of $\overline{f}_{n_1{\bf K}_1}(t)$ and $\overline{f}_{n_2{\bf K}_2}(t)$; as well, the lowest order contribution to magnetic breakdown is contained in the $(\partial \hat{U}_1/\partial t)$ part of $W_{n_1{\bf K}_1n_2{\bf K}_2}$; 3) through the definition of $Y_{n_1{\bf K}_1n_2{\bf K}_2}$ in Eq.~(\ref{Sec IV a81c}), we observe the presence of a {\it magnetic-induced} electric Zener tunneling term,
\begin{equation}\label{Sec IV a82b}
  {\bold F}(t)\cdot{\bold R}_{n_1n_2}({\bf K}_1)\delta_{{\bf K}_1{\bf K}_2} {\cal G}_{n_1n_2}({\bf K}_1)
  \big(\overline{f}_{n_1{\bf K}_1}(t_0) - \overline{f}_{n_2{\bf K}_2}(t_0)\big)    \,,  \nonumber
\end{equation}
\end{subequations}
which depends on the initial conditions for $\overline{f}_{n_1{\bf K}_1}(t_0)$ and $\overline{f}_{n_2{\bf K}_2}(t_0)$, as well as ${\cal G}_{n_1n_2}({\bf K}_1)$, a magnetic field-dependent variable defined in Eq.~(\ref{AppF13b}).

Lastly, in developing the reduced WDF equation for Eq.~(\ref{Sec IV a81a}), we first utilize ${\varepsilon}_{n{\bf K}}$ in Eq.~(\ref{Sec IV 68c}), with $n = (n_1,n_2)$, ${\bf K} = ({\bf K}_1,{\bf K}_2)$, and then let ${\bf K}_1 = {\bf K} + {\bf u}/2, {\bf K}_2 = {\bf K} - {\bf u}/2$ while using the transformations (\ref{Sec II 25b}), and form $f_{n_1 n_2}^0({\bf x},{\bf K},t)$ of Eq.~(\ref{Sec IV 55b}); likewise, we also transform coefficients in Eq.~(\ref{Sec IV a81a}) using
\begin{multline}\label{Sec IV a83}
  \omega_{n_1 n_2}({\bf x},{\bf K},t) = \Omega^{-1} \sum_{\bf u}
  \langle n_1, {\bf K} + \frac{\bf u}{2};t|W|n_2, {\bf K} - \frac{\bf u}{2};t\rangle   \\
  \times e^{i{\bf u}\cdot{\bf x}},
\end{multline}
with inverse
\begin{equation}\label{Sec IV a84}
  \langle n_1, {\bf K} + \frac{\bf u}{2};t|W|n_2, {\bf K} - \frac{\bf u}{2};t\rangle = \int_{\Omega} d{\bf x} e^{-i{\bf u}\cdot{\bf x}}
  \omega_{n_1 n_2}({\bf x},{\bf K},t)    \,.
\end{equation}
Thus, the equation for the reduced WDF in Eq.~(\ref{Sec IV a81a}) is of the form
\begin{multline}\label{Sec IV a85}
 i\hbar\frac{\partial }{\partial t} \overline{f}^{\,0}_{n_1n_2}({\bf x},{\bf K},t) = \Big\{\varepsilon^0_{n_1}({\bf K} + {\bf k}_c) - \varepsilon^0_{n_2}({\bf K} + {\bf k}_c)       \\
 -\frac{i\hbar}{2}\Big[\big({\bf v}^0_{n_1}({\bf K}+{\bf k}_c) + {\bf v}^0_{n_2}({\bf K}+{\bf k}_c)\big)\cdot\nabla_{{\bf x}}    \\
 + \frac{e}{2c} \Big( \big({\bf v}^0_{n_1}({\bf K}+{\bf k}_c) + {\bf v}^0_{n_2}({\bf K}+{\bf k}_c)\big)\times{\bf B}\Big) \cdot\nabla_{{\bf K}} \Big]   \\
 - \sum_{l,m=1}^3 \frac{\partial^2}{\partial {\bf k}_l \partial {\bf k}_m} \big[\varepsilon^0_{n_1}({\bf K}+{\bf k}_c) - \varepsilon^0_{n_2}({\bf K}+{\bf k}_c)\big] \times   \\
  \big(\nabla_{{\bf x}} + \frac{e}{2\hbar c}{\bf B}\times\nabla_{{\bf K}}\big)_l \big(\nabla_{{\bf x}} + \frac{e}{2\hbar c}{\bf B}\times\nabla_{{\bf K}}\big)_m \Big\} \overline{f}^{\,0}_{n_1n_2}({\bf x},{\bf K},t)    \\
 + \int d{\bf x}^{\prime} \big[\omega_{n_1 n_2}({\bf x}-{\bf x}^{\prime},{\bf K})\overline{f}^{\,0}_{n_2}({\bf x}^{\prime},{\bf K},t)   \\
 - \overline{f}^{\,0}_{n_1}({\bf x}^{\prime},{\bf K},t) \omega_{n_1n_2}({\bf x}-{\bf x}^{\prime},{\bf K})\big] \\
 - \int d{\bf x}^{\prime} Z_{n_1n_2}({\bf x}-{\bf x}^{\prime},{\bf K})\overline{f}^{\,0}_{n_1n_2}({\bf K},t_0)  \\
 + \int d{\bf x}^{\prime} \Big\{\big[\Psi_{n_1 n_2}({\bf x}-{\bf x}^{\prime},{\bf K}) - Y_{n_1 n_2}({\bf x}-{\bf x}^{\prime},{\bf K})\big]
 \overline{f}^{\,0}_{n_1}({\bf K},t_0)    \\
 + [\Psi_{n_1 n_2}({\bf x}-{\bf x}^{\prime},{\bf K}) + Y_{n_1 n_2}({\bf x}-{\bf x}^{\prime},{\bf K})\big] \overline{f}^{\,0}_{n_2}({\bf K},t_0) \Big\} \\
 + O({\bf B}^3).
\end{multline}
Here, ${\bf v}^0_{n}({\bf K}+{\bf k}_c) = \hbar^{-1} \nabla_{{\bf K}} \varepsilon^0_{n}({\bf K} + {\bf k}_c)$,
$\omega_{n_1 n_2}({\bf x},{\bf K}),  \Psi_{n_1 n_2}({\bf x},{\bf K})$, $Y_{n_1 n_2}({\bf x},{\bf K})$, and $Z_{n_1 n_2}({\bf x},{\bf K})$ are the Wigner-reduced transforms of $W_{n_1{\bf K}_1n_2{\bf K}_2}, X_{n_1{\bf K}_1n_2{\bf K}_2}$, $Y_{n_1{\bf K}_1n_2{\bf K}_2}$, and $Z_{n_1{\bf K}_1n_2{\bf K}_2}$ as governed by the protocol for transforming from $({\bf K}_1, {\bf K}_2)$ to $({\bf K}, {\bf u})$ along with the transform defined by Eq.~(\ref{Sec IV a84}). For the gauge invariant form of Eq.~(\ref{Sec IV a85}), we change variables from ${\bf K}$ to ${\bf k} = {\bf K} - (e/c){\bf A}$ with $\overline{f}^{\,0}({\bf x},{\bf K},t) \rightarrow \overline{F}^{\,0}({\bf x},{\bf k},t)$, while using the transformation properties from Eqs.~(\ref{Sec II 34a})-(\ref{Sec II 34c}), to find
\begin{multline}\label{Sec IV a86}
 \Big\{\frac{\partial}{\partial t} + \frac{1}{2} \big({\bf v}^{\,0}_{n_1} + {\bf v}^{\,0}_{n_2}\big)\cdot\nabla_{{\bf x}}   \\
 + e\big[{\bf E}(t) + \frac{1}{2c} \big({\bf v}^{\,0}_{n_1} + {\bf v}^{\,0}_{n_2}\big)\times{\bf B}\big]\cdot\hbar^{-1}\nabla_{{\bf k}} \Big\}\overline{F}^{\,0}_{n_1n_2}({\bf x},{\bf k},t) \\
 = \frac{1}{i\hbar} \Big\{\varepsilon^{\,0}_{n_1} - \varepsilon^{\,0}_{n_2}
 - \sum_{l,m=1}^3 \frac{\partial^2 (\varepsilon^{\,0}_{n_1} - \varepsilon^{\,0}_{n_2})}{\partial k_l \partial k_m} \times   \\
  \big(\nabla_{{\bf x}} + \frac{e}{\hbar c}{\bf B}\times\nabla_{{\bf k}}\big)_l \big(\nabla_{{\bf x}} + \frac{e}{\hbar c}{\bf B}\times\nabla_{{\bf k}}\big)_m \Big\} \overline{F}^{\,0}_{n_1n_2}({\bf x},{\bf k},t)    \\
 + \frac{1}{i\hbar} \int d{\bf x}^{\prime} \Big\{\tilde{\omega}_{n_1n_2}({\bf x}-{\bf x}^{\prime},{\bf k}) \overline{F}^{\,0}_{n_2}({\bf x}^{\prime},{\bf k},t)   \\
 - \overline{F}^{\,0}_{n_1}({\bf x}^{\prime},{\bf k},t) \tilde{\omega}_{n_1n_2}({\bf x}-{\bf x}^{\prime},{\bf k})  \\
 - \tilde{Z}_{n_1n_2}({\bf x}-{\bf x}^{\prime},{\bf k})\overline{F}^{\,0}_{n_1n_2}({\bf k}(t_0),t_0)  \\
 + \big[\tilde{\Psi}_{n_1n_2}({\bf x}-{\bf x}^{\prime},{\bf k}) - \tilde{Y}_{n_1n_2}({\bf x}-{\bf x}^{\prime},{\bf k})\big]
 \overline{F}^{\,0}_{n_1}({\bf k}(t_0),t_0)        \\
 + \big[\tilde{\Psi}_{n_1n_2}({\bf x}-{\bf x}^{\prime},{\bf k}) + \tilde{Y}_{n_1n_2}({\bf x}-{\bf x}^{\prime},{\bf k})\big]
 \overline{F}^{\,0}_{n_2}({\bf k}(t_0),t_0) \Big\}    \\
 + O({\bf B}^3);
\end{multline}
here, $``$tilda$"$ indicates transformed variables ${\bf K} \rightarrow {\bf k} = {\bf K} - (e/\hbar c){\bf A}$. It is noted that when $n_1 = n_2 = n$, Eq.~(\ref{Sec IV a86}) reduces to Eq.~(\ref{Sec IV 80}), the single-band equation.

\subsection{Results for multiband WDF to first order in $({\bf K}-{\bf K}_0)$}\label{SubSecF IV}

In developing Eq.~(\ref{Sec IV 55a}) to first order in $\delta{\bf K} = {\bf K}-{\bf K}_0$, we note from Eq.~(\ref{Sec IV 53}) that $c_{nn^{\prime}}({\bf K}-{\bf K}_0) = (u_{n^{\prime}{\bf K}_0}, u_{n{\bf K}})$, and using the ${\bf k}\cdot{\bf p}$ method for $u_{n{\bf K}}$ in Eq.~(\ref{App-B11a}), we determine that
\begin{equation}\label{App-B17}
 c_{nn^{\prime}}({\bf K}-{\bf K}_0) = \delta_{nn^{\prime}} + ({\bf K}-{\bf K}_0) \cdot {\bf L}_{n^{\prime}n}({\bf K}_0)
 + O[({\bf K}-{\bf K}_0)^2]    \,,
\end{equation}
where
\begin{equation}\label{App-B18}
  {\bf L}_{n^{\prime}n({\bf K}_0}) = \frac{\hbar}{m}
  \frac{{\bf p}_{n^{\prime}n}({\bf K}_0)}{\varepsilon_n({\bf K}_0) - \varepsilon_{n^{\prime}}({\bf K}_0)} = -i {\bf R}_{n^{\prime}n}({\bf K}_0) \,,
\end{equation}
In keeping with Eq.~(\ref{Sec IV 54b}), it follows that
\begin{multline}\label{App-B19}
 c^{\ast}_{n_2n^{\prime\prime}}({\bf k}(t) - {\bf K}_0 - \frac{{\bf u}}{2}) c_{n_1n^{\prime}}({\bf k}(t) - {\bf K}_0 + \frac{{\bf u}}{2}) =   \\
 \delta_{n_2n^{\prime\prime}} \delta_{n_1n^{\prime}}
 + ({\bf k}(t) - {\bf K}_0) \cdot \Big[\delta_{n_2n^{\prime\prime}} {\bf
 L}_{n^{\prime}n_1}({\bf K}_0)   \\
 + \delta_{n_1n^{\prime}} {\bf L}^{\ast}_{n^{\prime\prime}n_2}({\bf K}_0) \Big]  \\
 + \frac{1}{2}{\bf u} \cdot \Big[\delta_{n_2 n^{\prime\prime}} {\bf L}_{n^{\prime}n_1}({\bf K}_0) - \delta_{n_1n^{\prime}} {\bf
 L}^{\ast}_{n^{\prime\prime}n_2}({\bf K}_0) \Big]   \\
 + O[({\bf k}-{\bf K}_0 \pm {\bf u})^2].
\end{multline}
Putting (\ref{App-B19}) with ${\bf u} \rightarrow -i\nabla_{{\bf x}}$ into Eq.~(\ref{Sec IV 55a}), we find
\begin{multline}\label{App-B20}
 f_{n_1 n_2}({\bf x}, {\bf p},t) = \sum_{{\bf K}} \Big[I^{(0)}_{n_1 n_2}({\bf x},{\bf p};{\bf K},{\bf K}_0) +   \\
 ({\bf k}(t) - {\bf K}_0) \cdot {\bf I}^{(1)}_{n_1 n_2}({\bf x},{\bf p};{\bf K},{\bf K}_0) - i {\bf I}^{(2)}_{n_1 n_2}({\bf x},{\bf p};{\bf K},{\bf K}_0) \cdot \nabla_{{\bf x}}\Big] \\
 \times f^0_{n_1 n_2}({\bf x},{\bf K},t)      \,,
\end{multline}
where $f^0_{n_1 n_2}({\bf x},{\bf K},t)$ is given in Eq.~(\ref{Sec IV 55b}).  $I^{(0)}_{n_1 n_2}, {\bf I}^{(1)}_{n_1 n_2}$, and ${\bf I}^{(2)}_{n_1 n_2}$ are explicitly given by
\begin{subequations}
\begin{multline}\label{App-B21a}
  I^{(0)}_{n_1 n_2}({\bf x},{\bf p};{\bf K},{\bf K}_0) = (2\pi\hbar)^{-3} \int d{\bf y}
  u_{n_2 {\bf K}_0}^{\ast}({\bf x} + \frac{{\bf y}}{2})    \\
  \times u_{n_1 {\bf K}_0}({\bf x} - \frac{{\bf y}}{2})
  e^{i({\bf p} - \hbar{\bf K})\cdot{\bf y}/\hbar}   \,,
\end{multline}
\begin{multline}\label{App-B21b}
  {\bf I}^{(1)}_{n_1 n_2}({\bf x},{\bf p};{\bf K},{\bf K}_0) = (2\pi\hbar)^{-3} \int d{\bf y}
  \Big[u_{n_2 {\bf K}_0}^{\ast}({\bf x} + \frac{{\bf y}}{2})   \\
  \times{\bf D}_{n_1 {\bf K}_0}({\bf x} - \frac{{\bf y}}{2}) + u_{n_1 {\bf K}_0}({\bf x} -
   \frac{{\bf y}}{2}) {\bf D}^{\ast}_{n_2 {\bf K}_0}({\bf x} + \frac{{\bf y}}{2}) \Big]     \\
  \times e^{i({\bf p} - \hbar{\bf K})\cdot{\bf y}/\hbar}   \,,
\end{multline}
and
\begin{multline}\label{App-B21c}
  {\bf I}^{(2)}_{n_1 n_2}({\bf x},{\bf p};{\bf K},{\bf K}_0) = (2\pi\hbar)^{-3} \int d{\bf y}
  \Big[u_{n_2 {\bf K}_0}^{\ast}({\bf x} + \frac{{\bf y}}{2})    \\
  \times {\bf D}_{n_1 {\bf K}_0}({\bf x} - \frac{{\bf y}}{2}) - u_{n_1 {\bf K}_0}({\bf x} - \frac{{\bf y}}{2}) {\bf D}^{\ast}_{n_2 {\bf K}_0}({\bf x} + \frac{{\bf y}}{2}) \Big]     \\
  \times e^{i({\bf p} - \hbar{\bf K})\cdot{\bf y}/\hbar}  \,,
\end{multline}
\end{subequations}
where ${\bf D}_{n {\bf K}_0}({\bf x})$ is given in Eq.~(\ref{App-B11b}). Thus, to first order in $\delta{\bf K}$, the explicit expression for $\hat{\Gamma}_{n_1 n_2}$ in Eq.~(\ref{Sec IV 55c}) is
\begin{multline}\label{App-B22}
 \hat{\Gamma}_{n_1 n_2}({\bf x},-i{\bf \nabla}_{\bf x},{\bf p};{\bf K},{\bf K}_0,t) = I^{(0)}_{n_1 n_2}({\bf x},{\bf p};{\bf K},{\bf K}_0)    \\
 + ({\bf K} - {\bf K}_0 + {\bf k}_c(t)) \cdot {\bf I}^{(1)}_{n_1 n_2}({\bf x},{\bf p};{\bf K},{\bf K}_0)  \\
 + {\bf I}^{(2)}_{n_1 n_2}({\bf x},{\bf p};{\bf K},{\bf K}_0) \cdot (-i\nabla_{{\bf x}})    \,.
\end{multline}
Here, we let ${\bf K}_0 = 0$ thereby defining the band edges. Then, the quantities $I^{(0)}_{n_1 n_2}$ and ${\bf I}^{(i)}_{n_1 n_2}$ ($i = 1,2$) can be evaluated with
\begin{equation}\label{App-B23}
  u_{n_{1,2} 0}({\bf x}) = \sum_{{\bf G}} A^{(n_{1,2})}_{{\bf G}}(0)e^{i{\bf G}\cdot{\bf x}} \,,
\end{equation}
where $(u_{n_1 0},u_{n_2 0}) = \delta_{n_1,n_2}$. This completes the derivation of $f_{n_1 n_2}({\bf x}, {\bf p},t)$ to order $\delta{\bf K}$
for ${\bf K}_0 = 0$. Lastly, we note that to order $\delta{\bf K}$, we have the time evolution of $f_{n_1 n_2}({\bf x},{\bf p},t)$ as
\begin{multline}\label{App-B24}
  \frac{\partial}{\partial t} f_{n_1 n_2}({\bf x},{\bf p},t) = \sum_{{\bf K}} \Big\{{\bf I}^{(1)}_{n_1 n_2}\cdot \frac{\partial}{\partial t} ({\bf k}_c f^0_{n_1n_2})    \\
  + \Big[I^{(0)}_{n_1 n_2} + ({\bf K} - {\bf K}_0) \cdot {\bf I}^{(1)}_{n_1 n_2} + {\bf I}^{(2)}_{n_1 n_2} \cdot (-i\nabla_{{\bf x}}) \Big] \frac{\partial}{\partial t} f^0_{n_1n_2}({\bf x},{\bf K},t) \Big\}     \,,
\end{multline}
where ${\bf k}_c(t) = (e/\hbar)\int_0^t {\bf E}(t^{\prime})dt^{\prime}$. Integrating Eq.~(\ref{App-B24}) allows us to introduce initial conditions
\begin{multline}\label{App-B25}
  f_{n_1 n_2}({\bf x},{\bf p},t) = f_{n_1 n_2}({\bf x},{\bf p},t_0) + \sum_{{\bf K}} \Big\{ \Big[I^{(0)}_{n_1 n_2} \\
  + ({\bf k}(t) - {\bf K}_0) \cdot {\bf I}^{(1)}_{n_1 n_2} + {\bf I}^{(2)}_{n_1 n_2} \cdot (-i\nabla_{{\bf x}}) \Big] f^0_{n_1n_2}({\bf x},{\bf K},t) \\
  - \Big[I^{(0)}_{n_1 n_2} + ({\bf K} - {\bf K}_0) \cdot {\bf I}^{(1)}_{n_1 n_2} + {\bf I}^{(2)}_{n_1 n_2} \cdot (-i\nabla_{{\bf x}}) \Big]    \\
  \times f^0_{n_1n_2}({\bf x},{\bf K},t_0) \Big\}  \,,
\end{multline}
where ${\bf k}(t) = {\bf K} + {\bf k}_c(t)$. Equation (\ref{App-B25}) shows explicitly that the multiband WDF depends directly upon the reduced
multiband WDF, $f^0_{n_1n_2}({\bf x},{\bf K},t)$, and its initial conditions as determined by the Liouville equation of Eq.~(\ref{Sec IV 69a}).

In the single-band case, with $n_1 = n_2 = n$ in Eq.~(\ref{App-B25}), the WDF $f_{n n}({\bf x},{\bf p},t) \equiv f_n({\bf x},{\bf p},t)$ can be transformed from variable ${\bf K}$ to ${\bf k}({\bf x},t)$ using the transformations of Eqs.~(\ref{Sec II 34a})-(\ref{Sec II 34c}); then the single-band reduced WDF, $F^0_n({\bf x},{\bf k},t)$, will satisfy the Boltzmann-like equation of Eq.~(\ref{Sec IV 80}) to $O({\bf B}^2)$. Thus, applying (\ref{Sec II 34a})-(\ref{Sec II 34c}) and (\ref{Sec IV 55c}) to Eq.~({\ref{App-B25}}) while keeping terms to $O({\bf B}^2)$ only, we find
\begin{multline}\label{App-B26}
  f_n({\bf x},{\bf p},t) = f_n({\bf x},{\bf p},t_0) + \sum_{{\bf k}} \Big[\tilde{I}^{(0)}_{nn}    \\
  + \big({\bf k}({\bf x},t) - {\bf K}_0\big) \cdot \tilde{{\bf I}}^{(1)}_{nn} + \tilde{{\bf I}}^{(2)}_{nn} \cdot (-i\nabla_{\bf x}) \Big] F^0_n({\bf x},{\bf k},t) \\
  - \Big[\tilde{I}^{(0)}_{nn} + ({\bf K} - {\bf K}_0) \cdot \tilde{{\bf I}}^{(1)}_{nn} + \tilde{{\bf I}}^{(2)}_{nn} \cdot (-i\nabla_{\bf x}) \Big] F^0_n({\bf x},{\bf k},t_0)   \,,
\end{multline}
where
\begin{subequations}
\begin{multline}\label{App-B27a}
  \tilde{I}^{(0)}_{nn}({\bf x},{\bf p};{\bf k},{\bf K}_0) = (2\pi\hbar)^{-3} \int d{\bf y}
  u_{n {\bf K}_0}^{\ast}({\bf x} + \frac{{\bf y}}{2})    \\
  \times u_{n {\bf K}_0}({\bf x} - \frac{{\bf y}}{2})
  e^{i[{\bf p} - \frac{e}{c}{\bf A}({\bf x},t) - \hbar{\bf k}]\cdot{\bf y}/\hbar}     \,,
\end{multline}
\begin{multline}\label{App-B27b}
  \tilde{{\bf I}}^{(1)}_{nn}({\bf x},{\bf p};{\bf k},{\bf K}_0) = (2\pi\hbar)^{-3} \int d{\bf y}
  \Big[u_{n {\bf K}_0}^{\ast}({\bf x} + \frac{{\bf y}}{2})    \\
  \times {\bf D}_{n {\bf K}_0}({\bf x} - \frac{{\bf y}}{2}) + u^{\ast}_{n {\bf K}_0}({\bf x} -
   \frac{{\bf y}}{2}) {\bf D}_{n {\bf K}_0}({\bf x} + \frac{{\bf y}}{2}) \Big]     \\
  \times e^{i[{\bf p} - \frac{e}{c}{\bf A}({\bf x},t) - \hbar{\bf k}]\cdot{\bf y}/\hbar}   \,,
\end{multline}
and
\begin{multline}\label{App-B27c}
  \tilde{{\bf I}}^{(2)}_{nn}({\bf x},{\bf p};{\bf k},{\bf K}_0) = (2\pi\hbar)^{-3} \int d{\bf y}
  \Big[u_{n {\bf K}_0}^{\ast}({\bf x} + \frac{{\bf y}}{2})    \\
  \times {\bf D}_{n {\bf K}_0}({\bf x} - \frac{{\bf y}}{2}) - u^{\ast}_{n {\bf K}_0}({\bf x} - \frac{{\bf y}}{2}) {\bf D}_{n {\bf K}_0}({\bf x} + \frac{{\bf y}}{2}) \Big]     \\
  \times e^{i[{\bf p} - \frac{e}{c}{\bf A}({\bf x},t) - \hbar{\bf k}]\cdot{\bf y}/\hbar}  \,.
\end{multline}
\end{subequations}
In Eq.~(\ref{App-B26}), $F^0_n({\bf x},{\bf k},t)$ is satisfied by Eq.~(\ref{Sec IV 80}).
We have already noted from Eq.~(\ref{App-B11b}) that ${\bf D}_{n{\bf K}_0}({\bf x})$ is expressed by Eq.~(\ref{App-B16a}). It then follows for the single-band case, when in the weak electric field limit ${\bf R}_{n^{\prime}n}({\bf K}_0) \approx 0$, that $\tilde{{\bf I}}^{(1)}_{nn} = \tilde{{\bf I}}^{(2)}_{nn} = 0$; therefore Eq.~(\ref{App-B26}) becomes, to lowest order in $\delta {\bf K}$,
\begin{multline}\label{}
  f_n({\bf x},{\bf p},t) = f_n({\bf x},{\bf p},t_0) + \sum_{{\bf k}} \Big[\tilde{I}^{(0)}_{nn}(t) F^0_n({\bf x},{\bf k},t)   \\
  - \tilde{I}^{(0)}_{nn}(t_0) F^0_n({\bf x},{\bf k},t_0) \Big]   \,,
\end{multline}
where $\tilde{I}^{(0)}_{nn}(t)$ is given by Eq.~(\ref{App-B27a}) and $F^0_n({\bf x},{\bf k},t)$ is given by Eq.~(\ref{Sec IV 80}).

\section{Summary}\label{}

Quantum transport and the associated Wigner phase space analog have been considered for Bloch electrons in homogeneous electric and magnetic
fields of arbitrary time dependence. We have specifically considered the case of collisionless or ballistic transport in this work so as
to focus mainly on electron kinematics and transport $``$streaming$"$ to second order in the magnetic field while treating the electric field exactly.
In the general formulation, starting from the Liouville equation for the density matrix, we define the first principles WDF in terms of
the instantaneous eigenstate basis and then transform to a new set of variables defined in terms of the position, kinetic momentum, and time to insure
the gauge invariance of the WDF for the uniform magnetic field.

Our methodology for constructing the WDF and the associated equation of motion is explicitly demonstrated by deriving the exact WDF equation
for a free electron in homogeneous electric and magnetic fields; this result is of the same form as that obtained for the collisionless Boltzmann transport equation,
except that all the consequences of the WDF approach, including the specification of WDF initial conditions and associated wave packet analysis, pertains
to the quantum regime. We further extend the methodology to the case of electrons described by an effective Hamiltonian for an arbitrary
energy-band function. An exact equation for the WDF is obtained, but results are reduced to second order in the magnetic field for
comparative analysis with the free electron case; here, we find the same form for the WDF equation as compared to the free-electron result,
except that the velocity is now defind in terms of the gradient with ${\bf k}$ of the energy dispersion instead of free particle velocity.
Lastly, we apply the methodology to the case of Bloch electrons in the presence of the electric and magnetic field. In using the ABR as our
instantaneous eigenstates, we develop a multiband WDF using the ${\bf K}_0$-representation outlined in Appendix A; the leading term
of the ${\bf k}\cdot{\bf p}$ method provides the lowest order term of the multiband WDF which we use to analyze the single-band and multiband picture.

We show that in order to obtain results correct to second order in the magnetic field, we have to introduce the method of unitary
transformations into the analysis to diagonalize the Hamiltonian using the ABR and simultaneously transform the Liouville equation to the
appropriate order to obtain results. The single-band analysis using ABR and neglecting interband effects gives rise to an energy dispersion
and WDF equation correct to second order in the magnetic field; the derived energy dispersion using the ABR is exactly what one would have
obtained if we assumed the effective Hamiltonian in the electric and magnetic field, and taken the matrix elements with respect to plane waves,
and then expanded the results to order ${\bf B}^2$, and replacing ${\bf x}$ by $\frac{1}{i}\nabla_{\bf K}$.
In multiband considerations, we examined the transition matrix elements appearing in the Liouville transport equation. It is found that, in addition to the usual electric Zener tunneling term, a magnetic interband tunneling term appears of $O({\bf B}^2)$ which shows the influence of the magnetic field on interband tunneling.

The results of this paper are considered to be the first of a two part effort. In a future companion paper, we will be extending results for the WDF in the combined electric and magnetic fields to include collisional field effects from impurities and phonons with application to valley dependent transport in low-dimensional materials. Further on, we will be considering the role of broken inversion symmetry and Berry phase corrections in this problem.

\newpage
\appendix

\section{The ${\bf K}_0$-representation~\cite{Kane}}\label{App-A}

The Schr\"{o}dinger equation for the Bloch wave is
\begin{equation}\label{App-B1a}
  \hat{H}_0 \psi_{n\bf K}({\bf x}) = \varepsilon_{n{\bf K}} \psi_{n\bf K}({\bf x})    \,.
\end{equation}
Here, $\hat{H}_0$ is the one-electron Hamiltonian which is periodic in the crystal. If we exclude spin-orbit interaction, the Hamiltonian is
\begin{equation}\label{App-B1b}
  \hat{H}_0 = \frac{{\bf p}^2}{2m} + V_c({\bf x})    \,, \nonumber
\end{equation}
where $V_c({\bf x})$ is the periodic potential of the crystal. Since $\psi_{n\bf K}({\bf x}) = \Omega^{-1/2} e^{i{\bf K}\cdot{\bf x}} u_{n\bf K}({\bf x})$, Eq.~(\ref{App-B1a}) may be written in terms of the cell periodic function $u_{n\bf K}({\bf x})$ as
\begin{equation}\label{App-B3}
  \hat{H}({\bf K}) u_{n\bf K}({\bf x}) = \varepsilon_{n{\bf K}} u_{n\bf K}({\bf x})    \,,
\end{equation}
where $\hat{H}({\bf K}) =    e^{-i{\bf K}\cdot{\bf x}} \hat{H}_0 e^{i{\bf K}\cdot{\bf x}}$, or
\begin{equation}\label{App-B5}
  \hat{H}({\bf K}) = \hat{H}_0 + \frac{\hbar}{m}{\bf K}\cdot{\bf p} + \frac{\hbar^2{\bf K}^2}{2m}   \,.
\end{equation}
Note that if spin-orbit interaction is included in $\hat{H}_0$, then $\hat{H}({\bf K})$ is still a polynomial of second degree in the components of ${\bf K}$. In letting ${\bf K} = {\bf K}_0$, we see that Eq.~(\ref{App-B3}) becomes
\begin{equation}\label{App-B6}
  \hat{H}({\bf K}_0) u_{n{\bf K}_0}({\bf x}) = \varepsilon_{n{\bf K}_0} u_{n{\bf K}_0}({\bf x})    \,,
\end{equation}
where
\begin{equation}\label{App-B7}
  \hat{H}({\bf K}_0) = \hat{H}_0 + \frac{\hbar}{m}{\bf K}_0\cdot{\bf p} + \frac{\hbar^2 K_0^2}{2m}   \,.
\end{equation}
In solving for $\hat{H}_0$ in Eq.~(\ref{App-B7}) and eliminating it from (\ref{App-B5}), we see that Eq.~(\ref{App-B3}) becomes
\begin{equation}\label{App-B8}
  [\hat{H}({\bf K}_0) + \frac{\hbar}{m}({\bf K}-{\bf K}_0)\cdot{\bf p} + \frac{\hbar^2}{2m}(K^2 - K_0^2)] u_{n\bf K}({\bf x})
  = \varepsilon_{n{\bf K}}  u_{n\bf K}({\bf x})  \,.
\end{equation}

It is well known~\cite{LuttKohn} that the $u_{n\bf K}({\bf x})$, for any value of ${\bf K}={\bf K}_0$, span a complete set of orthonormal functions for any function having the same periodicity of the lattice. Therefore, we can express $u_{n\bf K}({\bf x})$ in Eq.~(\ref{App-B3}) as
\begin{equation}\label{App-B9a}
 u_{n\bf K}({\bf x}) = \sum_{n^{\prime}} c_{nn^{\prime}}({\bf K}-{\bf K}_0)u_{n^{\prime}{\bf K}_0}({\bf x})   \,,
\end{equation}
where $(u_{n{\bf K}_0}, u_{n^{\prime}{\bf K}_0}) = \delta_{nn^{\prime}}$ and $c_{nn^{\prime}}(0) = \delta_{nn^{\prime}}$; the functions $u_{n{\bf K}_0}({\bf x})$ are presumed known and satisfy Eq.~(\ref{App-B6}).
Putting (\ref{App-B9a}) into Eq.~(\ref{App-B8}), multiplying both sides by $u^{\ast}_{n{\bf K}_0}({\bf x})$ and integrating over the unit cell, we obtain
\begin{subequations}
\begin{multline}\label{App-B10a}
  \sum_{n^{\prime}} c_{nn^{\prime}}({\bf K} - {\bf K}_0) \Big\{ [\varepsilon_{n^{\prime}{\bf K}_0} - \varepsilon_{n{\bf K}} + \frac{\hbar^2}{2m}({\bf K}^2 - {\bf K}_0^2)] \delta_{nn^{\prime}}  \\
  + \frac{\hbar}{m}({\bf K} - {\bf K}_0) \cdot {\bf p}_{nn^{\prime}}({\bf K}_0)\Big\} = 0   \,,
\end{multline}
where the subindex $n^{\prime}$ sums over all bands and
\begin{equation}\label{App-B10b}
   {\bf p}_{nn^{\prime}}({\bf K}_0)  = \frac{1}{\Omega_c} \int_{\Omega_c} u^{\ast}_{n{\bf K}_0}({\bf x}) {\bf p} u_{n^{\prime}{\bf K}_0}({\bf x}) d {\bf x}   \,.
\end{equation}
\end{subequations}
 Equation (\ref{App-B10a}) is the matrix eigenvalue equation for the point ${\bf K}$ in ${\bf K}$ space in the so-called ${\bf K}_0$-representation~\cite{Kane}; although the equation for $c_{nn^{\prime}}({\bf K} - {\bf K}_0)$ is exact for any ${\bf K}$, it is most amenable to approximate solution when ${\bf K}$ is chosen near ${\bf K}_0$, for then, the off-diagonal term can be treated as a perturbation using the ${\bf k}\cdot{\bf p}$ formalism. For simplicity here, we will consider ${\bf K}_0$ to be an extremum point in ${\bf K}$ space such that $(\nabla_{{\bf K}} \varepsilon_{n{\bf K}})_{{\bf K} = {\bf K}_0} = 0$ for all bands; the specific case of ${\bf K}_0 = 0$ will be considered in Sec. IV F.

Although the sum $\sum_{n^{\prime}}(...)$ in Eq.~(\ref{App-B10a}) is over all bands, and is thus an infinite sum, the equation is amenable to perturbation theory, with $\delta{\bf K} \cdot {\bf p}_{nn^{\prime}}({\bf K}_0)$ as a perturbation. Then, we find to first order in $\delta{\bf K} = {\bf K} - {\bf K}_0$ that
\begin{subequations}
\begin{equation}\label{App-B11a}
  u_{n{\bf K}}({\bf x}) = u_{n{\bf K}_0}({\bf x}) + \delta{\bf K} \cdot {\bf D}_{n{\bf K}_0}({\bf x})  \,,
\end{equation}
where
\begin{equation}\label{App-B11b}
  {\bf D}_{n{\bf K}_0}({\bf x}) = \frac{\hbar}{m} \sum_{n^{\prime} \neq n}
  \frac{{\bf p}_{n^{\prime}n}({\bf K}_0)}{\varepsilon_{n{\bf K}_0} - \varepsilon_{n^{\prime}{\bf K}_0}} u_{n^{\prime}{\bf K}_0}({\bf x})
\end{equation}
\end{subequations}
and
\begin{multline}\label{App-B12}
  \varepsilon_{n{\bf K}} = \varepsilon_{n{\bf K}_0} + \frac{\hbar}{m} \delta{\bf K} \cdot[{\bf p}_{nn}({\bf K}_0) + \hbar {\bf K}_0] + \frac{\hbar^2}{2m} (\delta{\bf K})^2  \\
  + \frac{\hbar^2}{m^2} \sum_{n^{\prime} \neq n} \frac{[\delta{\bf K} \cdot {\bf p}_{nn^{\prime}}({\bf K}_0)]
  [\delta{\bf K} \cdot {\bf p}_{n^{\prime}n}({\bf K}_0)]}{\varepsilon_{n{\bf K}_0} - \varepsilon_{n^{\prime}{\bf K}_0}}  \,.
\end{multline}
Note that the linear term vanishes since $(\psi_{n{\bf K}_0},{\bf p}\psi_{n{\bf K}_0}) = {\bf p}_{nn}({\bf K}_0) + \hbar {\bf K}_0 = m\hbar^{-1} (\nabla_{{\bf K}} \varepsilon_{n{\bf K}})_{{\bf K} = {\bf K}_0}$, which is zero since ${\bf K}_0$ is an extremum point in ${\bf K}$ space. Equations (\ref{App-B11a})-(\ref{App-B12}) can be simplified. First, noting that the inverse effective-mass tensor can be expressed as
\begin{equation}\label{App-B13}
  m^{-1}_{ij} = m^{-1}{\delta_{ij}} + 2{m^{-2}}
  \sum_{n^{\prime} \neq n} \frac{p ({\bf K}_0)_{nn^{\prime},i} p ({\bf K}_0)_{n^{\prime}n,j}}
  {\varepsilon_{n{\bf K}_0} - \varepsilon_{n^{\prime}{\bf K}_0}}   \nonumber
\end{equation}
through the $f$-sum rule, then Eq.~(\ref{App-B12}) takes the form
\begin{equation}\label{App-B14}
  \varepsilon_{n{\bf K}} = \varepsilon_{n{\bf K}_0} + \frac{\hbar^2}{2}
  \sum_{i,j=1}^3 \frac{\delta K_i \delta K_j}{m^{\ast}_{ij}}   \,.
\end{equation}
Now, noting that $[{\bf x},\hat{H}_0] = (i\hbar/m){\bf p}$, we find that the off-diagonal matrix elements of ${\bf x}$ and ${\bf p}$ are related by
\begin{subequations}
\begin{equation}\label{App-B15a}
  (\psi_{n^{\prime}{\bf K}_0},{\bf x}\psi_{n{\bf K}_0}) = \frac{i\hbar}{m} \frac{(\psi_{n^{\prime}{\bf K}_0},{\bf p}\psi_{n{\bf K}_0})}{\varepsilon_{n{\bf K}_0} - \varepsilon_{n^{\prime}{\bf K}_0}} \,, \; n \neq n^{\prime}  \,.
\end{equation}
But since, in the Bloch representation,
\begin{eqnarray}\label{}
  \nonumber
  (\psi_{n^{\prime}{\bf K}^{\prime}},{\bf x} \psi_{n{\bf K}}) = \big(\delta_{n^{\prime}n}\frac{1}{i}\nabla_{\bf K} + {\bf R}_{n^{\prime}n}({\bf K})\big) \delta_{{\bf K}^{\prime}{\bf K}},  \\
  (\psi_{n^{\prime}{\bf K}^{\prime}},{\bf p}\psi_{n{\bf K}}) =  \big(\hbar {\bf K} \delta_{n^{\prime}n}
  + {\bf p}_{n^{\prime}n}({\bf K}) \big) \delta_{{\bf K}^{\prime}{\bf K}} \,,   \nonumber
\end{eqnarray}
where ${\bf R}_{n^{\prime}n}({\bf K})$ is given in Eq.~(\ref{Sec IV 58b}),
it follows that Eq.~(\ref{App-B15a}) becomes
\begin{equation}\label{App-B15b}
  {\bf R}_{n^{\prime}n}({\bf K}_0)  = \frac{i\hbar}{m} \frac{{\bf p}_{n^{\prime}n}({\bf K}_0)}{\varepsilon_{n{\bf K}_0} - \varepsilon_{n^{\prime}{\bf K}_0}}   \,, \; n \neq n^{\prime}  \,.
\end{equation}
\end{subequations}
Thus, ${\bf D}_{n{\bf K}_0}({\bf x})$ in  Eq.~(\ref{App-B11b}) becomes
\begin{subequations}
\begin{equation}\label{App-B16a}
  {\bf D}_{n{\bf K}_0}({\bf x}) = -i \sum_{n^{\prime} \neq n}
  {\bf R}_{n^{\prime}n}({\bf K}_0) u_{n^{\prime}{\bf K}_0}({\bf x})   \,.
\end{equation}
Since $\nabla_{\bf K} u_{n{\bf K}}({\bf x})$ is a periodic function of ${\bf x}$, we can write
\begin{equation}\label{App-B16b}
  i \nabla_{\bf K} u_{n{\bf K}}({\bf x}) = \sum_{n^{\prime} \neq n}
  {\bf R}_{n^{\prime}n}({\bf K}) u_{n^{\prime}{\bf K}}({\bf x})  \,;
\end{equation}
[note that, in this work, the phases of $\psi_{n{\bf K}}({\bf x})$ are chosen so that ${\bf R}_{nn}({\bf K}) = 0$]. Then, ${\bf D}_{n{\bf K}_0}$
of Eq.~(\ref{App-B16a}) becomes ${\bf D}_{n{\bf K}_0}({\bf x}) = \nabla_{{\bf K}_0} u_{n{\bf K}_0}({\bf x})$.
Therefore, $u_{n{\bf K}}({\bf x})$ of Eq.~(\ref{App-B11a}) can be formally expressed as
\begin{equation}\label{App-B16d}
  u_{n{\bf K}}({\bf x}) = u_{n{\bf K}_0}({\bf x}) + ({\bf K} - {\bf K}_0) \cdot \nabla_{{\bf K}_0} u_{n{\bf K}_0}({\bf x})  \,,
\end{equation}
\end{subequations}
a first-order Taylor series expansion of $u_{n{\bf K}}({\bf x})$ about $({\bf K} - {\bf K}_0)$; using (\ref{App-B16b}), Eq.~(\ref{App-B16d}) can always be expressed explicitly in terms of ${\bf R}_{n^{\prime}n}({\bf K}_0)$.

In using (\ref{App-B15b}) in Eq.~(\ref{App-B10a}), the matrix equation for $c_{nn^{\prime}}({\bf K}-{\bf K}_0)$ can be written in a form amenable to high-order perturbation theory in $({\bf K} - {\bf K}_0)$ as
\begin{subequations}
\begin{multline}\label{App-B17a}
  \sum_{n^{\prime}} c_{nn^{\prime}}({\bf K} - {\bf K}_0) \Big\{ \big(\varepsilon_{n^{\prime}{\bf K}_0} - \varepsilon_{n{\bf K}_0}\big)   \\
  \times [\delta_{nn^\prime} - i({\bf K} - {\bf K}_0) \cdot {\bf R}_{nn^{\prime}}({\bf K}_0)]  \\
  + \big[\frac{\hbar^2}{2m}({\bf K} - {\bf K}_0)^2
  - (\varepsilon_{n{\bf K}} - \varepsilon_{n{\bf K}_0})\big] \delta_{nn^{\prime}} \Big\} = 0   \,,
\end{multline}
where
\begin{multline}\label{App-B17b}
  c_{nn^{\prime}}({\bf K} - {\bf K}_0) = c_{nn^{\prime}}(0) +  \nabla_{{\bf K}}c_{nn^{\prime}}({\bf K} - {\bf K}_0)|_{{\bf K}={\bf K}_0}\cdot({\bf K} - {\bf K}_0)   \\
  + \frac{1}{2!} \sum_{i,j=1}^3 \frac{\partial^2 c_{nn^{\prime}}({\bf K} - {\bf K}_0)}{\partial K_i \partial K_j}\Big|_{{\bf K}={\bf K}_0}   \\
  \times (K - K_0)_i (K - K_0)_j + O[({\bf K} - {\bf K}_0)^3]
\end{multline}
and
\begin{multline}\label{App-B17c}
  \varepsilon_{n{\bf K}} - \varepsilon_{n{\bf K}_0} = \frac{1}{2!} \sum_{i,j=1}^3 \frac{\partial^2 \varepsilon_{n{\bf K}}}{\partial K_i \partial  K_j}\Big|_{{\bf K}={\bf K_0}}    \\
  \times (K - K_0)_i (K - K_0)_j + O[({\bf K} - {\bf K}_0)^3]      \,,
\end{multline}
\end{subequations}
with $(\nabla_{{\bf K}} \varepsilon_{n{\bf K}})_{{\bf K}={\bf K}_0}= 0$.

As an alternative to this perturbation approach, one can expand on the previous result of Eq.~(\ref{App-B16d}) and expand $u_{n{\bf K}}({\bf x})$ in a Taylor series about ${\bf K}_0$ as
\begin{subequations}
\begin{multline}\label{App-B18a}
  u_{n{\bf K}}({\bf x}) = \Big[1 + \sum_{i=1}^3 (K - K_0)_i \frac{\partial}{\partial K_{0i}}   \\
  + \frac{1}{2!} \sum_{i\,j=1}^3
  (K - K_0)_i (K - K_0)_j \frac{\partial^2}{\partial K_{0i} \partial K_{0j}}\Big] u_{n{\bf K}_0}({\bf x})   \\
   + O[({\bf K} - {\bf K}_0)^3];
\end{multline}
throughout the use of $c_{nn^{\prime}}({\bf K}-{\bf K}_0) = (u_{n^{\prime}{\bf K}_0}, u_{n{\bf K}})$ and the repeated use of Eq.~(\ref{App-B16b}) to establish the coefficients of $c_{nn^{\prime}}({\bf K} - {\bf K}_0)$, the Taylor series of $u_{n{\bf K}}$ about $({\bf K}-{\bf K}_0)$ can be found to any desired order. Putting $u_{n{\bf K}}$ of Eq.~(\ref{App-B18a}) into $c_{nn^{\prime}}({\bf K}-{\bf K}_0) = (u_{n^{\prime}{\bf K}_0}, u_{n{\bf K}})$, we get
\begin{multline}\label{App-B18b}
  c_{nn^{\prime}}({\bf K} - {\bf K}_0) = \delta_{nn^{\prime}} + ({\bf K} - {\bf K}_0)\cdot (u_{n^\prime{\bf K}_0},\nabla_{{\bf K}_0}u_{n{\bf K}_0})  \\
  + \frac{1}{2!} \sum_{i,j=1}^3 (K - K_0)_i (K - K_0)_j \big(u_{n^\prime{\bf K}_0}, \frac{\partial^2 u_{n{\bf K}_0}}{\partial K_{0i} \partial K_{0j}} \big)    \\
  + O[({\bf K} - {\bf K}_0)^3].
\end{multline}
\end{subequations}
Using (\ref{App-B16b}), we find
\begin{equation}\label{App-B18c}
  (u_{n^\prime{\bf K}_0},\nabla_{{\bf K}_0}u_{n{\bf K}_0}) = -i {\bf R}_{n^{\prime}n}({\bf K}_0)  \,, \nonumber
\end{equation}
\begin{multline}\label{App-B18d}
  \Big(u_{n^\prime{\bf K}_0}, \frac{\partial^2 u_{n{\bf K}_0}}{\partial K_{0l} \partial K_{0m}} \Big) =
  - \sum_{n^{\prime \prime}} R^l_{n^{\prime}n^{\prime\prime}}({\bf K}_0) R^m_{n^{\prime\prime}n}({\bf K}_0)   \\
  - i \frac{\partial}{\partial K_{0l}} R^m_{n^{\prime}n}({\bf K}_0)  \,, \nonumber
\end{multline}
where $R^l_{n^{\prime}n}({\bf K}_0)$ is the $l$th Cartesian component of ${\bf R}_{n^{\prime}n}({\bf K}_0)$.
Of course, once $c_{nn^{\prime}}({\bf K} - {\bf K}_0)$ is determined to a desired order of $({\bf K} - {\bf K}_0)$, then $(\varepsilon_{n{\bf K}} - \varepsilon_{n{\bf K}_0})$ immediately follows from Eq.~(\ref{App-B17a}).

From Eq.~(\ref{App-B9a}), we use $u_{{n\bf k}(t)}({\bf x}) = \sum_{n^{\prime}} c_{nn^{\prime}}({\bf k}(t)-{\bf K}_0)u_{n^{\prime}{\bf K}_0}({\bf x})$ in Eqs.~(\ref{Sec IV 52a}), (\ref{Sec IV 52b}), and we find the expressions for $f_{n_1 n_2}$, $f_{n_1 n_2}^0$, $\hat{\Gamma}_{n_1 n_2}$, and ${\cal I}_{n^{\prime\prime}n^{\prime}}$ given in Eqs.~(\ref{Sec IV 55a})-(\ref{Sec IV 55c}) and (\ref{Sec IV 54bb}), respectively.

\section{Hamiltonian diagonalization by unitary transformation}\label{App-C}

\subsection{The general scheme}
In our problem, as noted in Eq.~(\ref{Sec IV 66}), we have a Hamiltonian of the form
\begin{equation}\label{C1}
 \hat{H} = \hat{H}_0 + \beta V_1 + \beta^2 V_2 \,,
\end{equation}
in which the unperturbed Hamiltonian $\hat{H}_0$ is diagonal in the accelerated Bloch state representation
\begin{equation}\label{C2}
 \langle n^{\prime},{\bf K}^{\prime};t|\hat{H}_0|n,{\bf K};t \rangle = \varepsilon_n^0({\bf k}(t)) \delta_{nn^{\prime}} \delta_{{\bf K}{\bf K}^{\prime}}  \,.
\end{equation}
At the same time, the full Hamiltonian $\hat{H}$ is not diagonal in this convenient basis due to the perturbation terms of the two magnetic field potentials, $V_1$ and $V_2$. We now use the unitary transformation
\begin{equation}\label{C3}
 \overline{\hat{H}} = e^{-i\hat{U}}\hat{H} e^{i\hat{U}}
\end{equation}
and
\begin{equation}\label{C4}
 \overline{|n,{\bf K};t \rangle} = e^{i\hat{U}} |n,{\bf K};t \rangle
\end{equation}
to diagonalize $\hat{H}$ in Eq.~(\ref{C1}) to second order in ${\bf B}$. To this end, we expand the Hermitian operator $\hat{U}$ $(\hat{U}^{\dag} = \hat{U})$ as follows
\begin{equation}\label{C5}
 \hat{U} = \beta \hat{U}_1 + \beta^2 \hat{U}_2  + \ldots \,,
\end{equation}
where the subindeces of $\hat{U}_i$ stand for the order of the appropriate perturbation.
We thus look for $\hat{U}$ to second order in ${\bf B}$, which diagonalizes the Hamiltonian of Eq.~(\ref{C1}). Using the well-known formula
\begin{equation}\label{C7}
 e^{-i\hat{U}}\hat{H} e^{i\hat{U}} = \hat{H} + i[\hat{H},\hat{U}] - \frac{1}{2} [[\hat{H},\hat{U}],\hat{U}] + O(\hat{U}^3),
\end{equation}
and putting (\ref{C1}) and (\ref{C5}) into Eq.~(\ref{C7}), we arrive at $\overline{\hat{H}}$ to $O({\bf B}^2)$ as
\begin{equation}\label{C8}
 \overline{\hat{H}} = \hat{H}_0 + \beta \hat{R}_1 + \beta^2 \hat{R}_2       \,,
\end{equation}
where
\begin{equation}\label{C9a}
 \hat{R}_1 = V_1 + i[\hat{H}_0,\hat{U}_1]  \,, \nonumber
\end{equation}
\begin{equation}\label{C9b}
 \hat{R}_2 =  V_2 + i[\hat{H}_0,\hat{U}_2] + i[V_1,\hat{U}_1] - \frac{1}{2} [[\hat{H}_0,\hat{U}_1],\hat{U}_1] \,.
\end{equation}
Since $\hat{H}_0$ is already diagonal in the ABR basis, we chose $\hat{U}_{i}$ such that the off-diagonal matrix elements of $\hat{R}_1$ and $\hat{R}_2$ in the ABR are zero term by term. Then, from (\ref{C9b}) after matrix elements are taken, we see that
\begin{equation}\label{C10a}
  [\hat{H}_{0},\hat{U}_1]  =  i V_1  \,, \nonumber
\end{equation}
\begin{equation}\label{C10b}
  [\hat{H}_0,\hat{U}_{2}]  =  iV_{2} - \frac{1}{2} [V_1,\hat{U}_{1}]   \,.
\end{equation}
These equations give rise to commutator relations for $\hat{U}_i$ with $\hat{H}_0$. The right-hand side of each equation depends on lower order terms in $\hat{U}_i$, so we thereby have a hierarchy of relations.
The off-diagonal matrix elements of each operator $\hat{U}_i$ can now be found by taking the matrix elements of (\ref{C10b}) with respect to $\langle n,{\bf K};t|\ldots|n^{\prime},{\bf K}^{\prime};t\rangle$. We note that commutators of the type $[\hat{H}_0,\hat{U}_i] = \hat{A}_i$ are such that the appropriate matrix elements
$\langle n,{\bf K};t|\hat{[H}_0,\hat{U}_i]|n^{\prime},{\bf K}^{\prime};t\rangle = \langle n,{\bf K};t|\hat{A}_i|n^{\prime},{\bf K}^{\prime};t\rangle$
are obtained as
\begin{equation}\label{C11}
 (\varepsilon_{n{\bf k}}^0 - \varepsilon^0_{n^{\prime}{\bf k}^{\prime}})(U_i)_{n{\bf K}n^{\prime}{\bf K}^{\prime}} = (A_i)_{n{\bf K}n^{\prime}{\bf K}^{\prime}}   \,,
\end{equation}
and for $n{\bf K} \neq n^{\prime}{\bf K}^{\prime}$
\begin{equation}\label{C12}
 (U_i)_{n{\bf K}n^{\prime}{\bf K}^{\prime}} = \frac{(A_i)_{n{\bf K}n^{\prime}{\bf K}^{\prime}}}{\varepsilon_{n{\bf k}}^0 - \varepsilon^0_{n^{\prime}{\bf k}^{\prime}}}   \,;
\end{equation}
here, $\hat{A}_i$ stands for the right-hand side of equations (\ref{C10b}). The explicit expressions for matrix elements $(A_i)_{n{\bf K}n^{\prime}{\bf K}^{\prime}}$ are evaluated below.

It is clear that for $n{\bf K} = n^{\prime}{\bf K}^{\prime}$ the equation (\ref{C11}) leaves the diagonal matrix elements $(U_i)_{n{\bf K}n{\bf K}}$ arbitrary and undecided. To determine the diagonal elements of $\hat{U}_i$ we look for the perturbed wave function for $\hat{H}$ such that $\overline{|n,{\bf K};t \rangle} = |n,{\bf K};t \rangle + |\Phi \rangle$, where the change due to the perturbation, $|\Phi \rangle$, is orthogonal to the unperturbed state, $|n,{\bf K};t\rangle$; then, it follows that
\begin{equation}\label{C17}
 \langle n,{\bf K};t|\Phi\rangle  = 0   \,;
\end{equation}
this is frequently called intermediate normalization. Then, making use of Eq.~(\ref{C4}), we expand the exponent in this equation into a series, with $\hat{U}$ given in Eq.~(\ref{C5}), and group terms according to their order in ${\bf B}$, and so on. The result for
$\overline{|n,{\bf K};t\rangle} - |n,{\bf K};t\rangle = |\Phi\rangle$ is
\begin{equation}\label{C18}
 |\Phi\rangle = \Big[\beta \,i \hat{U}_1 + \beta^2 (i\hat{U}_2 - \frac{1}{2}\hat{U}_1^2 ) \Big] |n,{\bf K};t\rangle  \,.
\end{equation}
The diagonal matrix elements of the transformation matrix can be found from Eq.~(\ref{C17}) with making use of the obtained expression for $|\Phi\rangle$ (\ref{C18}).

\subsection{Off-diagonal elements of the transformation matrix}

To find the explicit expressions for off-diagonal elements of the transformation matrix $U_{n{\bf K}n^{\prime}{\bf K}^{\prime}}$, we use Eqs. (\ref{C10b}) and (\ref{C12}). Then, we obtain for $O({\bf B})$
\begin{equation}\label{offU1}
 (U_1)_{n{\bf K}n^{\prime}{\bf K}^{\prime}} = i \frac{(V_1)_{n{\bf K}n^{\prime}{\bf K}^{\prime}}}{\varepsilon^0_{n{\bf k}} - \varepsilon^0_{n^{\prime}{\bf k}^{\prime}}}
\end{equation}
and for $O({\bf B}^2)$
\begin{multline}\label{offU2}
 (U_2)_{n{\bf K}n^{\prime}{\bf K}^{\prime}} = \frac{i}{\varepsilon^0_{n{\bf k}} - \varepsilon^0_{n^{\prime}{\bf k}^{\prime}}}
 \Big[(V_2)_{n{\bf K}n^{\prime}{\bf K}^{\prime}} +    \\
 \frac{1}{2} {\sum_{n^{\prime\prime}{\bf K}^{\prime\prime}}}^{\prime}
 (V_1)_{n{\bf K}n^{\prime\prime}{\bf K}^{\prime\prime}}
 (V_1)_{n^{\prime\prime}{\bf K}^{\prime\prime}n^{\prime}{\bf K}^{\prime}}   \\
 \times \Big( \frac{1}{\varepsilon^0_{n{\bf k}} - \varepsilon^0_{n^{\prime\prime}{\bf k}^{\prime\prime}}}  +
 \frac{1}{\varepsilon^0_{n^{\prime}{\bf k}^{\prime}} - \varepsilon^0_{n^{\prime\prime}{\bf k}^{\prime\prime}}} \Big)\Big] \,,
\end{multline}
where $``$prime$"$ in the sum means that the summation is over $(n^{\prime\prime}{\bf K}^{\prime\prime}) \neq (n{\bf K}, n^{\prime}{\bf k}^{\prime})$.

\subsection{Diagonal elements of the transformation matrix}

The diagonal elements of the transformation matrix $U_{n{\bf K}n{\bf K}}$ are evaluated from Eqs.~(\ref{C17}) and (\ref{C18}) to the considered order in interaction with magnetic field. We reproduce them term by term in order according to Eq.~(\ref{C5}). So, in the lowest order in ${\bf B}$, we find
\begin{equation}\label{DiagU1}
 (U_1)_{n{\bf K}n{\bf K}} = 0  \,;
\end{equation}
in the second order in ${\bf B}$, we obtain
\begin{equation}\label{C19}
  (U_2)_{n{\bf K}n{\bf K}} = - \frac{i}{2} (U_1^2)_{n{\bf K}n{\bf K}}   \,.
\end{equation}
Clearly, it is seen that the term $(U_1)_{n{\bf K}n{\bf K}}$ is zero, whereas $(U_2)_{n{\bf K}n{\bf K}}$ depends on diagonal elements of $\hat{U}_1^2$. Then, making use of Eq.~(\ref{offU1}), we obtain the diagonal matrix elements for $\hat{U}_2$, in the second order in ${\bf B}$, as
\begin{equation}\label{DiagU2}
 (U_2)_{n{\bf K}n{\bf K}} = - \frac{i}{2} \sum_{n^{\prime}{\bf K}^{\prime} \neq n{\bf K}}\left|\frac{(V_1)_{n{\bf K}n^{\prime}{\bf K}^{\prime}}}{\varepsilon^0_{n{\bf k}} - \varepsilon^0_{n^{\prime}{\bf k}^{\prime}}}\right|^2   \,.
\end{equation}

\section{Determination of key matrix elements}\label{App-D}

The required matrix elements of $\hat{U}$ are now analyzed in terms of their perturbation theory contributions defined in Eqs.~(\ref{Sec IV 66})
and (\ref{Sec IV 67}). The particular $\hat{U}_i(\beta)$ are derived for each off-diagonal and diagonal term of perturbation [Eq.~(\ref{Sec IV 66})] in the ABR and can be found in Appendix B, Eqs.~(\ref{offU1})-(\ref{offU2}) and (\ref{DiagU1})-(\ref{DiagU2}), respectively. For each $\hat{U}_i$, the key matrix elements depend on terms in the Hamiltonian of Eq.~(\ref{Sec IV 48a}) [See also Eq.~(\ref{Sec II 9})],
\begin{equation}\label{App-D1}
 V_1 = -\frac{e}{mc} {\bf A}_2 \cdot ({\bf p} - \frac{e}{c}{\bf A}_1) ,    \;\;\;
 V_2 = \frac{e^2}{2mc^2} {\bf A}_2^2    \,.
\end{equation}
Here, $V_1$ is the interaction of the magnetic field with the dynamic electron, and $V_2$ is the second order term in the magnetic field. We consider the matrix elements of $V_1$ and $V_2$ in the ABR. This allowing for the determination of $\hat{U}_i$ for each perturbation term.

\subsection{Matrix elements of $V_1({\bold x},t)$}

The matrix elements of $V_1({\bold x},t)$,
\begin{equation}\label{Sec IV 84}
  (V_1)_{n{\bf K}n^{\prime}{\bf K}^{\prime}} = -\frac{e}{mc} \sum_{n^{\prime\prime}{\bf K}^{\prime\prime}} ({\bf A}_2)_{n{\bf K}n^{\prime\prime}{\bf K}^{\prime\prime}} \cdot ({\bf p} - \frac{e}{c}{\bf A}_1)_{n^{\prime\prime}{\bf K}^{\prime\prime}n^{\prime}{\bf K}^{\prime}}    ,
\end{equation}
where ${\bf A}_1$ and ${\bf A}_2$ are defined in Eqs.~(\ref{Sec II 7}) and (\ref{Sec II 8}), respectively, are evaluated as follows:
\begin{equation}\label{Sec IV 85}
 \frac{1}{m} ({\bf p} - \frac{e}{c}{\bf A}_1)_{n^{\prime\prime}{\bf K}^{\prime\prime}n^{\prime}{\bf K}^{\prime}} =
 {\bf v}_{n^{\prime\prime}n^{\prime}}({\bf k}^{\prime}(t)) \delta_{{\bf K}^{\prime\prime}{\bf K}^{\prime}} ,
\end{equation}
where  ${\bf v}_{n^{\prime\prime}n^{\prime}}({\bf k}^{\prime})$ is well known,~\cite{Krieger} that is
\begin{equation}\label{}
 {\bf v}_{n^{\prime\prime}n^{\prime}}({\bf k}^{\prime}) = \frac{1}{\hbar} {\bf \nabla}_{{\bf k}^{\prime}} \varepsilon^0_{n^{\prime}{\bf k}^{\prime}} ,  \;\;\;\; n^{\prime\prime} = n^{\prime} ,  \nonumber
\end{equation}
\begin{equation}\label{voffdiag}
 {\bf v}_{n^{\prime\prime}n^{\prime}}({\bf k}^{\prime}) = \frac{i}{\hbar}
 \left(\varepsilon^0_{n^{\prime\prime}{\bf k}^{\prime}} - \varepsilon^0_{n^{\prime}{\bf k}^{\prime}}\right) {\bold R}_{n^{\prime\prime}n^{\prime}}({\bf k}^{\prime}),  \;\;\; n^{\prime\prime} \neq n^{\prime}    \,;
\end{equation}
here
\begin{equation}\label{}
 {\bf R}_{n^{\prime\prime}n^{\prime}}({\bf k}) = \frac{i}{\Omega_c} \int_{\Omega_c}
 u^{\ast}_{n^{\prime\prime}{\bf k}}({\bf x}) {\bf \nabla}_{\bf k} u_{n^{\prime}{\bf k}}({\bf x}) d{\bf x}  = {\bf R}^{\ast}_{n^{\prime}n^{\prime\prime}}({\bf k})   \,.
\end{equation}
For ${\bf A}_2$ from Eq.~(\ref{Sec II 8}), the matrix elements 
are reduced to
\begin{multline}\label{Sec IV 86}
 \left({\bf A}_2 \right)_{n{\bf K}n^{\prime\prime}{\bf K}^{\prime\prime}} = \frac{i}{2}{\bf B}\times \Big[{\bf \nabla}_{\bf k} \delta_{nn^{\prime\prime}}\delta_{{\bf K}{\bf K}^{\prime\prime}}   \\
 - \frac{1}{\Omega} \int d{\bf x}{\bf \nabla}_{\bf k}
 u^{\ast}_{n{\bf k}}({\bf x})  u_{n^{\prime\prime}{\bf k}^{\prime\prime}}({\bf x}) e^{-i({\bf K} - {\bf K}^{\prime\prime})\cdot {\bf x}} \Big]   \,.
\end{multline}
Since
\begin{multline}\label{Sec IV 87}
 \frac{1}{\Omega} \int d{\bf x}{\bf \nabla}_{\bf k} u^{\ast}_{n{\bf k}}({\bf x})  u_{n^{\prime\prime}{\bf k}^{\prime\prime}}({\bf x})
 e^{-i({\bf K} - {\bf K}^{\prime\prime})\cdot {\bf x}}   \\
 = \frac{\delta_{{\bf K}{\bf K}^{\prime\prime}}}{\Omega_c}
 \int_{\Omega_c} d{\bf x}({\bf \nabla}_{\bf k} u^{\ast}_{n{\bf k}}({\bf x}))  u_{n^{\prime\prime}{\bf k}}({\bf x})              \,,
\end{multline}
Eq.~(\ref{Sec IV 86}) becomes
\begin{multline}\label{Sec IV 88}
 \left({\bf A}_2 \right)_{n{\bf K}n^{\prime\prime}{\bf K}^{\prime\prime}} = \frac{1}{2} {\bf B}\times \Big[i{\bf \nabla}_{\bf k} \delta_{nn^{\prime\prime}}   \\
  - \frac{i}{\Omega_c}
 \int_{\Omega_c} d{\bf x}({\bf \nabla}_{\bf k} u^{\ast}_{n{\bf k}}({\bf x}))  u_{n^{\prime\prime}{\bf k}}({\bf x}) \Big] \delta_{{\bf K}{\bf K}^{\prime\prime}} \,.
\end{multline}
Taking into account that ${\bf \nabla}_{\bf k} \int u^{\ast}_{n{\bf k}}({\bf x}) u_{n^{\prime\prime}{\bf k}}({\bf x})d{\bf x} = 0$, we can express (\ref{Sec IV 88}) as
\begin{equation}\label{Sec IV 89}
 ({\bf A}_2)_{n{\bf K}n^{\prime\prime}{\bf K}^{\prime\prime}} = \frac{1}{2} {\bf B}\times \Big[i{\bf \nabla}_{\bf k}
 \delta_{nn^{\prime\prime}} + {\bold R}_{nn^{\prime\prime}}({\bf k}) \Big] \delta_{{\bf K}{\bf K}^{\prime\prime}} \,.
\end{equation}
Using (\ref{Sec IV 85}) and (\ref{Sec IV 89}) in Eq.~(\ref{Sec IV 84}), it follows that
\begin{multline}\label{Sec IV 91}
 (V_1)_{n{\bf K}n^{\prime}{\bf K}^{\prime}} = - \frac{e}{2c} \Big\{({\bf B}\times i{\bf \nabla}_{\bf k}) \cdot {\bf v}_{nn^{\prime}}({\bf K})   \\
 + \sum_{n^{\prime\prime} \neq n} [{\bf B} \times {\bold R}_{nn^{\prime\prime}}({\bf k})] \cdot
 {\bf v}_{n^{\prime\prime}n^{\prime}}({\bf K}^{\prime}) \Big\} \delta_{{\bf K}{\bf K}^{\prime}}  \,.
\end{multline}
Noting that
the second term on the right-hand side of Eq.~(\ref{Sec IV 91}) can be written as
\begin{multline}\label{Sec IV 93}
 [{\bf B} \times {\bold R}_{nn^{\prime\prime}}({\bf k})] \cdot
 {\bf v}_{n^{\prime\prime}n^{\prime}}({\bf K}^{\prime}) = \\
 \frac{i}{\hbar} (\varepsilon^0_{n^{\prime\prime}{\bf k}} - \varepsilon^0_{n^{\prime}{\bf k}}) [{\bold R}_{nn^{\prime\prime}}({\bf k}) \times {\bold R}_{n^{\prime\prime}n^{\prime}}({\bf k})]
 \cdot {\bf B}   \,, \nonumber
\end{multline}
then the sum $\sum_{n^{\prime\prime} \neq n}(...)$ becomes $\sum_{n^{\prime\prime} \neq  n,n^{\prime}}(...)$ because of the properties of ${\bold R}_{nn^{\prime\prime}}({\bf k})$; therefore, within the WWA, we keep only terms in ($n,n^{\prime}$), so that Eq.~(\ref{Sec IV 91}) becomes
\begin{equation}\label{Sec IV 94}
 (V_1)_{n{\bf K}n^{\prime}{\bf K}^{\prime}} = \frac{e}{2ic} ({\bf B} \times {\bf \nabla}_{\bf K}) \cdot
 {\bf v}_{nn^{\prime}}({\bf K}) \delta_{{\bf K}{\bf K}^{\prime}}    \,.
\end{equation}

\subsection{Matrix elements of $V_2({\bold x},t)$}

We consider the matrix elements of $V_2({\bold x},t)$,
\begin{equation}\label{Sec IV 92a}
 (V_2)_{n{\bf K}n^{\prime}{\bf K}^{\prime}} = \frac{e^2}{2mc^2} \sum_{n^{\prime\prime}{\bf K}^{\prime\prime}}
 ({\bf A}_2)_{n{\bf K}n^{\prime\prime}{\bf K}^{\prime\prime}}
 ({\bf A}_2)_{n^{\prime\prime}{\bf K}^{\prime\prime}n^{\prime}{\bf K}^{\prime}}   \,,
\end{equation}
where the matrix elements $({\bf A}_2)_{n{\bf K}n^{\prime\prime}{\bf K}^{\prime\prime}}$ are given in Eq.~(\ref{Sec IV 89}). Using these matrix elements in Eq.~(\ref{Sec IV 92a}), we see that
\begin{multline}\label{Sec IV 93a}
 (V_2)_{n{\bf K}n^{\prime}{\bf K}^{\prime}} = \frac{e^2}{8mc^2} \Big\{({\bf B} \times i{\bf \nabla}_{\bf K})^2
 \delta_{nn^{\prime}}    \\
 + \Big[({\bf B} \times i{\bf \nabla}_{\bf K}) \cdot [{\bf B} \times {\bold R}_{nn^{\prime}}({\bf k})] +
 [{\bf B} \times {\bold R}_{nn^{\prime}}({\bf k})] \cdot ({\bf B} \times i{\bf \nabla}_{\bf K}) \Big]     \\
 + \sum_{n^{\prime\prime} \neq n,n^{\prime}} [{\bf B} \times {\bold R}_{nn^{\prime\prime}}({\bf k})] \cdot
 [{\bf B} \times {\bold R}_{n^{\prime\prime}n^{\prime}}({\bf k})] \Big\}  \delta_{{\bf K}{\bf K}^{\prime}}   \,.
\end{multline}
Dropping the terms with $n^{\prime\prime} \neq (n,n^{\prime})$ in the spirit of the WWA, Eq.~(\ref{Sec IV 93a}) results in
\begin{multline}\label{Sec IV 94a}
 (V_2)_{n{\bf K}n^{\prime}{\bf K}^{\prime}} = \frac{e^2}{8mc^2} \Big\{({\bf B} \times i{\bf \nabla}_{\bf K})^2
 \delta_{nn^{\prime}}  +   \\\Big[({\bf B} \times i{\bf \nabla}_{\bf K}) \cdot [{\bf B} \times {\bold R}_{nn^{\prime}}({\bf k})] +
 [{\bf B} \times {\bold R}_{nn^{\prime}}({\bf k})] \cdot ({\bf B} \times i{\bf \nabla}_{\bf K}) \Big]
 \Big\}    \\
 \times \delta_{{\bf K}{\bf K}^{\prime}}   \,.
\end{multline}
The matrix elements $(V_1)_{n{\bf K}n^{\prime}{\bf K}^{\prime}}$ and $(V_2)_{n{\bf K}n^{\prime}{\bf K}^{\prime}}$, reported in
(\ref{Sec IV 94}) and (\ref{Sec IV 94a}), retain only the contributions connecting ($n{\bf K},n^{\prime}{\bf K})$
and neglect contributions for $n^{\prime\prime} \neq (n,n^{\prime})$.

\subsection{Matrix elements of the Hamiltonian}

Having established all of the relevant matrix elements for our problem, we are now in a position to determine key physical quantities of the energy, $\varepsilon_{n{\bf K}}$. In order to express $\varepsilon_{n{\bf K}}(\beta)$ of Eq.~(\ref{Sec IV 68b}) in terms of the physical kinematic variables, we use the matrix elements $(V_1)_{n{\bf K}n^{\prime}{\bf K}^{\prime}}$ and $(V_2)_{n{\bf K}n^{\prime}{\bf K}^{\prime}}$ which have been evaluated in Eqs.~(\ref{Sec IV 94}) and (\ref{Sec IV 94a}), respectively. In particular, it follows from Eq.~(\ref{Sec IV 94a}) that
\begin{equation}\label{Sec IV 96}
 (V_2)_{n{\bf K}n{\bf K}} = \frac{e^2}{8mc^2} ({\bf B} \times i{\bf \nabla}_{\bf K})^2
 \delta_{{\bf K}{\bf K}^{\prime}}  \,.
\end{equation}
Thus, all terms in Eq.~(\ref{Sec IV 68b}) are straightforward to calculate except the term of the order of ${\bf B}^2$, which can be expressed as
\begin{multline}\label{Sec VI 99}
 \frac{e^2}{2mc^2}\Big[(A_2)^2_{n{\bf K}n{\bf K}} - \frac{2}{m} \sum_{n^{\prime} \neq n}
 \sum_{l} (A_{2l})_{n{\bf K}n{\bf K}} (p_l({\bf K}))_{nn^{\prime}}   \\
 \times \sum_{m} \frac{(A_{2m})_{n{\bf K}n{\bf K}} (p_m({\bf K}))_{nn^{\prime}}}{\varepsilon^{0}_{n^{\prime}{\bf k}} - \varepsilon^{0}_{n{\bf k}}} \Big]   \,;
\end{multline}
here, $A_{2l}$ and $p_l({\bf K})$ are the $l$th components of ${\bf A}_2$ and ${\bf p}({\bf K})$. Using the $f$-sum rule
\begin{multline}\label{}
 \frac{1}{m} \sum_{n^{\prime} \neq n} \frac{(p_i)_{nn^{\prime}} (p_j)_{n^{\prime}n} + (p_j)_{nn^{\prime}}(p_i)_{n^{\prime}n}}
 {\varepsilon^{0}_{n^{\prime}{\bf k}} - \varepsilon^{0}_{n{\bf k}}}  \\
 = \delta_{ij} - \frac{m}{\hbar^2} \frac{\partial^2 \varepsilon^{0}_{n{\bf k}}}{\partial k_i \partial k_j}\Big|_{{\bf k} = {\bf k}(t)},
\end{multline}
we see that the expression in (\ref{Sec VI 99}) reduces to
\begin{equation}\label{Sec VI 101}
 \frac{e^2}{2\hbar^2 c^2} \sum_{l,m =1}^3 A_{2l} A_{2m}
 \frac{\partial^2 \varepsilon^{0}_{n{\bf k}}}{\partial k_l \partial k_m}\Big|_{{\bf k} = {\bf k}(t)}    \,,
\end{equation}
where ${\bf A}_2 = (1/2i)({\bf B}\times \nabla_{{\bf k}})$.

\section{Evaluating of $({\partial U}/{\partial t})_{n{\bf K}n^{\prime}{\bf K}^{\prime}}$}\label{App-E}

In evaluating the matrix elements in question, one must consider the time dependence of the ABR [Eq.~(\ref{Sec IV 49})] with which the matrix elements are being taken. As such
\begin{multline}\label{AppE1}
 \left(\frac{\partial U}{\partial t}\right)_{n{\bf K}n^{\prime}{\bf K}^{\prime}} \equiv
 \int d{\bf x} \psi^{\ast}_{n{\bf K}}({\bf x},t) \frac{\partial {\hat U}}{\partial t} \psi_{n^{\prime}{\bf K}^{\prime}}({\bf x},t) =    \\
 \frac{\partial}{\partial t} \int d{\bf x} \psi^{\ast}_{n{\bf K}}({\bf x},t) {\hat U}  \psi_{n^{\prime}{\bf K}^{\prime}}({\bf x},t) - \Delta(t)  \,,
\end{multline}
where
\begin{equation}\label{AppE2}
 \Delta(t) =
 \int d{\bf x} \left (\frac{\partial \psi^{\ast}_{n{\bf K}}}{\partial t} {\hat U} \psi_{n^{\prime}{\bf K}^{\prime}} + \psi^{\ast}_{n{\bf K}} {\hat U} \frac{\partial \psi_{n^{\prime}{\bf K}^{\prime}}}{\partial t}
 \right)  \,.
\end{equation}
Now, since the explicit time dependence of $\psi_{n{\bf K}}({\bf x},t)$ gives
\begin{multline}\label{}
 i\hbar \frac{\partial}{\partial t}\psi_{n{\bf K}}({\bf x},t)
 = {\bf F}(t)\cdot {\bf R}_{n^{\prime}n}({\bf k}) \psi_{n^{\prime}{\bf K}}({\bf x},t)    \\
 + {\bf F}(t)\cdot \sum_{n^{\prime\prime} \neq n,n^{\prime}} {\bf R}_{n^{\prime\prime}n}({\bf k}) \psi_{n^{\prime\prime}{\bf K}}({\bf x},t), \nonumber
\end{multline}
then $\Delta(t)$ in Eq.~(\ref{AppE2}) becomes
\begin{multline}\label{AppE5}
 \Delta(t) = - \frac{1}{i\hbar}\int d{\bf x} {\bf F}(t)\cdot \big[ {\bf R}_{nn^{\prime}}({\bf k}) \psi^{\ast}_{n^{\prime}{\bf K}}
 {\hat U}\psi_{n^{\prime}{\bf K}^{\prime}}     \\
 + {\bf R}_{nn^{\prime}}({\bf k}^{\prime}) \psi^{\ast}_{n{\bf K}}
 {\hat U}\psi_{n{\bf K}^{\prime}} \big] + \sum_{n^{\prime\prime} \neq n,n^{\prime}} (...)   \,.
\end{multline}
Here, in the spirit of the WWA used throughout, we drop the sum over $n^{\prime\prime} \neq (n,n^{\prime})$
and retain only term connecting $(n,n^{\prime})$. Thus, Eq.~(\ref{AppE1}) reduces to
\begin{multline}\label{AppE6}
 \left(\frac{\partial U}{\partial t}\right)_{n{\bf K}n^{\prime}{\bf K}^{\prime}}
 = \frac{\partial}{\partial t} (U)_{n{\bf K}n^{\prime}{\bf K}^{\prime}} +   \\
 \frac{1}{i\hbar} {\bf F}(t)\cdot \big[ {\bf R}_{nn^{\prime}}({\bf k})
 U_{n^{\prime}{\bf K}n^{\prime}{\bf K}^{\prime}} - {\bf R}_{nn^{\prime}}({\bf k}^{\prime})
 U_{n{\bf K}n{\bf K}^{\prime}} \big]    \,.
\end{multline}

\section{Matrix elements of $(\hat{h}_1 \pm i\hat{h}_2)$}\label{App-F}

In considering the matrix elements of $(\hat{h}_1 \pm i\hat{h}_2)$, where $\hat{h}_1$ and $\hat{h}_2$ are given by Eq.~(\ref{Sec IV 71b}), and noting that $\hat{U}$ of Eq.~(\ref{Sec IV 67}) is expressed in orders of perturbation theory in the magnetic field parameter, $\beta$,
it follows that we can express $\hat{h}_1$ and $\hat{h}_2$ in terms of $\hat{U}$ to $O({\bf B}^2)$ as
\begin{eqnarray}\label{AppF2}
  \nonumber
  \hat{h}_1 =  2\Big(\beta \frac{\partial \hat{U}_1}{\partial t}
  + \beta^2 \frac{\partial \hat{U}_2}{\partial t} \Big),    \\
  \hat{h}_2 =  \frac{\beta^2}{2} \Big(\hat{U}_1 \frac{\partial \hat{U}_1}{\partial t}
  + \frac{\partial \hat{U}_1}{\partial t} \hat{U}_1 \Big)          \,.
\end{eqnarray}
We showed in Appendix D that $\left(\partial U/\partial t\right)_{n{\bf K}n^{\prime}{\bf K}^{\prime}}$ can be expressed, to within the WWA, by Eq.~(\ref{AppE6}) which is used below.

In considering the diagonal matrix elements of $\hat{h}_1$ and $\hat{h}_2$ in (\ref{AppF2}), while using
Eqs.~(\ref{AppE6}), (\ref{DiagU1}) and (\ref{C19}), we find
\begin{eqnarray}\label{AppF4}
  \nonumber
  (h_1)_{n{\bf K}n{\bf K}} = -i\beta^2 \frac{\partial}{\partial t}(U_1^2)_{n{\bf K}n{\bf K}}   \,,  \\
  (h_2)_{n{\bf K}n{\bf K}} = \frac{\beta^2}{2}  \frac{\partial}{\partial t}(U_1^2)_{n{\bf K}n{\bf K}}   \,.
\end{eqnarray}
Thus, from (\ref{AppF4}), we obtain
\begin{equation}\label{AppF6}
  (h_1 \pm i h_2)_{n{\bf K}n{\bf K}} = -i \frac{\beta^2}{2}
  \left(\begin{array}{c} a \\ b \end{array} \right)
  \frac{\partial}{\partial t}(U_1^2)_{n{\bf K}n{\bf K}}   \,,
\end{equation}
where $a = 1$ and $b = 3$ refer to $``$$+$$"$ and $``$$-$$"$, respectively. Note that in (\ref{AppF6}),
\begin{multline}\label{}
  (U_1^2)_{n{\bf K}n{\bf K}} = \sum_{n^{\prime}{\bf K}^{\prime}} (U_1)_{n{\bf K}n^{\prime}{\bf K}^{\prime}}
  (U_1)_{n^{\prime}{\bf K}^{\prime}n{\bf K}}     \\
  = \sum_{n^{\prime}\neq n} |(U_1)_{n{\bf K}n^{\prime}{\bf K}}|^2 , \nonumber
\end{multline}
where $(U_1)_{n{\bf K}n^{\prime}{\bf K}^{\prime}}$ is given by Eq.~(\ref{offU1}) in Appendix B and $(V_1)_{n{\bf K}n^{\prime}{\bf K}^{\prime}}$,
which appears in this equation, is given by Eq.~(\ref{Sec IV 94}) in Appendix C.
Details of the calculations are found in Appendices B and C. Thus, $(U_1^2)_{n{\bf K}n{\bf K}}$ can be written as
\begin{subequations}
\begin{equation}\label{AppF8a}
  (U_1^2)_{n{\bf K}n{\bf K}} = \left(\frac{e}{2c}\right)^2 \sum_{n^{\prime}\neq n}
  \left|\frac{[{\bf B} \times {\bf \nabla}_{\bf K}] \cdot{\bf v}_{nn^{\prime}}({\bf k})}
  {\varepsilon^0_{n{\bf k}} - \varepsilon^0_{n^{\prime}{\bf k}}} \right|^2    \,,
\end{equation}
where ${\bf v}_{nn^{\prime}}({\bf k})$ is defined in Eq.~(\ref{voffdiag}) and depends on ${\bold R}_{nn^{\prime}}({\bf k})$, the interband coupling matrix element. Also, the time dependence of $(U_1^2)_{n{\bf K}n{\bf K}}$ in Eq.~(\ref{AppF8a}) is governed by
${\bf v}_{nn^{\prime}}({\bf k}) / ({\varepsilon^0_{n{\bf k}} - \varepsilon^0_{n^{\prime}{\bf k}}})$. Hence, the diagonal matrix elements of
$(\hat{h}_1 \pm i\hat{h}_2)$ depend on all states $n^{\prime}\neq n$.
Since $(U_1^2)_{n{\bf K}n{\bf K}}$ of Eq.~(\ref{AppF8a}) is a key operator expression in Eqs.~(\ref{Sec IV 75}) and (\ref{Sec IV 76}), we note that $(U_1^2)_{n{\bf K}n{\bf K}} F$ [$F = F({\bf x},{\bf K},t)$ is arbitrary] can be written as
\begin{equation}\label{AppF8b}
  (U_1^2)_{n{\bf K}n{\bf K}}F = F(U_1^2)_{n{\bf K}n{\bf K}} + \sum_{n^{\prime}\neq n} \Pi_{nn^{\prime}}\{F\} + O({\bf B}^3),
\end{equation}
where
\begin{equation}\label{AppF8c}
 \Pi_{nn^{\prime}}\{F\} = {\bf \omega}_{nn^{\prime}}\cdot {\bf \nabla}_{\bf K}F  + \alpha_{nn^{\prime}}F + {\bf \gamma}^0_{nn^{\prime}} \cdot {\bf \nabla}_{\bf K}({\bf \gamma}_{nn^{\prime}}\cdot {\bf \nabla}_{\bf K}F)  \,;
\end{equation}
\end{subequations}
here,
\begin{subequations}
\begin{multline}\label{AppF9a}
 \omega_{nn^{\prime}} = \Big(\frac{e}{2\hbar c}\Big)^2 \frac{1}{g_{nn^{\prime}}^2}({\bf B}\times {\bf \nabla}_{\bf K}) \cdot \big(g_{nn^{\prime}}{\bf R}_{nn^{\prime}}^{\ast}{\bf \gamma}_{nn^{\prime}} \\
 + g_{nn^{\prime}}{\bf R}_{nn^{\prime}}{\bf \gamma}^{\ast}_{nn^{\prime}} \big) \,,      \\
 \alpha_{nn^{\prime}} = \Big(\frac{e}{2\hbar c}\Big)^2 \frac{1}{g_{nn^{\prime}}^2}
 {\bf \nabla}_{\bf K} \Big[({\bf B}\times {\bf \nabla}_{\bf K}) \cdot (g_{nn^{\prime}}{\bf R}_{nn^{\prime}})\Big]\cdot{\bf \gamma}^{\ast}_{nn^{\prime}}    \,,   \\
 {\bf \gamma}^0_{nn^{\prime}} = \Big(\frac{e}{2\hbar c}\Big)^2 \frac{1}{g_{nn^{\prime}}^2} {\bf \gamma}^{\ast}_{nn^{\prime}}, \;
 {\bf \gamma}_{nn^{\prime}} = g_{nn^{\prime}}{\bf R}_{nn^{\prime}}\times{\bf B}\,,
\end{multline}
with $g_{nn^{\prime}} = \varepsilon^0_{n{\bf K}} - \varepsilon^0_{n^{\prime}{\bf K}}$. From Eq.~(\ref{AppF8b}), it follows
\begin{multline}\label{AppF9b}
 \Big(\frac{\partial}{\partial t}\left(U_1^2 \right)_{n{\bf K}n{\bf K}}\Big) F = F \frac{\partial}{\partial t}\left(U_1^2 \right)_{n{\bf K}n{\bf K}}   \\
 + \sum_{n^{\prime}\neq n} \Big[\frac{\partial}{\partial t} \Pi_{nn^{\prime}}\{F\} - \Pi_{nn^{\prime}}\{\frac{\partial F}{\partial t}\}  \Big]  \,.
\end{multline}
\end{subequations}

In considering the off-diagonal matrix elements of $(\hat{h}_1 \pm i\hat{h}_2)$, we again analyze $\hat{h}_1$ and $\hat{h}_2$ of Eq.~(\ref{AppF2}) by utilizing matrix elements $\left(\partial U/\partial t \right)_{n{\bf K}n^{\prime}{\bf K}^{\prime}}$ of  Eq.~({\ref{AppE6}}) in Appendix D.
First, in the evaluation of $(h_1)_{n{\bf K}n^{\prime}{\bf K}^{\prime}}$, we need to evaluate $(\partial U_{1,2}/\partial t)_{n{\bf K}n^{\prime}{\bf K}^{\prime}}$. We observe, to lowest order in the WWA, that
\begin{multline}\label{AppF9}
 \left(\frac{\partial U_l}{\partial t}\right)_{n{\bf K}n^{\prime}{\bf K}^{\prime}}
 = \frac{\partial}{\partial t} (U_l)_{n{\bf K}n^{\prime}{\bf K}^{\prime}} +   \\
 \frac{1}{i\hbar} {\bf F}(t)\cdot \big[ {\bf R}_{nn^{\prime}}({\bf k})
 (U_l)_{n^{\prime}{\bf K}n^{\prime}{\bf K}^{\prime}} - {\bf R}_{nn^{\prime}}({\bf k}^{\prime})
 (U_l)_{n{\bf K}n{\bf K}^{\prime}} \big]    \,,
\end{multline}
where $l = 1, 2$. Since $(U_1)_{n{\bf K}n^{\prime}{\bf K}^{\prime}}$ is given in (\ref{offU1}) and (\ref{Sec IV 94}), which gives $(U_1)_{n{\bf K}n{\bf K}^{\prime}} = 0$, it then follows
\begin{equation}\label{AppF10}
 \left(\frac{\partial U_1}{\partial t}\right)_{n{\bf K}n^{\prime}{\bf K}^{\prime}}
 = \frac{\partial}{\partial t} (U_1)_{n{\bf K}n^{\prime}{\bf K}^{\prime}}    \,.
\end{equation}
Now, from normalization condition, $(U_2)_{n{\bf K}n{\bf K}} = -(i/2)(U_1^2)_{n{\bf K}n{\bf K}}$, and
\begin{multline}\label{AppF12}
  (U_1^2)_{n{\bf K}n{\bf K}^{\prime}} = \sum_{n_1{\bf K}_1} (U_1)_{n{\bf K}n_1{\bf K}_1}
  (U_1)_{n_1{\bf K}_1n{\bf K}^{\prime}}    \\
  = \sum_{n_1\neq n} |(U_1)_{n{\bf K}n_1{\bf K}}|^2 \delta_{{\bf K}{\bf K}^{\prime}}   \,.
\end{multline}
It then follows
\begin{subequations}
\begin{multline}\label{AppF13a}
 \left(\frac{\partial U_2}{\partial t}\right)_{n{\bf K}n^{\prime}{\bf K}^{\prime}}
 = \frac{\partial}{\partial t} (U_2)_{n{\bf K}n^{\prime}{\bf K}^{\prime}}   \\
 + \frac{1}{\hbar} {\bf F}(t)\cdot {\bf R}_{nn^{\prime}}({\bf k}) {\cal G}_{nn^{\prime}}({\bf K}) \delta_{{\bf K}{\bf K}^{\prime}}    \,,
\end{multline}
where
\begin{equation}\label{AppF13b}
 {\cal G}_{nn^{\prime}}({\bf K}) = \frac{1}{2} \sum_{n_1 \neq n,n^{\prime}}
 \left[|(U_1)_{n{\bf K}n_1{\bf K}}|^2 - |(U_1)_{n^{\prime}{\bf K}n_1{\bf K}}|^2\right]   \,.
\end{equation}
\end{subequations}

Thus, we obtain making use of (\ref{AppF10}), (\ref{AppF13a})
\begin{subequations}
\begin{multline}\label{AppF14a}
 (h_1)_{n{\bf K}n^{\prime}{\bf K}^{\prime}} = 2\Big[\beta \frac{\partial}{\partial t} (\hat{U}_1)_{n{\bf K}n^{\prime}{\bf K}^{\prime}} + \beta^2 \frac{\partial}{\partial t}
 (U_2)_{n{\bf K}n^{\prime}{\bf K}^{\prime}}    \\
 + \frac{\beta^2}{\hbar} {\bf F}\cdot {\bf R}_{nn^{\prime}}({\bf k}) {\cal G}_{nn^{\prime}}({\bf K}) \delta_{{\bf K}{\bf K}^{\prime}} \Big]
\end{multline}
and
\begin{equation}\label{AppF14b}
 (h_2)_{n{\bf K}n^{\prime}{\bf K}^{\prime}} = \frac{\beta^2}{2} \delta_{{\bf K}{\bf K}^{\prime}}
 \frac{\partial}{\partial t} N_{nn^{\prime}}({\bf K})  \,;
\end{equation}
\end{subequations}
here,
\begin{equation}\label{AppF15}
 N_{nn^{\prime}}({\bf K}) = \sum_{n_1 \neq n,n^{\prime}}
 (U_1)_{n{\bf K}n_1{\bf K}} (U_1)^{\ast}_{n^{\prime}{\bf K}n_1{\bf K}}   \,.
\end{equation}
It then follows from (\ref{AppF14a}) and (\ref{AppF14b}) that
\begin{multline}\label{AppF16}
 (\hat{h}_1 \pm i\hat{h}_2)_{n{\bf K}n^{\prime}{\bf K}^{\prime}} = \frac{\partial}{\partial t}
 \Big[2\beta ({U}_1)_{n{\bf K}n^{\prime}{\bf K}^{\prime}}
 + 2\beta^2 ({U}_2)_{n{\bf K}n^{\prime}{\bf K}^{\prime}}    \\
 \pm i\frac{\beta^2}{2} \delta_{{\bf K}{\bf K}^{\prime}} N_{nn^{\prime}}({\bf K}) \Big]  \\
 + \frac{2\beta^2}{\hbar} {\bf F}\cdot {\bf R}_{nn^{\prime}}({\bf k}) {\cal G}_{nn^{\prime}}({\bf K}) \delta_{{\bf K}{\bf K}^{\prime}} \,.
\end{multline}



\end{document}